%% file: draft.tex
\pdfoutput=1
\documentclass[12pt,hyperref,compress,notoc]{JHEP3}
\input{preamble.tex}
\title{NNLO QCD corrections to $pp \to \gamma^* \gamma^*$ in the large
$N_F$ limit}
\author{Charalampos Anastasiou$^a$, Juli\'an Cancino$^a$, Federico Chavez$^a$, Claude Duhr$^b$, Achilleas Lazopoulos$^a$, Bernhard Mistlberger$^a$, Romain M\"uller$^a$  
\\
  ${}^a$Institute for Theoretical Physics, ETH Zurich, \\
  8093 Zurich, Switzerland\\ 
${}^b$Institute for Particle Physics Phenomenology,  \\
  University of Durham, Durham, DH1 3LE, 
  U.K.
}

\input{abstract.tex}
\preprint{IPPP/14/76, DCPT/14/152}
\begin{document}

\input{intro.tex}

\input{general_setup.tex}
\input{reduction.tex}
\input{masters.tex}
\input{real_emission.tex}

\input{numerical_results.tex}

\input{conclusions.tex}
\input{acknowledgments.tex}

\appendix
\input{Appendix_Basis.tex}

\input{Appendix_Masters.tex}

%
%%%%%%	bibliography	%%%%%%
\input{citations.tex}
\end{document}

%% file: preamble.tex
\let\ifpdf\relax
\usepackage{ifpdf}
\usepackage{multirow}
\usepackage{color}
\let\normalcolor\relax
\usepackage{mathtools}
\usepackage[pdftex]{graphicx}
\usepackage{pdfpages}
\usepackage{shuffle}
\usepackage{xspace}

\usepackage{cite}
\usepackage{caption}
\usepackage{subcaption}
\allowdisplaybreaks[1]

\def\beq{\begin{equation}}   
\def\eeq{\end{equation}}
\def\bea{\begin{eqnarray}}  
\def\eea{\end{eqnarray}} 
\def\nn{\nonumber}
\def\eps{\epsilon}

\def\zb{\bar{z}}
\def\l{\lambda}

\def\lbar{\bar{\lambda}}

\def\e{\epsilon}

\newcommand{\be}{\begin{equation}}
\newcommand{\ee}{\end{equation}}
\newcommand{\eq}[1]{\begin{align}#1\end{align}}
\newcommand{\bs}{\begin{split}}
\newcommand{\es}{\end{split}}
\newcommand{\bg}{\begin{gathered}}
\newcommand{\eg}{\end{gathered}}

\newcommand{\ord}{\mathcal{O}}

\newcommand{\ud}{\mathrm{d}}

\newcommand{\ep}{\epsilon}
\newcommand{\zp}{\bar{r}}
\newcommand{\zz}{r}

\newcommand{\Li}{\mathrm{Li}}

\newcommand{\cP}{\mathcal{P}}
\newcommand{\cQ}{\mathcal{Q}}
\newcommand{\cR}{\mathcal{R}}

\newcommand{\cB}{\mathcal{B}}

\def\pic#1{\begin{picture}(90,60)(0,20)\includegraphics[scale=0.13]{figures/#1.png}\end{picture}}
\def\picx#1#2{\begin{picture}(#2,60)(0,20)\includegraphics[scale=0.13]{figures/#1.png}\end{picture}}

\renewcommand{\ln}{\log}

\newcommand{\DD}[2]{\mathcal{D}_{#1}(#2)}
\def\beq{\begin{equation}}   
\def\eeq{\end{equation}}
\def\bea{\begin{eqnarray}}  
\def\eea{\end{eqnarray}} 
\def\bal{\begin{align}}
\def\eal{\end{align}}
\def\nn{\nonumber}
\def\e{\epsilon}

\def\zbar{\bar{z}}
\def\l{\lambda}
\def\lbar{{\bar \lambda}}

\def\Nf{$N_F$ }

\def\msbar{\overline{\text{MS}}}

\newcommand{\order}[1]{\mathcal O \left( {#1} \right)}

%% file: abstract.tex
\abstract{
We compute the NNLO QCD corrections for the hadroproduction of a pair
of off-shell photons in the limit of a large number of quark flavors.
We perform a reduction of the two-loop amplitude to master
integrals and calculate the latter analytically as a Laurent series in the
dimensional regulator using modern integration methods. Real radiation
corrections are evaluated numerically with a direct subtraction of
infrared limits which we cast in a simple factorized form. 
The results presented here
constitute a gauge invariant part of the full NNLO corrections but are not necessarily dominant. 
We view this calculation as a step towards a complete computation. 
Our partial corrections to the total cross-section are about $1\%-3\%$ and
vary with the virtuality of the two off-shell photons. 
}

%% file: intro.tex
\section{Introduction}
\label{sec:introduction}

The Tevatron and the LHC have performed studies on a wide spectrum of processes which probe the 
electroweak sector of the Standard Model.  In particular, the production processes of a pair of electroweak 
gauge bosons \cite{Aaltonen:2009aa,Abazov:2013opa,Aad:2011kk,CDF:2011ab,Abazov:2012cj,D0:2013rca,Aad:2011xj,ATLAS:2012mec,Aad:2012twa,Aad:2011cx,Chatrchyan:2012sga,Chatrchyan:2014aqa} are of great interest as they allow to test the electroweak theory, 
constrain physics beyond the Standard Model 
and are background to signals of the Higgs boson decaying into $H \to WW, H \to ZZ$. 
While the bulk of the cross-sections is due to on-shell production of the $W$ or $Z$ bosons, 
off-shell production is interesting especially for the background estimation in Higgs 
searches.  

Diboson production has been studied theoretically in detail within perturbation theory, 
including next-to-leading-order (NLO) perturbative QCD effects~\cite{Ohnemus:1990za,Ohnemus:1991gb,Ohnemus:1991kk,Mele:1990bq,Frixione:1992pj,Frixione:1993yp,Baur:1995uv,Dixon:1998py,Dixon:1999di,Campbell:1999ah,Frixione:2002ik,Nason:2006hfa,Hamilton:2010mb,Hoche:2010pf,Melia:2011tj,Frederix:2011ss}, 
electroweak corrections~\cite{Billoni:2013aba,Accomando:2004de,Accomando:2005xp,Kuhn:2011mh,Bierweiler:2012kw,Bierweiler:2013dja,Campbell:2011bn,Baglio:2013toa} and resummation~\cite{Grazzini:2005vw,Dawson:2013lya,Wang:2013qua,Wang:2014mqt}.  The gluon initiated partonic cross-section which emerges for the first time 
at next-to-next-to-leading-order (NNLO) from the square of one-loop amplitudes has been singled out 
due to its numerical importance and it was computed in refs.~\cite{Dicus:1987dj,Glover:1988rg,Binoth:2005ua,Binoth:2006mf,Binoth:2008pr}.
Recently, a complete NNLO computation for $pp \to ZZ$ in the double pole approximation  was performed for the first time in ref.~\cite{Cascioli:2014yka}. 

In this publication we make a first step towards the computation of NNLO corrections for diboson production in the case of two off-shell electroweak gauge bosons.  
We restrict ourselves to computing the NNLO cross-section for an idealized process $pp \to \gamma^* \gamma^*$ in the limit 
of a large number of massless quark flavors \Nf. 
 
While the large-\Nf limit is not necessarily dominant it provides the opportunity of obtaining a gauge invariant part of the cross section and serves as an excellent means to treat and develop analytic and numeric methods.
We generate and  reduce the required amplitudes to master integrals using established methods~\cite{Nogueira:1991ex,Kuipers:2012rf,Laporta:2001dd,Anastasiou:2004vj}. We evaluate the latter by directly performing the integrations over the Feynman parameter following methods similar to the ones introduced in refs.~\cite{Brown:2008um,Ablinger:2012qm, Bogner:2012dn, Chavez:2012kn,Anastasiou:2013srw,Panzer:2014gra,Panzer:2014caa,Ablinger:2014yaa,Bogner:2014mha}. As a by-product, we construct a set of basis functions up to transcendental weight four with the correct branch cut structures which are sufficient to write down the answer for the class of integrals studied in this paper. Moreover,  the master integrals presented here have been computed independently and agree numerically with the results of refs.~\cite{Gehrmann:2014bfa,Papadopoulos:2014lla,Henn:2014lfa}.
For the calculation of real radiation corrections we apply a subtraction scheme based on a hierarchical parameterization of the phase-space and the universal collinear and infrared limits of the squared matrix-elements. 
All singularities cancel after adding the partonic cross-sections together and performing UV renormalization. 

This article is organized as follows.  In section~\ref{sec:setup} we present our notation and setup of the calculation. In section~\ref{sec:virtuals} we present the calculation of the two-loop amplitude in the large \Nf limit and we outline the computations of the relevant master integrals in section~\ref{sec:Fede}.  The computation of corrections due to 
real radiation  and our subtraction scheme are presented in sections~\ref{sec:reals} and~\ref{sec:reals2}.  We demonstrate the numerical impact of the contributions that we have computed here in section~\ref{sec:results}. We conclude in section~\ref{sec:conclusions}.

%% file: general_setup.tex
\section{Setup and notation}
\label{sec:setup}

In this article, we compute the fully differential cross-section at the LHC for the process of producing 
two idealized off-shell photons,
\begin{equation*}
P(P_1) + P(P_2) \to  \gamma^*(p_3) + \gamma^*(p_4) + X, 
\end{equation*}
where $P$ denotes a proton and $X$ is a shorthand notation for the associated QCD final-state radiation. 
In parentheses we indicate the momenta of the external particles. 

We compute cross sections which are fully differential in the momenta $p_3$ and $p_4$ of the photons, as well as in the momenta of the associated QCD jet radiation. The hadronic cross section for a generic 
observable  ${\mathcal J}$ is given by 
\begin{equation}
\sigma_{P_1 P_2 \to \gamma^* \gamma^*X}[\mathcal J] =  \sum_{i, j} \int_0^1 dx_1 dx_2\; f^b_i(x_1) f^b_j(x_2)\; 
 \sigma_{ij \to  \gamma^* \gamma^*X}[\mathcal J],
\end{equation}
with $\sigma_{ij \to  \gamma^* \gamma^*X}\left[\mathcal J\right]$ denoting the differential cross section for the process
\[
	i(p_1) + j(p_2) \to \gamma^* (p_3) + \gamma^*(p_4) + X,
\]
where $i$ and $j$ run over the parton flavors $g, u, \bar u, d, \bar d, \ldots$ relevant to this process, $p_1 = x_1 P_1$ and $p_2 = x_2 P_2$ are the momenta of the initial-state partons and $f^b_i(x)$  the bare 
parton distribution functions (PDFs). The function ${\mathcal J}$ depends on the final-state momenta and 
restricts the phase-space to the desired infrared-safe observable.   

The partonic cross sections are computed as a  perturbative expansion in the bare strong coupling constant $\alpha_s^b$, 
\begin{align}
	\sigma_{ij \to \gamma^* \gamma^* X}[\mathcal J] &= \sigma^{(0)}_{i j \to \gamma^* \gamma^*}[\mathcal J] \phantom{\sum_k}  & \left(\propto (\alpha^b_s)^0 \right)\nn \\ 
			  & \quad + \sigma^{(1)}_{i j \to \gamma^* \gamma^*}[\mathcal J] +  \sum_k \sigma^{(0)}_{i j \to \gamma^* \gamma^* k}[\mathcal J] & \left(\propto (\alpha^b_s)^1 \right) \nn \\
			& \quad + \sigma^{(2)}_{i j \to \gamma^* \gamma^*}[\mathcal J] + \sum_{k} \sigma^{(1)}_{i j \to \gamma^* \gamma^* k}[\mathcal J] + \sum_{k, l} \sigma^{(0)}_{i j \to \gamma^* \gamma^* kl}[\mathcal J] & \left(\propto (\alpha^b_s)^2 \right) \nn \\
			& \quad + \mathcal O((\alpha_s^b)^3)  \label{eq:partonicexp},
\end{align}
where $k$ and $l$ run over the final-state parton flavors.
The partonic cross sections with definite final state $\gamma^* \gamma^*$, $\gamma^* \gamma^* q$, $\gamma^* \gamma^* q' \bar q'$, etc, are given by:
\begin{align}
	\sigma^{(m)}_{ij \to \gamma^* \gamma^* \ldots} [\mathcal J] 
	& = \frac{1}{2 s } \int d \Phi_{12\to \gamma^* \gamma^* \ldots}\; \mathcal J (p_3, p_4, \ldots)\, |M_{q\bar q \to \gamma^* \gamma^* \ldots }|^2_{(m)} ,
\end{align}
where $s = 2 p_1 \cdot p_2$ is the partonic center-of-mass energy squared and 
$|M_{ij\to \gamma^*\gamma^* \ldots}|^2_{(m)}$ is the $m-$loop contribution to the 
$i j \to \gamma^*\gamma^* \ldots $ amplitude squared, summed over spin and colour and averaged over initial state quantum numbers. 
We compute the matrix elements using conventional dimensional regularization in $d=4-2\epsilon$ space-time dimensions. We assume that the photons do not decay and use the polarization sum:
\begin{align}
\sum_\lambda \epsilon_\lambda^\mu(p)^* \epsilon_\lambda^\nu(p) =& -g^{\mu\nu}+\frac{p^\mu p^\nu}{p^2},
\end{align}
where $p$ denotes the photon-momentum. We consider $N_F=5$ light quark flavours
and we ignore the effects of the top-quark both in the loops and the evolution of the strong coupling.  

In the present article, we compute the complete ${\cal O}(\alpha_s)$ corrections, while at 
${\cal O}(\alpha_s^2)$ we retain only the gauge-invariant terms which contribute in the 
$N_F \to \infty$ limit.
\begin{figure}[h!]
\centering
\begin{subfigure}{.3\textwidth}
\centering
\includegraphics[width=.7\textwidth]{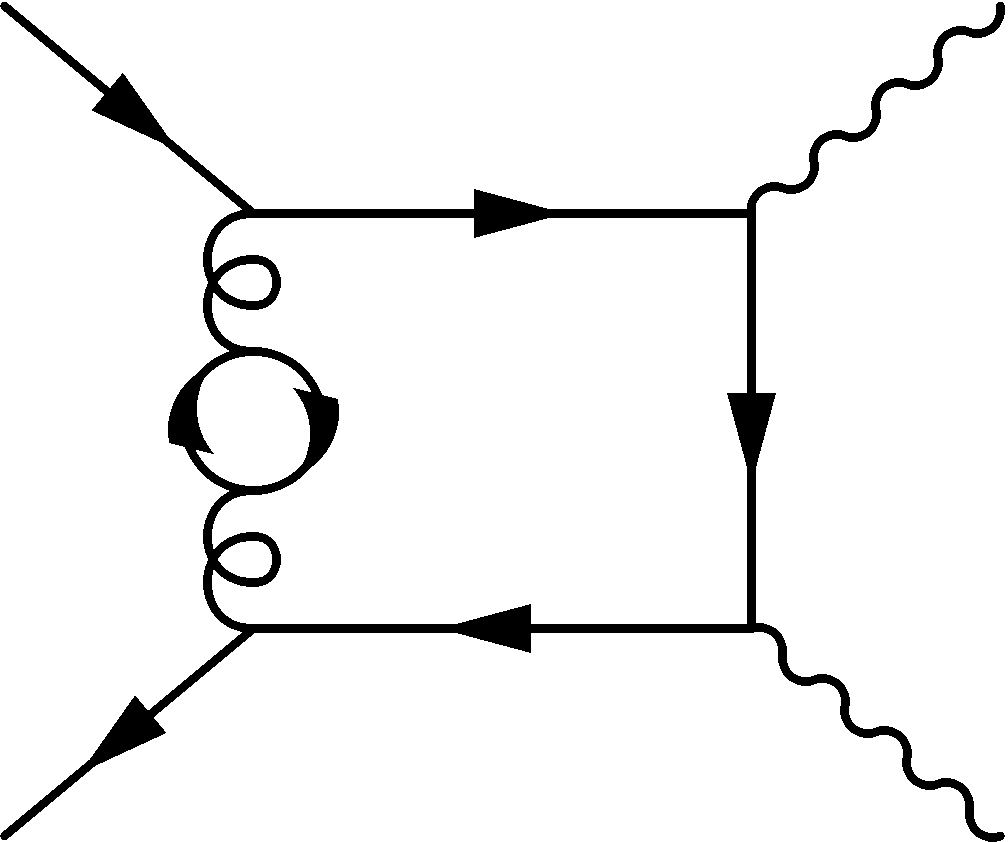} \hfill
\end{subfigure}
\begin{subfigure}{.3\textwidth}
\centering
\includegraphics[width=.9\textwidth]{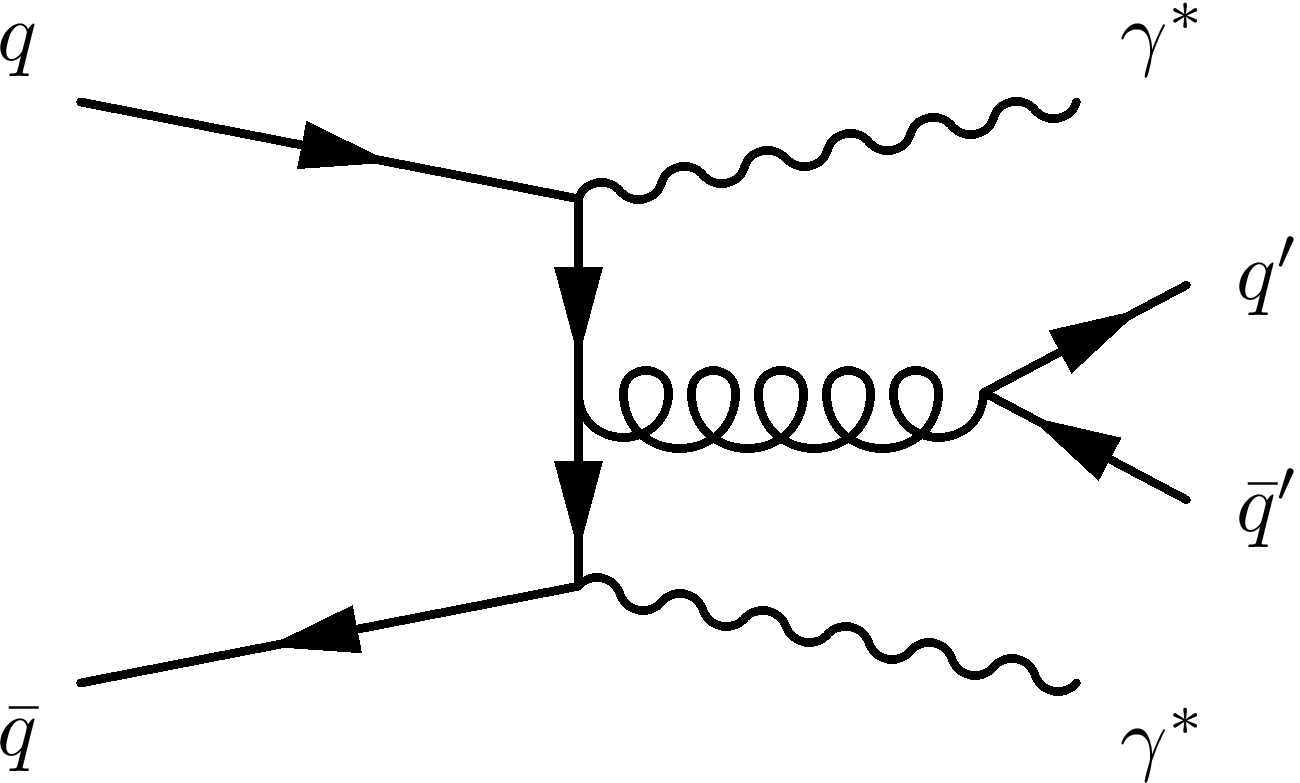}
\end{subfigure}
\caption{Sample tree and two-loop diagrams contributing to the NNLO corrections for 
$q\bar q\to\gamma^*\gamma^*$ in the large-\Nf limit.}
\label{fig:diagstypes}
\end{figure}
Some tree and two-loop diagrams that contribute to the NNLO large-\Nf correction are shown 
in ~figure~\ref{fig:diagstypes}. The two-loop diagrams contributing to the large \Nf limit 
are in one-to-one correspondence with the one-loop diagrams appearing at NLO, 
by replacing the gluon propagator by its one-loop self energy graph.
At NNLO, the partonic processes which contribute to the correction are
$q \bar{q} \to \gamma^* \gamma^* q^{\prime} {\bar q}^\prime$ and $q \bar{q} \to \gamma^* \gamma^* q {\bar q}$. In the latter process, we retain only the interference terms with two spin lines.

The Lorentz invariant phase space is given by
\beq
	d \Phi_{12\to \gamma^* \gamma^* \ldots} = \frac{d^d p_3}{(2 \pi)^{d-1}} \delta^+(p_3^2-m_3^2) \frac{d^d p_4}{(2 \pi)^{d-1}} \delta^+(p_4^2-m_4^2) \ldots (2\pi)^d \delta (p_1 + p_2 -p_3 -p_4 - \ldots),
\eeq
where `$\ldots$' indicates the phase-space measure of the massless final state partons. 
The virtualities of the external particles are
\beq 
	p_1^2 = 0, \qquad p_2^2 = 0, \qquad p_3^2 = m_3^2, \qquad p_4^2 = m_4^2,
\eeq
and we define the following Mandelstam variables and their ratios:
\eq{\bg
s = (p_1+p_2)^2, \qquad t  = (p_1-p_3)^2,  \qquad Q^2 = (p_3+p_4)^2, \\
u = \frac{m_3^2}{s}, \qquad v = \frac{m_4^2}{s}, \qquad w = \frac{t}{s}, \qquad z = \frac{Q^2}{s} .
\eg}

Ultraviolet renormalization is performed in the $\msbar$ scheme.
The bare  strong coupling constant 
$\alpha^b_s$ is given in terms of the renormalized coupling $\alpha_s(\mu)$ as
\begin{equation}\label{equ:catani_ren}
	\alpha^b_s  \, S_{\epsilon}  = \alpha_s(\mu)\, \left[ 1- \frac{\alpha_s(\mu)}{\pi} \frac{\beta_0}{\epsilon} + \left(\frac{\alpha_s(\mu)}{\pi} \right)^2 \left( \frac{\beta_0^2}{\epsilon^2} - \frac{\beta_1}{2 \epsilon}\right) \right] + \mathcal{O} \left(\alpha_s^4 (\mu)\right) ,
\end{equation}
where $\beta_0$ and $\beta_1$ are the first and second coefficients of the QCD beta function
\[
	\beta_0 = \frac{11 N_C - 4 T_R N_F}{12}, \qquad \beta_1 = \frac{17 N_C^2-10 N_C T_R N_F - 6 C_F T_R N_F}{24},
\]
with $C_F= \frac{N_c^2-1}{2 N_c}$, $N_C=3$, $T_R = \frac 1 2$; and $S_{\epsilon} = \mathrm{e}^{ \epsilon(\log 4 \pi - \gamma_E)}$. 
Since the Born cross section  is independent of $\alpha^b_s$, only the  $\alpha_s^2 (\mu)$ term of eq.~\eqref{equ:catani_ren} is required for renormalization.  

We absorb the initial-state collinear singularities into the parton densities in 
the $\msbar$-factorization scheme. The bare PDFs $f^b_i(x)$ are written in terms of the 
renormalized PDFs $f_j(x, \mu)$ as  
\beq
f^b_i(x) = f_i(x, \mu) + \left(\frac{\alpha_s(\mu)}{\pi}\right) [\; \Delta^{(1)}_{ij}\otimes f_j\; ] (x, \mu) + \left(\frac{\alpha_s(\mu)}{\pi}\right)^2 [\;\Delta^{(2)}_{ij} \otimes f_j\; ] (x, \mu) + \mathcal{O}(\alpha_s^3),
\eeq
where implicit summation over $j$ is understood, and the convolution integral is defined as
\beq
  [\;g \otimes f_j\; ](x, \mu) \equiv \int_0^1 d y d z\, \delta(x - y z) g(y) f_j(z, \mu)  .
\eeq
The kernels $\Delta^{(1,2)}_{ij}$ can be written in terms of the  Altarelli-Parisi splitting kernels as
\beq
\Delta^{(1)}_{ij}(z) = \frac{P^{(0)}_{ij}(z)}{\epsilon},
\label{eq:delta1}
\eeq
\beq
\Delta^{(2)}_{ij}(z) = \frac{P_{ij}^{(1)}(z)}{2\epsilon} + \frac{1}{2\e^2} \left( [\, P_{ik}^{(0)} \otimes P_{kj}^{(0)}\, ] (z) - \beta_0 P_{ij}^{(0)}(z) \right).
\label{eq:delta2}
\eeq
The splitting kernels relevant for this computation are
\beq
P_{qq}^{(0)}(z) = C_F \left( \DD{0}{1-z} + \frac{3}{4}\delta(1-z) - \frac 1 2 (1+z)\right),
\label{eq:apP0}
\eeq
\beq
P_{qg}^{(0)}(z) = \frac{1}{4}\left(z^2+(1-z)^2\right),
\label{eq:APqg}
\eeq
\beq
P_{qq}^{(1)}|_{N_F} = -\frac{N_FC_F}{18}\left[\delta(1-z)\left(\pi^2+\frac{3}{4}\right) + 10 \DD{0}{1-z} + 3\ln z\,\frac{1+z^2}{1-z}-11z+1\right].
\label{eq:apP1}
\eeq
The $\DD{n}{1-z}$ plus-distributions are defined as
\beq
  \int_0^1 \DD{n}{1-z} \phi(z) = \int_0^1 \log^n(1-z) \frac{\phi(z)-\phi(1)}{1-z}.
\eeq
For $P_{qq}^{(1)}$ we need only the terms proportional to $N_F$. We remark, however, that in the 
numerical evaluation of the PDFs and the strong coupling from their values at their initial scales we 
use the complete $\beta-$function and Altarelli-Parisi kernels and not just their \Nf parts.  

In the rest of this article we will set the renormalization and factorization scales to be equal, $\mu_f=\mu_r\equiv\mu$. The generic dependence on both scales can be easily restored by 
first setting $\mu=\mu_f$ and writing: 
\beq
\alpha_s(\mu_f)  =  \alpha_s(\mu_r)\left[ 1 + \frac{\alpha_s(\mu_r)}{\pi} \beta_0\, \log\frac{\mu_r^2}{\mu_f^2} \right] + \mathcal{O}\left(\alpha_s^2(\mu_r)\right) \,.
\eeq

%% file: reduction.tex
\section{Virtual corrections
}\label{sec:virtuals}

Ingredients of the NLO and NNLO corrections are the one-loop and two-loop amplitudes 
for the partonic process $q \bar{q} \to \gamma^* \gamma^*$.  We generate the required Feynman diagrams 
using QGRAF~\cite{Nogueira:1991ex} and then compute the interference of the one-loop amplitude and the tree-amplitude as well as the interference of the two-loop amplitude and the tree amplitude, summing 
over external-state colours and polarizations. We perform the Dirac and colour algebra with programs 
implemented in the FORM~\cite{Kuipers:2012rf} programming language.

From the interference of the tree and two-loop amplitudes, we keep only the terms which contribute to the large \Nf limit. These are expressed in terms of two-loop integrals of the form: 
\begin{align}
T_2(n_1,\dots,n_9,q_1,q_2,q_3)&\equiv \int\frac{d^dk}{i \pi^{\frac d 2}}\frac{d^dl}{i\pi^{\frac d 2}}\prod_{i=1}^9 D_i^{-n_i},
\end{align}
with 
\begin{align*}
	&D_1=k^2,&
	&D_2=(k+q_1)^2,&
	&D_3=(k+q_{12})^2,&
	&D_4=(k+q_{123})^2,&\\
	&D_5=l^2,&
	&D_6=(l+q_1)^2,&
	&D_7=(l+q_{12})^2,&
	&D_8=(l+q_{123})^2,\\
&D_9=(k-l)^2.
\end{align*}
where we have used the shorthand notation  $q_{1\cdots n}\equiv q_1+\cdots+q_n$ and the external momenta 
$q_i$ take the values: $(q_1, q_2, q_3) \in \left\{ (p_1,p_2,p_3), (p_1, p_2, p_4) \right\}$. The powers $n_i$ take 
integer values in the range $n_i \in [-4, 2]$. These integrals are not independent and they can be reduced to a basis of six master integrals. We use the program AIR~\cite{Anastasiou:2004vj} based on the Laporta algorithm~\cite{Laporta:2001dd}, and obtain the following two-loop master integrals:
\begin{align}
T_2(0,1,0,0,0,0,0,1,1,p_1,p_2,p_4)&\equiv
\pic{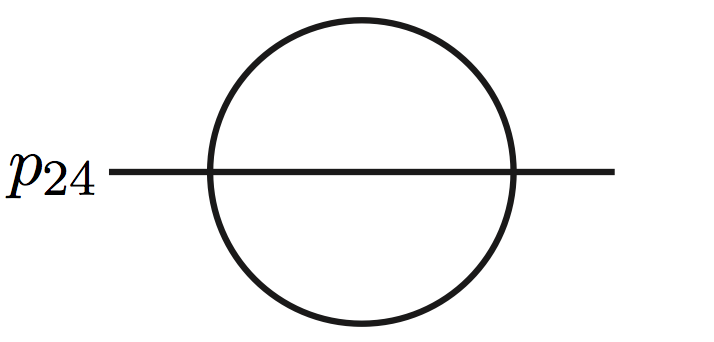}\\
T_2(0,1,0,0,1,0,1,0,1,p_1,p_2,p_4)&\equiv
\pic{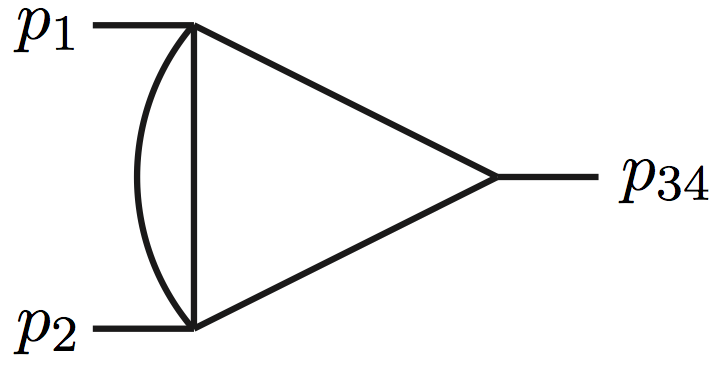}\\
T_2(0,1,0,0,1,0,0,1,1,p_1,p_2,p_4)&\equiv
\pic{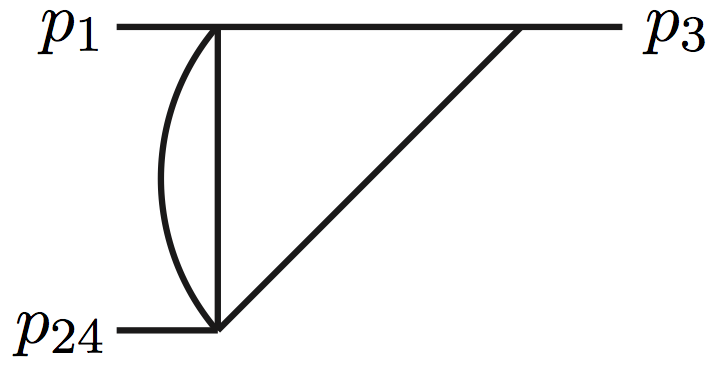}\\
T_2(0,1,0,0,0,0,1,1,1,p_1,p_2,p_4)&\equiv
\pic{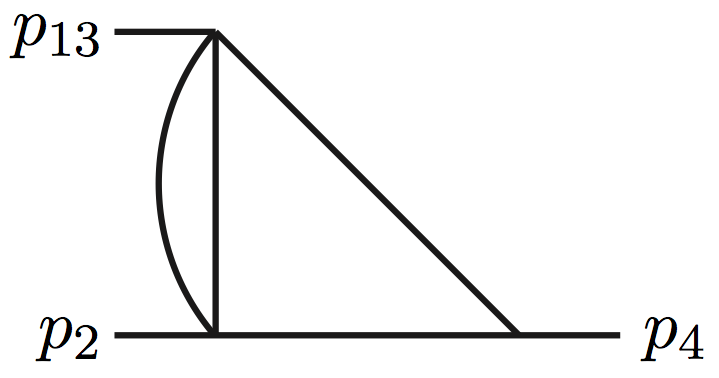}\\
T_2(0,1,0,0,1,0,1,1,1,p_1,p_2,p_4)&\equiv
\pic{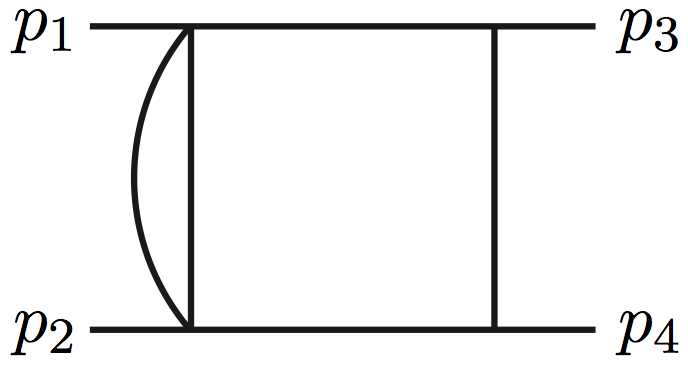}\\
T_2(0,1,0,0,1,0,1,1,2,p_1,p_2,p_4)&\equiv
\pic{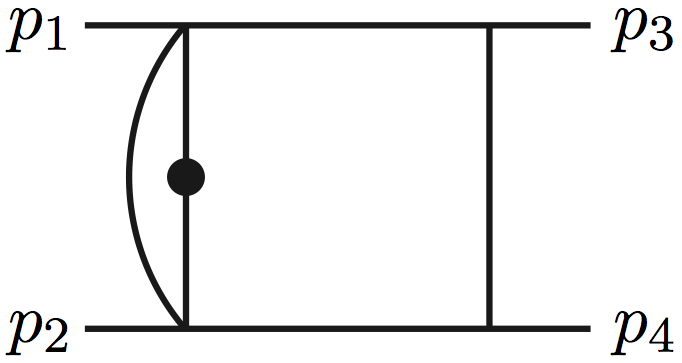}\\&\nonumber
\end{align}
The same integrals with $p_3$ and $p_4$ exchanged also appear in the two-loop amplitude.

Similarly, the interference of the tree and one-loop amplitudes can be expressed in terms of integrals of the form: 
\begin{align}
T_1(n_1,\dots,n_4,q_1,q_2,q_3)&\equiv \int\frac{d^dk}{i \pi^{\frac d 2}}\prod_{i=1}^4 D_i^{-n_i},
\end{align}
where the integer powers $n_i$ range in $[-4,1]$. The one-loop integrals are reduced to the following master 
integrals :
\begin{align}
T_1(1,0,1,0,p_1, p_2, p_4)  & \equiv \pic{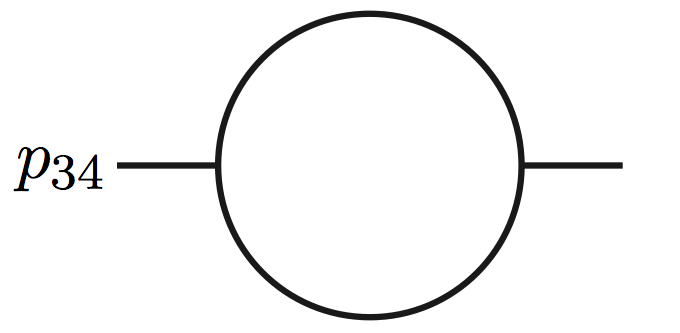} \\
T_1(1,0,1,1,p_1, p_2, p_4)  & \equiv \picx{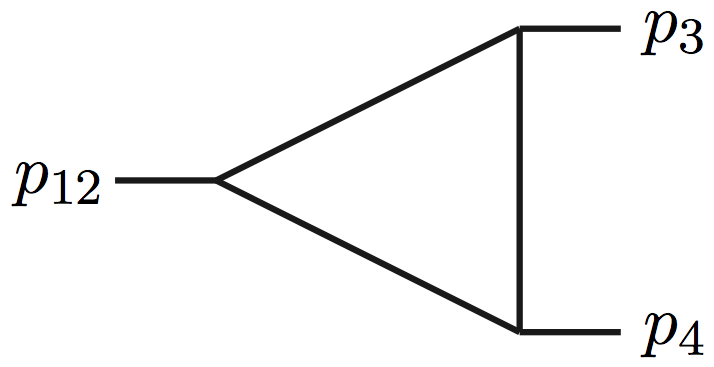}{100} \\ 
T_1(1,1,1,1,p_1, p_2, p_4)  & \equiv \pic{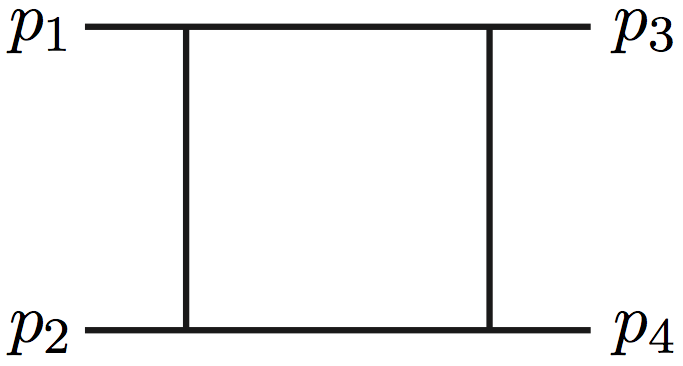} \\
& \nonumber 
\end{align}
The master integrals $T_1(1,0,1,1,p_1, p_2, p_3), T_1(1,1,1,1,p_1, p_2, p_3)$ also appear in the one-loop amplitude.

In the following section, we present a computation of the required master integrals, as well as of some master integrals which are needed for the full calculation beyond the large \Nf limit.   
The complete set of master integrals contributing to diboson production at two-loop order was recently computed in ref.~\cite{Henn:2014lfa}. We have performed an independent computation and confirm these results. 

%% file: masters.tex
\section{Master Integrals}
\label{sec:Fede}

In this section we present the analytic results for all master integrals that enter the $N_F$-part of the amplitude for $q\,\bar{q}\to\gamma^*\,\gamma^*$ up to two-loop order. 

\subsection{Analytic results in the Euclidean region}
We start by giving the analytic results for the master integrals in the Euclidean region where all consecutive Mandelstam invariants are negative. Note that in this region the variables $u$, $v$ and $w$ defined in section~\ref{sec:setup} are all positive. The results with the two virtualities $p_3^2$ and $p_4^2$ exchanged can easily be obtained from the replacement
\be
(u, v, w) \leftrightarrow
%\quad z \rightarrow 1-\zp, \quad \zp \rightarrow 1-z, 
(v,u, u+v-1-w)\,.
\ee
Before presenting our results, we first discuss some general properties of the integrals. 

In dimensional regularization with $d=4-2\eps$, every master integral is computed as a Laurent series in $\eps$, whose coefficients are expressed in terms of polylogarithmic functions. The simplest possible representatives of this class of functions are the ordinary logarithm and classical polylogarithms, defined by
\beq
\log x = \int_1^x\frac{\ud t}{t} {\rm~~and~~} \textrm{Li}_n(x) = \int_0^x\frac{\ud t}{t}\,\textrm{Li}_{n-1}(t)\,,
\eeq
with $\textrm{Li}_1(x) = -\log(1-x)$. However, more general functions can also appear. These are the \emph{multiple} polylogarithms~\cite{Goncharov:1998, Goncharov:2001}, defined by
\eq{\bg \label{G_def}
G(\vec{0}_n;\zz) \equiv \frac{1}{n!}\ln^n{\zz}{\rm~~and~~}G(a_1,\ldots,a_n;\zz) = \int\limits_0^\zz\frac{\ud t}{t-a_1}G(a_2,\ldots,a_n;t)\,,
\eg}
with $G(\zz) = 1$ and the arguments $a_i,\zz \in \mathbb{C}$. 
The number of elements of the vector $\vec{a}=(a_1\ldots,a_n)$ is called the \emph{weight} of the multiple polylogarithm. 
Note that up to weight three, all multiple polylogarithms can be expressed in terms of classical polylogarithms and ordinary logarithms. In particular, the two-loop  amplitude for $q\,\bar{q}\to\gamma^*\,\gamma^*$ in the large $N_F$ limit only involves polylogarithmic functions up to weight three (up to $\ord(\eps^0)$), and hence we can always express our two-loop amplitudes in terms of classical polylogarithms only. This greatly facilitates the numerical evaluation. This point will be discussed in more detail in section~\ref{sec:analytic_cont} when discussing the analytic continuation from the Euclidean region to the Minkowski region.

The arguments of the polylogarithms are in general algebraic functions of the Mandelstam invariants, and in particular they involve the square root $\sqrt{\lambda(1,u,v)}$, where $\lambda(a,b,c) = a^2 + b^2 + c^2 -2ab - 2ac - 2bc$ denotes the K\"all\'en function. 
A convenient parameterization which rationalises this square root is given by
\beq\label{eq:uvdef}
u = \zz\zp {\rm~~and~~} v = (1-\zz)(1-\zp)\,,
\eeq
or equivalently
\eq{\bg \label{\zz,zp}
\zz= \frac{1}{2}\left(1+u-v+\sqrt{\lambda(1,u,v)}\right) {\rm~~and~~} 
\zp= \frac{1}{2}\left(1+u-v-\sqrt{\lambda(1,u,v)}\right)\,.
\eg}
This choice of parameterization is inspired by ref.~\cite{Chavez:2012kn}, where it was argued that the variables $(\zz,\zp)$ define a natural set of variables for parameterizing the kinematics of a massless three-point function with all external legs off shell. These integrals naturally appear as master integrals in our case. Furthermore, it was shown in ref.~\cite{Chavez:2012kn} (see also refs.~\cite{Drummond:2012bg,Schnetz:2013hqa}) that in the region where $\lambda(1,u,v)<0$, such that $\zz$ and $\zp$ are complex conjugate to each other, massless three-point functions are described by single-valued functions in the complex $\zz$ plane. Indeed, it is well-known that massless loop integrals can only have branch cuts starting at points where one of the Mandelstam variables vanishes. The single-valuedness condition is equivalent to the condition that these functions have the correct physical branch cuts. The advantage of this approach is that for every weight, there is only a very limited set of single-valued functions. In ref.~\cite{Chavez:2012kn} a method was presented to construct these functions explicitly up to weight four in the case of massless three-point functions (see also refs.~\cite{Dixon:2013eka} for similar ideas). In particular, up to weight three only three functions can appear besides the ordinary logarithms, $\log u=\log(\zz\zp)$ and $\log v=\log(1-\zz)(1-\zp)$. Following ref.~\cite{Chavez:2012kn}, we denote these functions
by $\cP_2(\zz)$, $\cP_3(\zz)$,$\cP_3(1-\zz)$ and $\cQ_3(\zz)$. The functions $\cP_n(\zz)$ are closely related to the so-called Bloch-Wigner function,
\beq\label{Zagier}
\cP_n(\zz) \equiv \left\{\begin{array}{ll}
2 P_n(\zz)\,,&\textrm{ if } n \textrm{ odd}\,,\\
2 i P_n(\zz)\,,&\textrm{ if } n \textrm{ even}\,,\\
\end{array}\right. 
\eeq
with
\beq\label{Zagier2}
P_n(\zz) = \mathfrak{R}_n\left\{\sum_{k=0}^{n-1}\frac{2^k\,B_k}{k!}\,\ln^k|\zz|\,\Li_{n-k}(\zz)\right\}\,,
\eeq
where $B_k$ denotes the $k$-th Bernoulli number and $ \mathfrak{R}_n$ denotes the real part for odd $n$ and the imaginary part otherwise. Note that the function defined by eq.~\eqref{Zagier2} is a combination of classical polylogarithms without branch cuts for $r\in\mathbb{C}$, and it is therefor natural to call the functions~\eqref{Zagier2} the \emph{single-valued versions of the classical polylogarithms}. The function $\mathcal{Q}_3(\zz)$ is defined by
\eq{\label{eq:Q3}
\cQ_3(\zz) &= \frac{1}{2} \left[G\left(0,\frac{1}{\zp},\frac{1}{\zz},1\right)-G\left(0,\frac{1}{\zz},\frac{1}{\zp},1\right)\right]+\frac{1}{2} \Big[\text{Li}_3(1-\zz)-\text{Li}_3(1-\zp)\Big]\\
&\,+\frac{1}{4} \log |\zz|^2 \left[G\left(\frac{1}{\zz},\frac{1}{\zp},1\right)-G\left(\frac{1}{\zp},\frac{1}{\zz},1\right)\right] +\text{Li}_3(\zz)-\text{Li}_3(\zp)\nn \\
&\,+\frac{1}{4} \Big[\text{Li}_2(\zz)+\text{Li}_2(\zp)\Big] \log \frac{1-\zz}{1-\zp} +\frac{1}{4} \Big[\text{Li}_2(\zz)-\text{Li}_2(\zp)\Big] \log |1-\zz|^2\nn \\
&\,+\frac{1}{16} \log \frac{\zz}{\zp} \log^2 \frac{1-\zz}{1-\zp}+\frac{1}{8} \log^2 |\zz|^2 \log \frac{1-\zz}{1-\zp} +\frac{1}{4} \log|\zz|^2\,\log|1-\zz|^2\, \log \frac{1-\zz}{1-\zp}\nn \\
&\,+\frac{1}{16} \log^2|1-\zz|^2\log\frac{\zz}{\zp}-\frac{\pi ^2}{12} \log \frac{1-\zz}{1-\zp}\,.\nn
}
Up to weight three and two loops, all massless three-point functions can be written as linear combinations of (products of) these functions~\cite{Chavez:2012kn} (with coefficients that are $\mathbb{Q}$-linear combinations of $\zeta$ values).

The previous considerations, however, only apply to massless three-point functions. It is nevertheless straightforward to generalise these ideas to four-point functions with two adjacent off-shell legs. In appendix~\ref{app:basis} we present a way to construct a set of basis functions up to weight four with the correct physical branch cuts contributing to the large $N_F$ limit of the the $q\,\bar{q}\to\gamma^*\,\gamma^*$ amplitude at two loops. In the following we only concentrate on the set of basis functions up to weight three, which is relevant in the present case. Besides the functions defined in eq.~(\ref{Zagier} - \ref{eq:Q3}), we find six possible classical polylogarithms,
\beq\begin{split}
\textrm{Li}_2\left(1-\frac{u}{w}\right)\,,\qquad\textrm{Li}_3\left(1-\frac{u}{w}\right)\,,\qquad\textrm{Li}_3\left(1-\frac{w}{u}\right)\,,\\
\textrm{Li}_2\left(1-\frac{v}{w}\right)\,,\qquad\textrm{Li}_3\left(1-\frac{v}{w}\right)\,,\qquad\textrm{Li}_3\left(1-\frac{w}{v}\right)\,,
\end{split}\eeq
and two new functions $\cR_3^{\pm}(\zz,w)\equiv\cR_3^{\pm}(\zz,\zp,w)$, where the superscript `$\pm$' refers to the parity of the functions under the exchange $\zz\leftrightarrow\zp$, 
\eq{\bs
\cR_3^+(\zz,w) &= G(0,v,\zp(-1+\zz);w) + G(0,v,(-1+\zp)\zz;w)  \\ &+ G(0,u,(-1+\zp)\zz;w) - G\left(0,\frac{1}{\zp},\frac{1}{\zp};1\right) - G\left(0,\frac{1}{\zz},\frac{1}{\zz};1\right) \\ &- \left[G(u,(-1+\zp)\zz;w) + G(u,\zp(-1+\zz);w)\right]\ln{\frac{w}{u}} \\ &- \left[G(v,(-1+\zp)\zz;w) + G(v,\zp(-1+\zz);w)\right]\ln{\frac{w}{v}} \\ &+ G(0,u,\zp(-1+\zz);w)+ \Li_3\left(\frac{w}{\zp(-1+\zz)}\right) + \Li_3\left(\frac{w}{\zz(-1+\zp)}\right) \\ &+ \Li_3(1-\zp) + \Li_3(1-\zz) + 2\left[\Li_3(\zz) + \Li_3(\zp)\right] \\ &+ \left[\Li_2\left(\frac{w}{\zp(-1+\zz)}\right)+\Li_2\left(\frac{w}{(-1+\zp)\zz}\right)\right]\ln{\frac{uv}{w}} \\ &+ \left[\Li_2\left(1-\frac{u}{w}\right) + \Li_2\left(1-\frac{v}{w}\right)\right]\ln{(w^2+(1-u-v)w+uv)} \\ &+ \left[\Li_2(\zp) - \Li_2(\zz)\right]\left[\ln{\frac{\zz}{\zp}} - \ln{\left(\frac{-w-\zz+\zz\zp}{-w-\zp+\zz\zp}\right)}\right] \\ &+ \left[\Li_2(\zz) + \Li_2(\zp)\right]\left(\frac{1}{2}\ln{v} - \ln{u}\right) \\ &+\frac{3}{8}\ln^2{\frac{1-\zz}{1-\zp}}\ln{u}+\frac{1}{2}\ln^2{w}\ln{(w^2+(1-u-v)w+uv)} \\ &- \frac{1}{2}\ln{u}\ln{\frac{\zz}{\zp}}\ln{\frac{1-\zz}{1-\zp}} + \frac{1}{4}\ln{v}\ln{\frac{\zz}{\zp}}\ln{\frac{1-\zz}{1-\zp}} \\ &+ \frac{1}{2}\ln{u}\ln{\frac{1-\zz}{1-\zp}}\ln{\left(\frac{-w-\zz+\zz\zp}{-w-\zp+\zz\zp}\right)} \\ &- \frac{1}{2}\ln{u}\ln{v}\ln{(w^2+(1-u-v)w+uv)}\,,
\es}

\eq{
\cR&_3^-(\zz,w) = G(0,v,\zp(-1+\zz);w) - G(0,v,(-1+\zp)\zz;w) + G(0,u,\zp(-1+\zz);w) \nn \\ &- G(0,u,(-1+\zp)\zz;w) + G\left(0,\frac{1}{\zp},\frac{1}{\zp};1\right) - G\left(0,\frac{1}{\zz},\frac{1}{\zz};1\right) \nn \\ &+ \left[G(u,(-1+\zp)\zz;w) - G(u,\zp(-1+\zz);w)\right]\ln{\frac{w}{u}} \nn \\ &+ \left[G(v,(-1+\zp)\zz;w) - G(v,\zp(-1+\zz);w)\right]\ln{\frac{w}{v}} \nn \\ &+ \Li_3\left(\frac{w}{\zp(-1+\zz)}\right) - \Li_3\left(\frac{w}{\zz(-1+\zp)}\right) + \Li_3(1-\zp) - \Li_3(1-\zz) \\ &+ \left[\Li_2\left(\frac{w}{\zp(-1+\zz)}\right)-\Li_2\left(\frac{w}{(-1+\zp)\zz}\right)\right]\ln{\frac{uv}{w}} \nn \\ &+ \left[\Li_2\left(1-\frac{u}{w}\right) + \Li_2\left(1-\frac{v}{w}\right)\right]\left[\ln{\frac{\zz}{\zp}} - \ln{\frac{1-\zz}{1-\zp}} - \ln{\left(\frac{-w-\zz+\zz\zp}{-w-\zp+\zz\zp}\right)}\right] \nn \\ &+ \left[\Li_2(\zz)-\Li_2(\zp)\right]\ln{(w^2+(1-u-v)w+uv)} + \Li_2(\zp)\ln{(1-\zp)} \nn \\ & - \Li_2(\zz)\ln{(1-\zz)} - \frac{1}{8}\ln^2{\frac{1-\zz}{1-\zp}}\ln{\frac{\zz}{\zp}} \nn\\
&+ \left(\frac{1}{2}\ln{u}\ln{v} - \frac{1}{2}\ln^2{w}\right)\ln{\left(\frac{-w-\zz+\zz\zp}{-w-\zp+\zz\zp}\right)} \nn \\ &+ \left[\frac{1}{4}\ln{u}\ln{v} - \frac{1}{2}\ln^2{w} - \frac{1}{2}\ln{u}\ln{(w^2+(1-u-v)w+uv)}\right]\ln{\frac{1-\zz}{1-\zp}} \nn \\ &+ \frac{1}{8}\ln{\frac{\zz}{\zp}}\left(4\ln^2{w} - \ln^2{v} - 4\ln{u}\ln{v}\right) + \zeta_2\ln{\frac{1-\zz}{1-\zp}} \nn\,.
}
%Some comments about these expressions. 
First,
we emphasise that each of these functions has the correct branch-cut structure corresponding to a massless four-point function with two adjacent off-shell legs, i.e., they have branch cuts at most starting at points where one of the external Mandelstam invariants vanishes. Second, this set of functions is \emph{linearly independent}, i.e., it is not possible to express any of these functions as a linear combination of (products of) all the others. It is therefore justified to call these functions a set of \emph{basis functions}. As a consequence, \emph{all} master integrals contributing to the large $N_F$ part of the $q\,\bar{q}\to\gamma^*\,\gamma^*$ two-loop amplitude can be expressed as a unique linear combination of (products of) basis functions. The construction of these functions, as well as the proof that they form a basis, is given in appendix~\ref{app:basis}.

In the rest of this section we collect our results for the master integrals contributing to the large $N_F$ part of the $q\,\bar{q}\to\gamma^*\,\gamma^*$ two-loop amplitude. Details about the computation can be found in appendix~\ref{app:masters_computation}. All the expressions are valid in the Euclidean region, and the results are given in terms of the basis functions we have just defined. We explicitly show the results up to weight three. Analytic results up to weight four are provided as ancillary files with the {\tt arXiv} submission. 

We checked that our results satisfy the differential equations for the master integrals. Moreover the results were checked numerically with {\tt FIESTA}~\cite{Smirnov:2013eza}, which is based on the method of sector decomposition \cite{Binoth:2000ps} (the multiple polylogarithms were evaluated using {\tt GiNaC}~\cite{Bauer:2000cp, Vollinga:2004sn}. In addition, we have compared our results with existing results in the literature whenever available~\cite{Usyukina:1994iw,'tHooft:1978xw,Birthwright:2004kk,Davydychev:1999mq,Chavez:2012kn,Gehrmann:2014bfa,Henn:2014lfa,Papadopoulos:2014lla}.

\paragraph{One-loop integrals.} We start by summarising the one-loop integrals. The relevant one-loop two, three and four-point functions are given by
\begin{align}
\pic{bubble} &\,= \frac{c_{\Gamma}}{\ep(1-2\ep)}(-s)^{-\ep},\\
\picx{1looptri}{100} &\,= -2c_{\Gamma}\frac{\Gamma(1-2\ep)}{\Gamma(1-\ep)^2}(-s)^{-1-\ep}\frac{u^{-\ep}v^{-\ep}}{\zz-\zp}\Big\{\cP_2(\zz) + 2\,\ep\,\cQ_3(\zz) %\\ & + \ep^2\left[\left(\frac{1}{6}\ln u\ln v-\zeta_2\right)\cP_2(\zz) + 2\cQ_4^-(\zz)\right] 
 + \ord(\ep^3) \Big\},\\
 \nn\pic{1loopbox}&\, =c_{\Gamma}\frac{(-s)^{-2-\ep}}{w}\Bigg\{\frac{1}{\ep^2} + \frac{1}{\ep}\ln{\frac{uv}{w^2}} - \Big[2\Li_2\left(1-\frac{u}{w}\right) \\
&\, + 2\,\Li_2\left(1-\frac{v}{w}\right)+ \frac{1}{2}\ln^2{\frac{u}{v}}\Big] + \ep\Big[-2\,\cR_3^+(\zz,w) + 4\,\cP_3(\zz) \nn\\
&\,\nn+ 4\,\cP_3(1-\zz) - 6\,\left(\Li_3\left(1-\frac{u}{w}\right) + \Li_3\left(1-\frac{v}{w}\right)  + \Li_3\left(1-\frac{w}{u}\right) \right.\\
&\,\nn\left.+ \Li_3\left(1-\frac{w}{v}\right)\right) + 2\,\Li_2\left(1-\frac{u}{w}\right)\,\ln{\frac{u^3v}{w}}  + \frac{7}{6}\,\ln^3{u}\\ &\,+ 2\,\Li_2\left(1-\frac{v}{w}\right)\,\ln{\frac{v^3u}{w}} - \frac{4}{3}\,\ln^3{w} + \frac{7}{6}\,\ln^3{v} + 8\,\zeta_3\nn\\
&\,  \nn- \frac{1}{6}\,\ln^2u\,(\ln v+18\,\ln w) -\frac{1}{12}\,\ln^2v\,(11\,\ln u+36\,\ln w)  \\ 
 &\nn +4\,\ln^2w\,(\ln u+\ln v) -2\,\ln u\,\ln v\,\ln w+2\,\zeta_2\,\ln u\\
 &\,+4\,\zeta_2\,\ln v  \Big] + \ord(\ep^2) \Bigg\},
\end{align}
where $\gamma_E=-\Gamma'(1)$ denotes the Euler-Mascheroni constant and we introduced the usual normalization factor
\beq
c_\Gamma=\frac{\Gamma(1-\eps)^2\,\Gamma(1+\eps)}{\Gamma(1-2\eps)}\,.
\eeq
Note that all results are entirely expressed in terms of the basis functions defined at the beginning of this section, as expected.

The one-loop box has been previously computed up to the finite part in the $\ep$-expansion in ref.~\cite{Ohnemus:1990za}.

\paragraph{Two-loop integrals.}
In this subsection we give the analytic expression for the two-loop integrals. Besides the loop integrals necessary for the amplitudes presented in this work, we also display all the boxes with bubble insertions with two adjacent off-shell legs. The master integrals  are  presented up to the order in $\ep$ that corresponds to coefficients of weight up to three. The full results including coefficients of weight four can be found in the file attached as ancillary files to the {\tt arXiv} submission of the paper.

\eq{
\pic{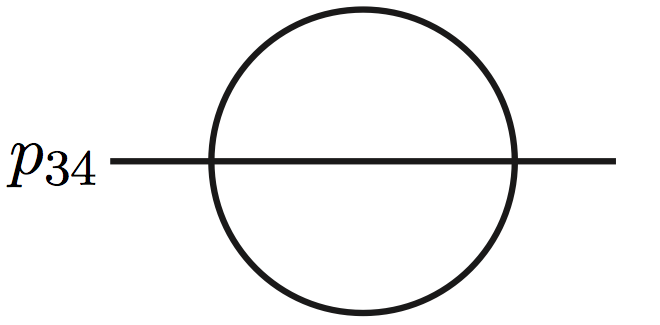} &= -c^2_{\Gamma}\frac{\Gamma(2\ep-1)\Gamma(1-2\ep)^2}{\Gamma(1-\ep)\Gamma(3-3\ep)\Gamma(1+\ep)}(-s)^{1-2\ep},
}
\eq{
\picx{1scaletri}{105} &= c^2_{\Gamma}(-s)^{-2\ep}\left\lbrace\frac{1}{\ep^2}\frac{1}{2} + \frac{1}{\ep}\frac{5}{2} + \frac{19}{2}+\zeta_2 + \ep\left[\frac{65}{2}+5\,\zeta_2-2\,\zeta_3\right]\right. \nn \\ &\quad \left. + \ep^2\left[\frac{211}{2}+19\,\zeta_2-10\,\zeta_3\right] + \ord(\ep^3)\right\rbrace,
}
\eq{
\picx{2scaletri}{98} &= -c_{\Gamma}^2(-su)^{-2\ep}\Big\{-\frac{1}{\ep^2}\frac{1}{2} - \frac{1}{\ep}\frac{5}{2} + \Li_2\left(1-\frac{u}{w}\right) +\frac{1}{2}\ln^2{\frac{u}{w}} -\frac{19}{2} \nn \\ &\quad +\ep\left[-2\Li_3\left(1-\frac{u}{w}\right) -\Li_3\left(1-\frac{w}{u}\right) + 2\Li_2\left(1-\frac{u}{w}\right)\ln{\frac{u}{w}} \right. \nn \\ & \qquad \left. + 5\Li_2\left(1-\frac{u}{w}\right) +\frac{2}{3}\ln^3{\frac{u}{w}} + \frac{5}{2}\ln^2{\frac{u}{w}} + 3\,\zeta_3 - \frac{65}{2}\right] \nn \\ & \qquad + \ord(\ep^2) \Big\},
}
\eq{\bs \label{B1}
&\pic{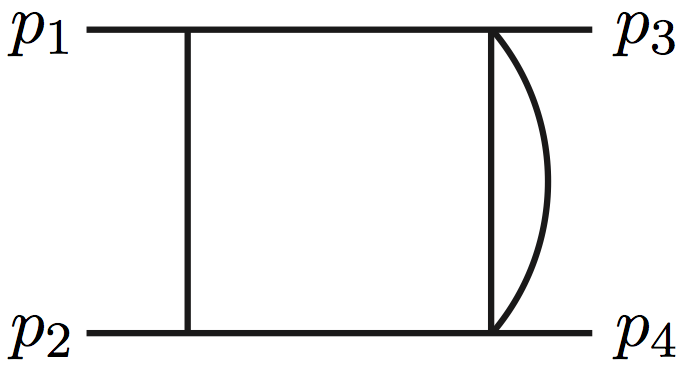} = \,c_{\Gamma}^2w^{-\ep}(1+2\ep)(-s)^{-1-2\ep} \left\lbrace -\frac{1}{\ep^3} + \frac{1}{\ep}\left[\Li_2\left(1-\frac{u}{w}\right) + \Li_2\left(1-\frac{v}{w}\right) - 4\right] \right. \\ \\ &\left. + \cR_3^+(\zz) + 4\left[\Li_3\left(1-\frac{u}{w}\right) + \Li_3\left(1-\frac{v}{w}\right) + \Li_3\left(1-\frac{w}{u}\right) + \Li_3\left(1-\frac{w}{v}\right)\right] + \right.  \\ &\left. +\Li_2\left(1-\frac{u}{w}\right)\ln{\left(\frac{w^2}{u^4v}\right)} + \Li_2\left(1-\frac{v}{w}\right)\ln{\left(\frac{w^2}{uv^4}\right)} -\frac{2}{3}\ln^3{u}+2\ln^2{u}\ln{w} \right. \\ &\left. +\frac{3}{8}\ln{u}\ln^2{v}+\ln{u}\ln{v}\ln{w}-\frac{5}{2}\ln{u}\ln^2{w} -\frac{2}{3}\ln^3{v}+2\ln^2{v}\ln{w} \right.  \\ &\left. -\frac{5}{2}\ln{v}\ln^2{w}+\ln^3{w}-\frac{1}{6}\pi^2\ln{u}-\frac{1}{3}\pi^2\ln{v}-4\zeta_3  + \ord(\ep) \right\rbrace ,
\es}
\eq{ \label{B1a}
\pic{B1a} = &\,c_{\Gamma}^2(uv)^{-\frac{3}{2}\ep}(1+2\ep)(-s)^{-1-2\ep}\frac{1}{\zz-\zp} \nn \\ &\times\Big\{-\frac{1}{\ep}2\cP_2(\zz) - 8\cQ_3(\zz) - \cR_3^-(\zz) + \mathcal{O}(\ep) \Big\},
}
\eq{ \label{B1b}
&\pic{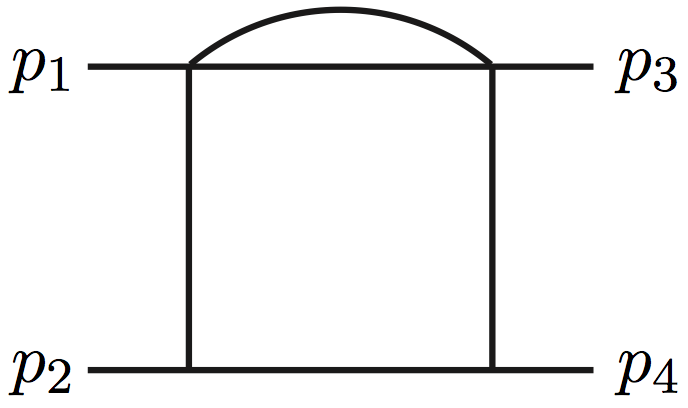} = \,c_{\Gamma}^2(vw)^{-\frac{1}{2}\ep}(1+2\ep)(-s)^{-1-2\ep}\frac{1}{v-w} \left\lbrace\frac{1}{\ep^2}\ln{\left(\frac{v}{w}\right)} - \frac{1}{\ep}\Li_2\left(1-\frac{v}{w}\right) \right. \nn \\ \nn \\ &\left. -\cR_3^+(\zz) + 2\cP_3(1-\zz) - 2\Li_3\left(1-\frac{u}{w}\right) - 4\Li_3\left(1-\frac{v}{w}\right) - 2\Li_3\left(1-\frac{w}{u}\right) \right. \nn \\ &\left. - 3\Li_3\left(1-\frac{w}{v}\right) + (2\ln{u}+\ln{v}-\ln{w})\Li_2\left(1-{\frac{u}{w}}\right) \right.  \\ &\left. + \left(\frac{5}{2}\ln{v}+\ln{u}-\frac{3}{2}\ln{w}\right)\Li_2\left(1-{\frac{v}{w}}\right) +(2\ln{v}+\ln{u})\zeta_2 \right. \nn \\ &\left. + \frac{1}{3}\ln^3{u}-\frac{17}{24}\ln^3{w}+\frac{13}{24}\ln^3{v}-\frac{13}{8}\ln^2{v}\ln{w} + \frac{17}{8}\ln{v}\ln^2{w}+4\ln{v}+4\,\zeta_3 \right. \nn \\ &\left. -\frac{5}{24}\ln{u}\ln^2{v}-\ln^2{u}\ln{w} + \frac{3}{2}\ln{u}\ln^2{w}-\ln{u}\ln{v}\ln{w}-4\ln{w} + \ord(\ep) \right\rbrace, \nn
}
\eq{ \label{B2}
&\pic{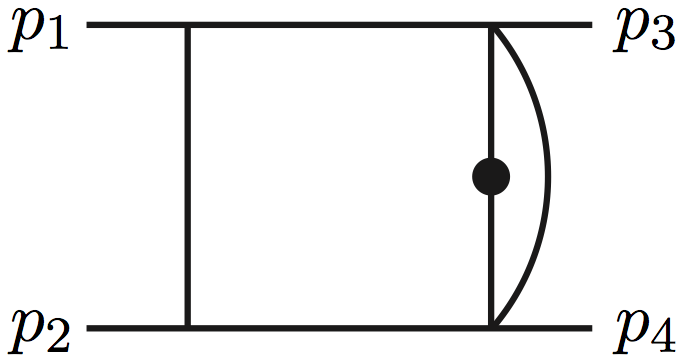} = \,c_{\Gamma}^2(-s)^{-2-2\ep}\frac{1}{w} \Big\{ -\frac{1}{\ep^3} + \frac{1}{\ep^2}[3\ln{w}-\ln{u}-\ln{v}] \nn \\ \nn \\ & +\frac{1}{\ep}\left[3\Li_2\left(1-{\frac{u}{w}}\right)+3\Li_2\left(1-{\frac{v}{w}}\right) -\frac{3}{2}\ln^2{w}+\ln^2{u}-\ln{u}\ln{v}+\ln^2{v} \right] \nn \\ & + 3\cR_3^+(\zz)-6\cP_3(\zz)-6\cP_3(1-\zz)+(3\ln{w}-12\ln{u}-3\ln{v})\Li_2\left(1-{\frac{u}{w}}\right) \nn \\ & +(-3\ln{u}-12\ln{v}+3\ln{w})\Li_2\left(1-{\frac{v}{w}}\right) +12\Li_3\left(1-{\frac{u}{w}}\right)+12\Li_3\left(1-{\frac{v}{w}}\right) \nn \\ & +12\Li_3\left(1-{\frac{w}{u}}\right)+12\Li_3\left(1-{\frac{w}{v}}\right)-(6\ln{v}+3\ln{u})\zeta_2+\frac{7}{2}\ln^3{w} \\ & -\frac{15}{2}\ln{u}\ln^2{w}-\frac{15}{2}\ln{v}\ln^2{w}+6\ln^2{u}\ln{w}+3\ln{u}\ln{v}\ln{w} + 6\ln^2{v}\ln{w} \nn \\ & -\frac{8}{3}\ln^3{u}+\frac{1}{2}\ln^2{u}\ln{v}+\frac{13}{8}\ln{u}\ln^2{v}-\frac{8}{3}\ln^3{v}-12\,\zeta_3 + \ord(\ep) \Big\}, \nn
}
\eq{ \label{B2a}
&\pic{B2a} = \,c_{\Gamma}^2(-s)^{-2-2\ep}\frac{1}{w} \left\lbrace -\frac{1}{\ep^3}\frac{1}{4} + \frac{1}{\ep^2}[\ln{w}-\frac{1}{2}\ln{u}-\frac{1}{2}\ln{v}] \right. \nn \\ \nn \\ &\left. +\frac{1}{\ep}\left[\frac{3}{2}\Li_2\left(1-\frac{v}{w}\right)+\frac{3}{2}\Li_2\left(1-\frac{u}{w}\right) + \frac{1}{2}\ln^2{\frac{u}{v}}-\frac{1}{4}\ln^2{\frac{u}{w}}-\frac{1}{4}\ln^2{\frac{v}{w}} -\frac{1}{2}\zeta_2\right] \right. \nn \\ &\left. +3\cR_3^+ -6\cP_3(\zz)-6\cP_3(1-\zz) +9\Li_3\left(1-\frac{u}{w}\right)+9\Li_3\left(1-\frac{v}{w}\right) \right. \\ &\left. +\frac{21}{2}\Li_3\left(1-\frac{w}{u}\right) +\frac{21}{2}\Li_3\left(1-\frac{w}{v}\right) + (-3\ln{v}+3\ln{w}-9\ln{u})\Li_2\left(1-\frac{u}{w}\right) \right. \nn \\ &\left. +(3\ln{w}-3\ln{u}-9\ln{v})\Li_2\left(1-\frac{v}{w}\right) + (-7\ln{v}+2\ln{w}-4\ln{u})\zeta_2 \right. \nn \\ &\left. +\frac{8}{3}\ln^3{w}-7\ln{u}\ln^2{w}-7\ln{v}\ln^2{w}+5\ln^2{u}\ln{w}+4\ln{u}\ln{v}\ln{w} \right. \nn \\ &\left. +5\ln^2{v}\ln{w}-\frac{11}{6}\ln^3{u}+\frac{9}{8}\ln{u}\ln^2{v}-\frac{11}{6}\ln^3{v} -11\zeta_3 + \ord(\ep) \right\rbrace, \nn
}
\eq{ \label{B2b}
&\pic{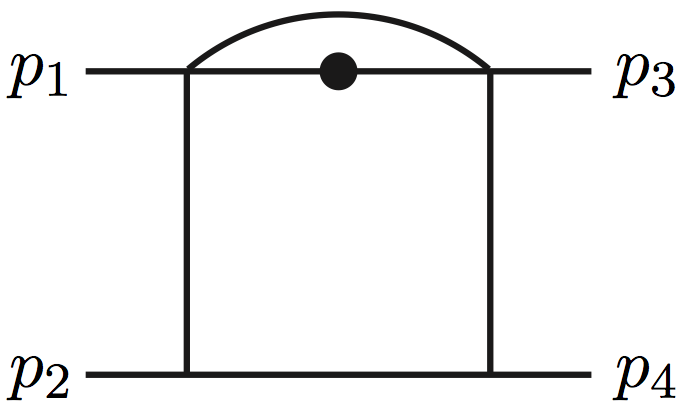} = \,c_{\Gamma}^2(-s)^{-2-2\ep}\frac{1}{w} \left\lbrace -\frac{1}{\ep^3}\frac{1}{2} + \frac{1}{\ep^2}\left[-\frac{1}{2}\ln{u}-\ln{v}+\frac{3}{2}\ln{w}\right] \right. \nn \\ \nn \\ &\left. +\frac{1}{\ep}\left[3\Li_2\left(1-\frac{v}{w}\right)+\frac{3}{2}\Li_2\left(1-\frac{u}{w}\right) + \frac{1}{2}\ln^2{\frac{u}{v}} \right] \right. \\ &\left. +3\cR_3^+ -6\cP_3(\zz)-6\cP_3(1-\zz) +\frac{15}{2}\Li_3\left(1-\frac{u}{w}\right)+12\Li_3\left(1-\frac{v}{w}\right) \right. \nn \\ &\left. + 9\Li_3\left(1-\frac{w}{u}\right) + 9\Li_3\left(1-\frac{w}{v}\right) + (-3\ln{v}+3\ln{w}-9\ln{u})\Li_2\left(1-\frac{u}{w}\right) \nn \right. \\ &\left. +(3\ln{w}-3\ln{u}-9\ln{v})\Li_2\left(1-\frac{v}{w}\right) + (-6\ln{v}-3\ln{u})\zeta_2 -9\,\zeta_3 \right. \nn \\ &\left. +2\ln^3{w}-6\ln{u}\ln^2{w}-6\ln{v}\ln^2{w}+\frac{9}{2}\ln^2{u}\ln{w}+3\ln{u}\ln{v}\ln{w} \right. \nn \\ &\left. +\frac{9}{2}\ln^2{v}\ln{w}-\frac{11}{6}\ln^3{u}+\frac{9}{8}\ln{u}\ln^2{v}-\frac{5}{3}\ln^3{v}  + \frac{1}{2}\ln^2{u}\ln{v} + \ord(\ep) \right\rbrace. \nn
}
After we completed the computation of these integrals, a complete basis of planar master integrals for the production of off-shell vector bosons was presented in ref.~\cite{Henn:2014lfa}. We have checked numerically that our results agree with the results of ref.~\cite{Henn:2014lfa}. An analytic expression for the integral~\eqref{B1b} was also published in ref.~\cite{Papadopoulos:2014lla}. In addition, we compared eqs.~\eqref{B1}, \eqref{B1a}, \eqref{B1b} and \eqref{B2a} numerically against the equal-mass results of ref.~\cite{Gehrmann:2014bfa}.

%%%%%%%%%%%%%%%%%%%%%%%%%%%%%%

\subsection{Analytic continuation into the physical region}
\label{sec:analytic_cont}

The results of the previous section are only valid in the Euclidean region, where $s,t,p_3^2,p_4^2<0$, such that $u,v,w>0$. In this section we perform the analytic continuation into the region defined by
\beq
s,p_3^2,p_4^2>0 {\rm~~and~~} t<0\,.
\eeq
The analytic continuation can be performed via the usual replacements
\be
-(s+i\varepsilon)\to s \,e^{-i\pi} {\rm~~and~~}
-(p_k^2+i\varepsilon)\rightarrow e^{-i\pi}\,p_k^2\,,\quad k=3,4\,.
\ee
This implies that the ratios $u$, $v$ and $w$ are analytically continued according to the prescription
\beq\label{eq:anal_cont}
u \to u {\rm~~and~~} v\to v{\rm~~and~~} w \to e^{+i\pi}\,\bar{w}\,,
\eeq
where we defined 
\beq
\bar{w} = -\frac{t}{s}>0\,.
\eeq
In ref.~\cite{Henn:2014lfa} it was shown how to analytically continue the multiple polylogarithmic functions to the physical region using this prescription. In the following, we present an alternative way of performing the analytic continuation, which will allow us in the end to express the large $N_F$ part of the amplitude entirely in terms of classical polylogarithms of weight three at most and with arguments in the interval $[0,1]$ everywhere in the physical phase space. Consequently, the classical polylogarithms are real and admit a convergent power series representation. The advantage of this representation is very fast and stable numerical evaluation. 

It turns out, however, that in order to obtain such a representation, we need to split the phase space into three different regions, such that a representation of the desired type exists in each region. We first discuss these different regions, and present our procedure to perform the analytic continuation in each region at the end of this section.

\paragraph{Definition of the regions.}
In the following we describe how to identify the parts of the physical region in which the results can be expressed in terms of classical polylogarithms with arguments inside the range $[0,1]$, where all the functions are convergent.
The results of the previous section were valid in the Euclidean region where $\lambda(1,u,v)<0$, and thus $\zz$ and $\zp$ complex conjugate to each other. Without loss of generality we may assume that in that region we have 
\be
\textrm{Im}\,\zz>0 \quad \text{and} \quad \textrm{Im}\,\zp<0.
\ee
It is easy to check that the physical phase space, however, corresponds to $\lambda(1,u,v)>0$, i.e. $\zz$ and $\zp$ real. In ref.~\cite{Chavez:2012kn} it was shown that the correct prescription for the analytic continuation from $\lambda(1,u,v)<0$ to $\lambda(1,u,v)>0$ while keeping $u$ and $v$ real is
\be
\zz \rightarrow \zz + i\ep \quad \text{and} \quad \zp \rightarrow \zp - i\ep.
\ee
It is sufficient to work out the analytic continuation for the basis functions. Moreover, since in the physical region $\sqrt{s} > m_3+m_4$, we must have $0<u,v<1$, which implies $0<\zp<\zz<1$~\cite{Chavez:2012kn}. In the following we show that we must furthermore have 
%$0<\bar{w}<\zz<1$ 
$\bar{r}<\bar{w}+r\bar{r}< r$ in the physical region.

In order to show this inequality, we parameterize the external momenta as
\eq{\bg
p_1 = \frac{\sqrt{s}}{2}\begin{pmatrix} 1 \\ \vec{e}_3 \end{pmatrix}, \qquad p_2 = \frac{\sqrt{s}}{2}\begin{pmatrix} 1 \\ -\vec{e}_3 \end{pmatrix}\,, \qquad
p_3 = \begin{pmatrix} E_3 \\ \vec{p}_3 \end{pmatrix}, \qquad p_4 = \begin{pmatrix} E_4 \\ \vec{p}_4 \end{pmatrix},
\eg}
where $\vec e_3=(0,0,1)^T$ and
\beq\begin{split}\label{energy}
E_3  &\,= \frac{\sqrt{s}}{2}(1+u-v) = \frac{\sqrt{s}}{2}(r+\bar{r}),\\
E_4 &\,= \frac{\sqrt{s}}{2}(1-u+v) = \frac{\sqrt{s}}{2}(2-r-\bar{r}).
\end{split}\eeq
We thus obtain for $\vec{p}_3$
\eq{
|\vec{p}_3|^2 = E_3^2 - m_3^2 = \frac{s}{4}\lambda(1,u,v) = \frac{s}{4}(\zz-\zp)^2 > 0,
}
and so
\eq{
 \vec{p}_3 = |\vec{p}_3| \begin{pmatrix} \sin\theta \\ 0 \\ \cos\theta \end{pmatrix} = \frac{\sqrt{s}}{2} (\zz-\zp) \begin{pmatrix} \sin\theta \\ 0 \\ \cos\theta \end{pmatrix}\,,
}
for some $\theta\in[0,\pi]$, and where we used rotational invariance to remove the dependence on the azimuthal angle.
At this point we can already conclude that $\lambda(1,u,v)>0$, i.e. $\zz$ and $\zp$ are indeed real and moreover we see from eq.~\eqref{energy} that $0<\zp<\zz<1$. Using this parameterization, we find
\eq{\bs
 \bar{w} &= \frac{1}{2}\left(\zz+\zp-2\zz\zp-(\zz-\zp)\cos\theta)\right) > 0\,,
%\bar{w} &= \frac{1}{2}\left[\zz+\zp-(\zz-\zp)\cos\theta\right] > 0\,,
\es}
and so $\zp < \bar{w} +\zz\zp< \zz$.

In the end we assert that the only non-trivial part in switching to the physical region is the analytic continuation of the basis functions depending on $w$. Besides the analytic continuation in $w$, some of the functions appearing in the basis functions are not defined for arbitrary values of $\zz,\zp$ and $w$. Consider for example
\be
\Li_n\left(-\frac{w}{\zp(1-\zz)}\right),
\ee
which develops an imaginary part if $-w>\zp(1-\zz)$. We find that we have to split some of the basis functions for physical values into three different regions
\eq{\bs
\tau > 1\,, \qquad\tau =1\,, \qquad\tau < 1\,,
\es}
where $\tau$ is defined through
\be
\bar{w}+\zz\zp = \zp + \tau(\zz-\zp)\,.
\ee
In deriving the analytic continuation of functions depending on $w$ we have to keep in mind these different regions.
Note that the physical phase space corresponds to $0 \le \tau \le 1$.

\paragraph{Analytic continuation of the functions.}
In this section we demonstrate how to perform the analytic continuation~\eqref{eq:anal_cont}. Our main goal is to obtain a representation of the amplitude in the physical region in terms of classical polylogarithms up to weight three with arguments lying in the range $[0,1]$, such that the polylogarithms admit a convergent power series representation.  
Technically speaking, we are looking for a functional equation which allows us to express the amplitude in terms of functions that are real in the physical region, and where all the imaginary parts are explicit. Functional equations among multiple polylogarithms are most conveniently described in terms the Hopf algebra of multiple polylogarithms (see appendix~\ref{app:basis}). In a nutshell, multiple polylogarithms admit a coproduct structure which allows to decompose a polylogarithm of weight $n$ into a sum of pairs of polylogarithms of weight $(k,n-k)$. It is then possible to find functional equations among multiple polylogarithms of weight $n$ recursively by first decomposing them into functions of lower weight, for which all relations are assumed to be known.

Let us illustrate this on some simple examples. First, we know that there are only three basis functions of weight one, and their analytic continuation follows immediately from eq.~\eqref{eq:anal_cont},
\beq
\log u \to\log u\,,\qquad \log v\to \log v\,,\qquad\log w \to \log\bar{w} + i\pi\,.
\eeq
Next, consider one of the basis functions of weight two, and let us consider $\textrm{Li}_2\left(1-\frac{u}{w}\right)$ as a representative example. In the physical region where $w=-\bar{w}<0$, the argument of the dilogarithm becomes greater than 1, and so the dilogarithm develops an imaginary part. Acting with the coproduct, we obtain,
\beq\begin{split}\label{eq:delta11}
\Delta_{1,1}\left[\Li_2\left(1-\frac{u}{w}\right)\right] &\,= -\log\frac{u}{w}\otimes\log{\left(1-\frac{u}{w}\right)}\\
&\,= -\log\frac{u}{\bar{w}}\otimes\log{\left(1+\frac{u}{\bar{w}}\right)} +i\pi\otimes\log{\left(1+\frac{u}{\bar{w}}\right)}\,,
\end{split}\eeq
where in the second line we used eq.~\eqref{eq:anal_cont}. Note that by construction every basis function of weight $n$ will only contain $\log u$, $\log v$ or $\log w$ in the first factor of its $(1,n-1)$ component of the coproduct (see appendix~\ref{app:basis} for details).
The imaginary part of this dilogarithm can immediately be read off from the second term,
\beq
i\pi\otimes\log{\left(1+\frac{u}{\bar{w}}\right)} = \Delta_{1,1}\left[i\pi\,\log{\left(1+\frac{u}{\bar{w}}\right)}\right]\,.
\eeq
At this point we need to find a real function whose coproduct matches the real part of eq.~\eqref{eq:delta11}. It is easy to check that
\beq
-\log\frac{u}{\bar{w}}\otimes\log{\left(1+\frac{u}{\bar{w}}\right)} = \Delta_{1,1}\left[-\Li_2\left(\frac{w}{w+u}\right) -\frac{1}{2}\log\left(1+\frac{u}{w}\right)\right]\,.
\eeq
Hence, we can conclude that
\beq
\Delta_{1,1}\left[\Li_2\left(1-\frac{u}{w}\right)\right] = \Delta_{1,1}\left[-\Li_2\left(\frac{w}{w+u}\right) -\frac{1}{2}\ln\left(1+\frac{u}{w}\right) + i\pi\log\left(1+\frac{u}{w}\right)\right]\,.
\eeq
We can thus conclude that the arguments of $\Delta_{1,1}$ are equal, up to (constant) terms that vanish when acting with $\Delta_{1,1}$. In order to determine this constant, we expand both side close to the branch point at $w=0$,
\beq\begin{split}
\Li_2&\left(1-\frac{u}{w}\right) = -\frac{1}{2}\log^2\frac{u}{w}-\zeta_2 + \ord(w)\\
&\,= -\frac{1}{2}\log^2\frac{u}{\bar{w}}+i\pi\,\log\frac{u}{\bar{w}} + 2\zeta_2+ \ord(\bar{w})\,,\\
-\Li_2&\left(\frac{\bar w}{\bar w+u}\right) -\frac{1}{2}\log\left(1+\frac{u}{\bar w}\right) + i\pi\log\left(1+\frac{u}{\bar w}\right)\\
&\,= -\frac{1}{2}\log^2\frac{u}{\bar{w}}+i\pi\,\log\frac{u}{\bar{w}}+\ord(\bar{w})\,.
\end{split}\eeq
where in the first line we have used the fact that $\log w=\log\bar{w}+i\pi$. Equating the two expressions, we see that
\beq
\Li_2\left(1-\frac{u}{w}\right) = -\Li_2\left(\frac{w}{w+u}\right) -\frac{1}{2}\ln\left(1+\frac{u}{w}\right) + i\pi\ln\left(1+\frac{u}{w}\right) + 2\zeta_2\,.
\eeq
Analogously we obtain the analytic continuation of all the basis function depending on $w$.

%% file: real_emission.tex
\section{Single-real contributions}\label{sec:reals}

We now turn our attention to the 
phase-space integrations over tree-level matrix elements for partonic processes, where the photon pair is produced in association with an additional parton in the final state: 
\begin{align*}
 q(p_1) + {\bar q}(p_2) &\to \gamma^*(p_3) + \gamma^*(p_4) + g(p_g), \\
 \bar q(p_1) + {q}(p_2) &\to \gamma^*(p_3) + \gamma^*(p_4) + g(p_g), \\
 q(p_1) + g(p_2)  &\to \gamma^*(p_3) + \gamma^*(p_4) + q(p_q),  \\
 \bar q(p_1) + g(p_2)  &\to \gamma^*(p_3) + \gamma^*(p_4) + \bar q(p_{\bar q}),  \\
 g(p_1) + q(p_2)  &\to \gamma^*(p_3) + \gamma^*(p_4) + q(p_q),  \\
 g(p_1) + \bar q(p_2)  &\to \gamma^*(p_3) + \gamma^*(p_4) + \bar q(p_{\bar q}).
\end{align*}
As before, we denote in brackets the momenta of the partons. These processes contribute at NLO to the hadronic process, and via renormalization and mass-factorization also 
at NNLO.  

\subsection{Quark-antiquark channels}

We first consider the channels with a $q \bar q$-pair in the initial state.
The corresponding tree-level cross section is given by
\beq
\label{eq:sigma_nlo}
%\sigma^R_{q \bar q}[ \mathcal J ] \equiv  
\sigma^{(0)}_{q \bar q \to \gamma^* \gamma^* g}[ \mathcal J ] = \frac{1}{2 s} \int d\Phi_{12\to \gamma^*\gamma^* g} \;\mathcal J (p_1, p_2, p_3, p_4, p_g)\, |M_{q \bar q\to \gamma^*\gamma^* g}|^2_{(0)},
\eeq
where $|M_{q \bar q\to \gamma^*\gamma^* g}|_{(0)}^2$ is the $q \bar q \to \gamma^*\gamma^* g$ tree matrix element squared, summed over spin and colour and averaged over initial-state quantum numbers. 
The case where the quark and the anti-quark are exchanged is identical.
The phase-space measure can be decomposed into a phase space producing a gluon and an intermediate off-shell particle in the final state with momentum $Q$ of virtuality $Q^2= z s$, and a phase space for the decay of the intermediate particle into two photons as
\beq
d\Phi_{12\to \gamma^*\gamma^* g} = \frac{s\, dz}{2\pi} d\Phi_{12\to Q g}d\Phi_{Q\to \gamma^*\gamma^*},
\eeq
where
\beq
	Q \equiv p_1 + p_2 - p_g = p_3 + p_4.
	\label{eq:cul}
\eeq
We parameterize the momentum of the gluon as
\begin{align} \label{eq:nlog}
p_g &= \zbar \bar{\lambda} \;p_1 + \zbar \lambda \;p_2 + \zbar \sqrt{s\lambda\bar{\lambda}}\; e_T, 
\intertext{such that}
Q &= (1-\zbar \bar \lambda) p_1 +(1-\zbar \lambda) p_2 - \zbar \sqrt{s\lambda\bar{\lambda}}\; e_T,
\end{align}
where $z, \lambda \in [0,1]$ and $e_T$ is a unit vector transverse to $p_1$ and $p_2$ in $d=4-2\epsilon$ dimensions. In this section and the following ones we use the shorthand notation
\beq
\bar{x} \equiv 1-x,
\eeq
for integration variables.
In the parameterization of eq.~\eqref{eq:nlog}, the phase space measure becomes
\beq
d\Phi_{12\to Qg}  =  \zbar \left( s \zbar^2\lambda\bar{\lambda} \right)^{-\epsilon}\frac{d\lambda d\Omega_{d-2}}{4 (2\pi)^{d-2}},
\label{eq:PS12togQ}
\eeq
where we use $d\Omega_{d-2}$ to denote the differential solid angle generating $e_T$.

The matrix element squared in the integrand of eq.~\eqref{eq:sigma_nlo} is singular in the collinear limits $p_g \parallel p_1$ and $p_g \parallel p_2$  
(corresponding to $\lambda \to 0$ and $\lambda \to 1$ respectively) 
and in the soft limit  $p_g \to 0$ (corresponding to $z \to 1$). 
The singular behaviour of matrix elements squared is universal \cite{Catani:1999ss}, in the sense that it is independent of the process under consideration. In particular, the formulae presented here are valid for any colourless final state in place of the two off-shell photons $\gamma^* \gamma^*$. Explicitly, we have
\beq
 |{M}_{q \bar q \to  \gamma^*\gamma^*g}|^2_{(0)} = 2 g_s^2 \mu^{2\epsilon}\frac{S_{qq}(z)}{\bar{z}^2\lambda} \frac{B_1(z)}{ z s} + \order{\lambda^0}, \qquad \text{as} \quad \lambda \to 0,
\label{eq:nlo_l0}
\eeq
\beq
 |{M}_{q \bar q \to  \gamma^*\gamma^*g}|^2_{(0)} = 2 g_s^2 \mu^{2\epsilon}\frac{S_{qq}(z)}{\bar{z}^2\bar{\lambda}} \frac{B_2(z)}{z s}+\order{\bar \lambda ^0}, \qquad \text{as} \quad \lambda \to 1,
 \label{eq:nlo_l1}
\eeq
where $g_s^2  = 4 \pi \alpha_s^b$, and the splitting kernel is given by
\beq
S_{qq}(z) = C_F\left( 2z + (1-\epsilon) \bar{z}^2 \right),
\label{eq:nloSplittingKernel}
\eeq
and
\begin{align}
B_1(z) &\equiv |M_{q \bar q\to \gamma^*\gamma^*}|^2_{(0)} (zp_1,p_2, p_3, p_4), \\ 
B_2(z) & \equiv |M_{q \bar q\to \gamma^*\gamma^*}|^2_{(0)} (p_1,zp_2, p_3, p_4).
\end{align}
In the above, the squared matrix elements for the Born process $q \bar q\to \gamma^*\gamma^*$ are evaluated with the momentum in the collinear direction rescaled by a factor of $z$.

Since we consider a colourless final state, the soft limit does not involve any colour correlations and can be simply written as
\beq \label{eq:soft_nlo}
	|{M}_{q \bar q \to  \gamma^*\gamma^*g}|^2_{(0)} =  2 g_s^2 \mu^{2\epsilon} \frac{2 C_F }{s \bar z^2 \lambda \bar \lambda}\, |{M}_{q \bar q \to  \gamma^*\gamma^*}|^2_{(0)}+ \order{\bar z ^{-1}}, \qquad \text{as} \quad z \to 1.
\eeq
Note that the sum of the collinear limits, given by equations \eqref{eq:nlo_l0} and \eqref{eq:nlo_l1}, reproduces eq.~\eqref{eq:soft_nlo} exactly in the limit where $z \to 1$. This means that although the matrix element squared is singular in the soft limit, no explicit subtraction of this singularity will be needed.

We now recast the partonic cross-section as:
\beq \label{eq:subNLO}
	\sigma^{(0)}_{q \bar q \to \gamma^* \gamma^* g}[ \mathcal J ] = \sigma^H_{q\bar{q}}[ \mathcal J ] + \sigma^{C_1}_{q\bar{q}}[ \mathcal J ] + \sigma^{C_2}_{q\bar{q}}[ \mathcal J ],
\eeq
where $\sigma^H_{q\bar{q}}$ has an integrand which is finite in all singular limits ($\lambda,\bar \lambda, \bar z \to 0$) 
as we take $\epsilon$ to zero and it is therefore allowed to perform a Taylor expansion 
in $\epsilon$, while $\sigma^{C_{1}}_{q\bar{q}}$ and $\sigma^{C_{2}}_{q\bar{q}}$ are divergent as $\e \to 0$.

The contributions read
\begin{align}
	\sigma^H_{q\bar{q}}[ \mathcal J ] &= \frac{1}{2 s} \int \frac{dz d\lambda d\Omega_{d-2}}{4 (2\pi)^{d-1}} s \bar z \left( s \zbar^2\lambda\bar{\lambda} \right)^{-\epsilon} \nn\\ 
	& \quad \times \Big[ |{M}_{q \bar q\to \gamma^*\gamma^* g}|^2_{(0)} \mathcal J (p_1, p_2, p_3, p_4, p_g) d\Phi_{Q(z ,\lambda)\to \gamma^*\gamma^*} \nn \\
	& \qquad \quad -  2 g_s^2 \mu^{2 \e} \frac{S_{qq}(z)}{\bar{z}^2 \lambda} \frac{B_{1} (z)}{z s} \mathcal J(z p_1, p_2, p_3, p_4) d\Phi_{Q(z, 0) \to \gamma^*\gamma^*} \nn \\ 
	& \qquad \quad -  2 g_s^2 \mu^{2 \e} \frac{S_{qq}(z)}{\bar{z}^2 \bar{\lambda} } \frac{B_{2} (z)}{z s} \mathcal J(p_1, z p_2, p_3, p_4) d\Phi_{Q(z, 1) \to \gamma^*\gamma^*} \Big], \label{eq:real}
\end{align}
and
\begin{align}
	\sigma^{C_{1}}_{q\bar{q}}[ \mathcal J ] &= 2 g_s^2 \mu^{2 \e}
   \int \frac{dz d\lambda d\Omega_{d-2}}{4 (2\pi)^{d-1}} \left( s \zbar^2\lambda\bar{\lambda} \right)^{-\epsilon} \frac{S_{qq}(z)}{\bar{z} \lambda} \frac{B_{1} (z)}{2 z s} \mathcal J(z p_1, p_2, p_3, p_4) d\Phi_{Q(z, 0) \to \gamma^*\gamma^*}   , \\
	\sigma^{C_{2}}_{q\bar{q}}[ \mathcal J ] &= 2 g_s^2 \mu^{2 \e}
  \int \frac{dz d\lambda d\Omega_{d-2}}{4 (2\pi)^{d-1}} \left( s \zbar^2\lambda\bar{\lambda} \right)^{-\epsilon}  \frac{S_{qq}(z)}{\bar{z} \bar \lambda} \frac{B_{2} (z)}{2 z s} \mathcal J(p_1, z p_2, p_3, p_4) d\Phi_{Q(z, 1) \to \gamma^*\gamma^*}.
\end{align}

We extract the pole in $\epsilon$ of $\sigma^{C_{1}}_{q\bar{q}}$ and $\sigma^{C_{2}}_{q\bar{q}}$ by integrating over the variables $\lambda$ and $e_T$. Since $\mathcal J$, $B_i$, and $Q$ do not depend on these variables anymore, this step is straightforward.
The result is still singular in the $z\to 1$ limit and we use an expansion in plus-distributions to extract this last singularity. We obtain
\beq
\sigma^{C_{1}}_{q\bar{q}}[ \mathcal J ] = \frac{\alpha^b_s  \, S_{\epsilon}  }{\pi}  \left( \frac{\mu^2}{s}  \right)^\epsilon\int dz\;  G^{(0)}_{qq} (z)\, \sigma^{(0)}_{q \bar q \to \gamma^* \gamma^*}[ \mathcal J ] (z p_1, p_2),
\label{eq:c11}
\eeq
and
\beq
\sigma^{C_{2}}_{q\bar{q}}[ \mathcal J ] = \frac{\alpha^b_s  \, S_{\epsilon}  }{\pi}  \left( \frac{\mu^2}{s}  \right)^\epsilon\int dz\;  G^{(0)}_{qq} (z)\, \sigma^{(0)}_{q \bar q \to \gamma^* \gamma^*}[ \mathcal J ] (p_1, z p_2),
\label{eq:c12}
\eeq
where the Born cross section $\sigma^{(0)}_{q \bar q \to \gamma^* \gamma^*}[ \mathcal J ]$ is evaluated with rescaled momenta in the collinear direction, 
and the integrated splitting kernel is
\begin{align}
	G^{(0)}_{qq}(z) &= \frac{C_F}{2} \left[  \delta(\bar{z})\left(\frac{1}{\epsilon^2}+\frac{3}{2\epsilon} - \frac 3 2 \zeta_2 \right) + 4\DD{1}{\bar{z}}+\bar{z} - 2(1+z)\ln \bar{z}\right] \nn \\ 
	& \quad - \frac{P^{(0)}_{qq}(z)}{\epsilon}  + \order{\epsilon},  \nonumber
\end{align}
with $P^{(0)}_{qq}(z)$ being the Altarelli-Parisi splitting kernel \eqref{eq:apP0}.

The partonic cross section can then be subtracted using eq.~\eqref{eq:subNLO}. Recalling that $p_1 = x_1 P_1$ and $p_2 = x_2 P_2$, and using eqs.~\eqref{eq:c11} and \eqref{eq:c12}, we obtain
\begin{align}
	\label{eq:justabove}
	& \int_0^1 d x_1 d x_2 \, f^b_q(x_1) f^b_{\bar q}(x_2) \sigma^{(0)}_{q\bar q \to \gamma^* \gamma^* g} [ \mathcal J ] = \int_0^1 d x_1 d x_2 \, f^b_q(x_1) f^b_{\bar q}(x_2) \sigma^H_{q\bar q} [ \mathcal J ] \nn \\
	& \qquad \qquad \qquad + \frac{\alpha_s^b S_\epsilon}{\pi} \left( \frac{\mu^2}{s}  \right)^\epsilon \int_0^1 d x_1 d x_2\, [\, f^b_q\otimes G^{(0)}_{q q}\,](x_1)  f^b_{\bar q}(x_2) \sigma^{(0)}_{q\bar q \to \gamma^* \gamma ^*} [ \mathcal J ] \nn \\
	& \qquad \qquad \qquad + \frac{\alpha_s^b S_\epsilon}{\pi} \left( \frac{\mu^2}{s}  \right)^\epsilon \int_0^1 d x_1 d x_2\, f^b_q(x_1) [\, f^b_{\bar q}\otimes G^{(0)}_{q q}\,](x_2)  \sigma^{(0)}_{q\bar q \to \gamma^* \gamma ^*} [ \mathcal J ],
\end{align}
where we used the trivial identity
\beq
	\int_0^1 d x d z\, f(x) g(z) h(x z) = \int_0^1 d y\, [\, f \otimes g\, ](y)\, h(y),
\eeq
with $y=x z$. The first term of eq.~\eqref{eq:justabove} is finite, while the second and third terms contain all the poles in $\e$.

\subsection{(Anti-)quark gluon  channels}
The remaining channels $qg\to \gamma^*\gamma^*q$, $gq\to \gamma^*\gamma^*q$, $\bar{q}g\to \gamma^*\gamma^*\bar{q}$, and $g\bar{q}\to \gamma^*\gamma^*\bar{q}$ are treated similarly, and it is only necessary to consider the channel $qg\to \gamma^*\gamma^* q$.

We parameterize, as before,
\beq
p_q = \zbar \bar{\lambda} \;p_1 + \zbar \lambda \;p_2 + \zbar \sqrt{s \lambda\bar{\lambda}}\; e_T.
\eeq
The matrix element squared is finite in in the limit where $p_q \parallel p_1$ but is singular when $p_q \parallel p_2$, with the asymptotic behaviour
\beq
 |{M}_{q g\to  \gamma^*\gamma^* q}|^2_{(0)} = 2 g_s^2 \mu^{2\e} \frac{S_{qg}(z)}{\bar{z}\bar{\lambda}} \frac{B_2(z)}{z s}
 + \order{\bar \lambda^0}
 , \qquad \text{as} \quad \lambda \to 1,
\eeq
where
\beq
S_{qg}(z) = \frac{1}{2( 1-\epsilon )}(z^2+\zbar^2-\epsilon).
\eeq
Note that since we consider averaged matrix elements squared (with $d-2$ polarizations for the gluons), we needed to compensate for averaging factors.

The matrix element squared has a simple pole at $z=1$, but since the phase-space measure \eqref{eq:PS12togQ} vanishes linearly in this limit, the cross section is free of soft singularities.

We subtract as before, writing
\beq
	\sigma^{(0)}_{qg \to \gamma^* \gamma^*} [ \mathcal J ] = \sigma_{qg}^H [ \mathcal J ] + \sigma_{qg}^C [ \mathcal J ],
\eeq
where
\begin{align}
\sigma^H_{qg}[ \mathcal J ] &= \frac{1}{2 s}
  \int \frac{dz d\lambda d\Omega_{d-2}}{4 (2\pi)^{d-1}} s \bar z \left( s \zbar^2\lambda\bar{\lambda} \right)^{-\epsilon}  \nn\\ 
  & \quad \times \Big[|{M}_{q g\to \gamma^*\gamma^* q}|^2_{(0)} \mathcal J (p_1, p_2, p_3, p_4, p_g) d\Phi_{Q(z ,\lambda)\to \gamma^*\gamma^*}  \\
& \qquad \quad -  2 g_s^2 \mu^{2 \e} \frac{S_{qg}(z)}{\bar{z} \bar{\lambda} } \frac{B_{2} (z)}{z s} \mathcal J(p_1, z p_2, p_3, p_4) d\Phi_{Q(z, 1) \to \gamma^*\gamma^*} \Big],\nn
\end{align}
and
\begin{align}
	\sigma^{C}_{qg}[ \mathcal J ] &= 2 g_s^2 \mu^{2 \e}
   \int \frac{dz d\lambda d\Omega_{d-2}}{4 (2\pi)^{d-1}} \left( s \zbar^2\lambda\bar{\lambda} \right)^{-\epsilon}  \frac{S_{qg}(z)}{\bar \lambda} \frac{B_{2} (z)}{2 z s} \mathcal J(p_1, z p_2, p_3, p_4) d\Phi_{Q(z, 1) \to \gamma^*\gamma^*} \nn \\
   & = \frac{\alpha_s^b S_\e}{\pi}  \left( \frac{\mu^2}{s}  \right)^\epsilon \int d z \;G^{(0)}_{qg} (z)\, \sigma^{(0)}_{q \bar q \to \gamma^* \gamma^*}[\mathcal J](p_1, z p_2) , 
\end{align}
with
\beq
G^{(0)}_{qg} (z) = - \frac{P^{(0)}_{qg}(z)}{\epsilon} + \frac 1 2 \Big( z\bar{z} +(z^2+\zbar^2)\ln\bar{z} \Big) + \order{\epsilon},
\eeq
where the Altarelli-Parisi splitting kernel is given by eq.~\eqref{eq:APqg}.

As before, the corresponding partonic cross-section can be written as
\begin{align}
	& \int_0^1 d x_1 d x_2 \, f^b_q(x_1) f^b_g(x_2) \sigma^{(0)}_{q g \to \gamma^* \gamma^* q} [ \mathcal J ] = \int_0^1 d x_1 d x_2 \, f^b_q(x_1) f^b_g(x_2) \sigma^H_{q g} [ \mathcal J ] \nn \\
	& \qquad \qquad \qquad + \frac{\alpha_s^b S_\epsilon}{\pi} \left( \frac{\mu^2}{s}  \right)^\epsilon \int_0^1 d x_1 d x_2\, f^b_q(x_1) [\, f^b_g\otimes G^{(0)}_{q g}\,](x_2)  \sigma^{(0)}_{q\bar q \to \gamma^* \gamma ^*} [ \mathcal J ],
\end{align}
where the second term contains all the poles in $\e$.

\section{Double-real contributions}\label{sec:reals2}
We now consider the double-real contributions to the partonic cross sections. As explained in section~\ref{sec:setup}, only the channels $q \bar q \to \gamma^* \gamma^* q' \bar q'$ and $\bar q q \to \gamma^* \gamma^* q' \bar q'$ contribute to the large $N_F$ limit.
We will first consider observables that do not involve differential information about the final-state quarks separately.
In the second part of this section we then present a fully differential subtraction scheme.
\subsection{Semi differential subtraction}
Restricting ourselves to observables that do not resolve any of the differential properties of the final state quarks allow us to write
%, such that we can write
\beq
	\mathcal J(p_1, p_2, p_3, p_4, p_{q'}, p_{\bar q'}) = \mathcal J(p_1, p_2, p_3, p_4, p_{g^*}),
\eeq
where $p_{g^*}$ is the momentum of the parent \emph{off-shell} gluon, $p_{g^*}=p_{q'} + p_{\bar{q}'}$.
The phase space of the final-state quarks can then be integrated out, simplifying the extraction of limits. 

Hence we first consider 
\beq \label{eq:sigma_nnlo_int}
	\sigma^{(0),int.}_{q\bar q \to \gamma^* \gamma^* q' \bar q'} [\mathcal J] = \frac{1}{2 s} \int d\Phi_{12\to \gamma^*\gamma^*q'\bar q '}\; \mathcal J (p_1, p_2, p_3, p_4, p_{g^*}) |M_{q \bar q \to \gamma^*\gamma^*q'\bar{q}'}|^2_{(0)},
\eeq
where $int.$ indicates that we restrict ourselves to the aforementioned observables. The phase space can be factorized as
\beq
d\Phi_{12\to \gamma^*\gamma^*q'\bar q '} = \frac{s \;dz}{2\pi}\frac{ds_{g^*}}{2\pi}d\Phi_{12 \to Q g^*}d\Phi_{g^*\to q'\bar{q}'} d\Phi_{Q\to\gamma^*\gamma^*},
\eeq
with $s_{g^*} = p_{g^*}^2$, and
since the function $\mathcal J$ does not depend on $p_{q'}$ and $p_{\bar q'}$, we can perform the integration over the decay phase space  of the off-shell gluon explicitly. We obtain
\beq
\int d\Phi_{g^*\to q'\bar{q}'}  |M_{q \bar q \to \gamma^*\gamma^* q' \bar q '}|^2_{(0)} = \frac{A(\epsilon)}{s_{g^*}^{1+\epsilon}} |M_{q \bar q \to  \gamma^*\gamma^* g^*}|^2_{(0)},
\eeq
where $|M_{q \bar q \to  \gamma^*\gamma^* g^*}|^2_{(0)}$ is the averaged tree-level matrix element squared for the production of two off-shell photons and an off-shell gluon\footnote{The choice of gauge for the off-shell gluon is irrelevant because of the Ward identities. We choose the Feynman gauge.}, and $A(\epsilon)$ is given by
\beq
A(\epsilon) = 2g_s^2 \mu^{2 \e} \frac{1}{4} \frac{d-2}{d-1} \frac{\Omega_{d-1}}{(4\pi)^{d-2}},
\eeq
such that eq.~\eqref{eq:sigma_nnlo_int} becomes
\begin{align}
	& \sigma^{(0), int.}_{q\bar q \to \gamma^* \gamma^* q' \bar q'} [\mathcal J]  =  \nn \\
  	& \qquad \frac{A(\epsilon)}{2s}  \int \frac{s\;dz}{2\pi}\frac{ds_{g^*}}{2\pi}d\Phi_{12\to Q g^*} d\Phi_{Q\to\gamma^*\gamma^*} \mathcal J (p_1, p_2, p_3, p_4, p_{g^*})\frac{|M_{q \bar q\to \gamma^*\gamma^* g^* }|^2_{(0)}}{s_{g^*}^{1+\epsilon}} .
\label{eq:RRprefinal}
\end{align}
%\subsection{Subtraction}
\label{sec:hierarchical}
To perform the subtraction for $\sigma^{(0), int.}_{q\bar q \to \gamma^* \gamma^* q' \bar q'}$ we parameterize the momentum of the off-shell gluon as

\beq \label{eq:nnlo_g}
p_{g^*} = \zbar \bar{\lambda} p_1 + \zbar \lambda \frac{1-\rho\zbar\bar{\lambda}}{1-\zbar\bar{\lambda}}  p_2+ \zbar\sqrt{s\rho\lambda\bar{\lambda}}\;e_T,
\eeq
where $z, \lambda, \rho \in [0,1]$ and $e_T$ is again a unit vector transverse to $p_1$ and $p_2$ in $d=4-2\epsilon$ dimensions. We obtain the invariants
\begin{gather}
	s_{1g^*} = (p_1-p_{g^*})^2 = - s \zbar\lambda, \qquad s_{2g^*} = (p_2-p_{g^*})^2 = - s \zbar \bar \lambda \left( 1- \frac{\bar z \lambda \bar \rho}{1-\bar \lambda \bar z} \right) \\
	s_{g^*} = p_{g^*}^2 = (p_{q'}+p_{\bar q'})^2 = s \frac{\zbar^2 \l \lbar \bar \rho }{1- \zbar \lbar}.
\end{gather}
Using this parameterization, the phase-space measure reads
\beq
d\Phi_{12\to Q g^*}  = \zbar \left( s\zbar^2\lambda\bar{\lambda} \rho\right)^{-\epsilon} \frac{d\lambda d\Omega_{d-2}}{4 (2\pi)^{d-2}} ,
\eeq
where $d\Omega_{d-2}$ denotes the differential solid angle parameterizing $e_T$, and eq.~\eqref{eq:RRprefinal} becomes
\begin{align}
\sigma^{(0), int.}_{q\bar q \to \gamma^* \gamma^* q' \bar q'} [\mathcal J] = 
\frac{A(\epsilon)}{2s}  & \int \frac{dz d\lambda ds_{g^*} d \Omega_{d-2}}{4 (2\pi)^{d}} \frac{s \bar z (s \bar z^2 \lambda \bar \lambda \rho)^{-\e}}{s_{g^*}^{1+\epsilon}}  \nn \\
 & \quad \times |M_{q \bar q \to \gamma^*\gamma^*g^* }|^2_{(0)} \mathcal J (p_1, p_2, p_3, p_4, p_g) d\Phi_{Q(z,\lambda, \rho)\to\gamma^*\gamma^*} .
\label{eq:RRprefinalparam}
\end{align}

The singular limits of the matrix element squared are once again universal but are asymmetric as a consequence of the asymmetry of the parameterization \eqref{eq:nnlo_g} under the exchange $p_1 \leftrightarrow p_2$. We have to consider the following singular limits:

\begin{itemize}
\item $p_{g^*}\parallel p_1$: This corresponds to $\lambda\to 0$, such that $p_{g^*} \to \zbar p_1$, and the matrix element squared has the asymptotic behaviour
\beq \label{eq:triplecoll1}
|M_{q \bar q \to \gamma^*\gamma^* g^*}|^2_{(0)}  = 2 g_s^2 \mu^{2\e} \frac{S_{qq;1}(z,\rho)}{\bar z^2 \lambda} \frac{B_1(z)}{z s} + \mathcal O (\lambda^0),
\eeq
with 
\beq
 S_{qq;1}(z,\rho) = C_F \left(2 z + (1-\epsilon)\bar{z}^2 \rho \right).
 \label{eq:nnloSplittingKernel1}
\eeq
\item $p_{g^*} \parallel p_2$: This corresponds to $\lambda\to 1$, such that $p_{g^*} \to \zbar p_2$, and the matrix element squared has the asymptotic behaviour
\beq \label{eq:triplecoll2}
|M_{q \bar q \to\gamma^*\gamma^* g^*}|^2_{(0)}  = 2 g_s^2 \mu^{2\e} \frac{S_{qq;2}(z,\rho)  }{\zbar^2\lbar(1-\zbar\bar{\rho})}\frac{B_2(z)}{z s} + \mathcal O (\bar \lambda^0),
\eeq
with 
\beq
 S_{qq;2}(z,\rho) = C_F \left( 2 z + (1-\epsilon)  \bar{z}^2  (1-\frac{z\bar{\rho}}{1-\zbar\bar{\rho}}) \right).
 \label{eq:nnloSplittingKernel2}
\eeq
\item $s_{g^*} =0$: This is the final state collinear singularity, when the gluon becomes on-shell, but remains in the hard region. It corresponds to $\rho\to 1$, and the matrix element squared has the smooth limit
\beq  \label{eq:doublecoll2}
	|M_{q \bar q\to \gamma^*\gamma^*g^*}|^2_{(0)}  = |M_{q\bar q\to \gamma^*\gamma^* g}|^2_{(0)} + \order{\bar \rho}.
\eeq
The corresponding singularity comes from the factor $s_{g^*}^{-1-\e}$ in \eqref{eq:RRprefinalparam}.
\end{itemize}
Note that in the limit where $\rho\to 1$, both splitting kernels \eqref{eq:nnloSplittingKernel1} and \eqref{eq:nnloSplittingKernel2} tend smoothly  to the splitting kernel we obtained in the previous section, eq.~\eqref{eq:nloSplittingKernel}. 

We proceed to the subtraction by writing
\beq \label{eq:NNLOsubtraction}
 \sigma^{(0), int.}_{q\bar q \to \gamma^* \gamma^* q' \bar q'} [\mathcal J] =  \sigma^{HH}_{q\bar q} [\mathcal J] + \sigma^{R;C}_{q\bar q} [\mathcal J] + \sigma^{CC_1}_{q\bar q} [\mathcal J] + \sigma^{CC_2}_{q\bar q} [\mathcal J],
\eeq
where $\sigma^{HH}_{q\bar q}$ has a Taylor expansion around $\e = 0$ and the other terms contain all the poles in $\e$. The different contributions are:
\begin{align}
\label{eq:sigmaHH}
\sigma^{HH}_{q\bar q} [\mathcal J] &= \frac{A(\epsilon)}{2s}  \int \frac{dz d\lambda ds_{g^*} d \Omega_{d-2}}{4 (2\pi)^{d}} \frac{s \bar z (s \bar z^2 \lambda \bar \lambda \rho)^{-\e}}{s_{g^*}^{1+\epsilon}} \nn\\
	& \quad\times\bigg[ |M_{q \bar q\to \gamma^*\gamma^*g^* } |^2_{(0)}  \mathcal J (p_1, p_2, p_3, p_4, p_g{g^*}) d \Phi_{Q(z, \lambda, \rho)\to\gamma^*\gamma^*}  \nn \\ 
	& \qquad \quad -  2 g_s^2 \mu^{2\e} \frac{S_{qq;1}(z,\rho)}{\bar z^2 \lambda} \frac{B_1(z)}{z s} \mathcal J (z p_1, p_2, p_3, p_4) d \Phi_{Q(z, 0, \rho)\to\gamma^*\gamma^*} \nn \\ 
	& \qquad \quad - 2 g_s^2 \mu^{2\e} \frac{S_{qq;2}(z,\rho)}{\zbar^2\lbar(1-\bar{\rho}\zbar)}\frac{B_2(z)}{z s} \mathcal J (p_1, z p_2, p_3, p_4) d \Phi_{Q(z, 1, \rho)\to\gamma^*\gamma^*} \nn\\
	& \qquad \quad - |M_{q \bar q\to \gamma^*\gamma^* g}|^2_{(0)} \mathcal J (p_1, p_2, p_3, p_4, p_g) d \Phi_{Q(z, \lambda, 1)\to\gamma^*\gamma^*} \nn \\
	& \qquad \quad + 2 g_s^2 \mu^{2\e} \frac{S_{qq}(z)}{\zbar^2 \lambda} \frac{B_1(z)}{z s} \mathcal J (z p_1, p_2, p_3, p_4) d \Phi_{Q(z, 0, 1)\to\gamma^*\gamma^*} \nn \\ 
	& \qquad \quad + 2 g_s^2 \mu^{2\e} \frac{S_{qq}(z)  }{\zbar^2\lbar}\frac{B_2(z)}{z s} \mathcal J (p_1, z p_2, p_3, p_4) d \Phi_{Q(z, 1, 1)\to\gamma^*\gamma^*} \bigg],
\intertext{and}
	\label{eq:sigmaCC1}
	\sigma^{CC_1}_{q \bar q} [\mathcal J] &= 2 g_s^2 \mu^{2\e} A(\epsilon) \int \frac{dz d\lambda ds_{g^*} d \Omega_{d-2}}{4 (2\pi)^{d}} \frac{(s \bar z^2 \lambda \bar \lambda \rho)^{-\e}}{s_{g^*}^{1+\epsilon}} \nn \\ 
	& \quad  \times\frac{S_{qq;1}(z,\rho)}{\bar z \lambda} \frac{B_1(z)}{2 z s} \mathcal J (z p_1, p_2, p_3, p_4) d \Phi_{Q(z, 0, \rho)\to\gamma^*\gamma^*},\\
	\label{eq:sigmaCC2}
	\sigma^{CC_2}_{q \bar q} [\mathcal J] &= 2 g_s^2 \mu^{2\e} A(\epsilon) \int  \frac{dz d\lambda ds_{g^*} d \Omega_{d-2}}{4 (2\pi)^{d}} \frac{(s \bar z^2 \lambda \bar \lambda \rho)^{-\e}}{s_{g^*}^{1+\epsilon}} \nn \\ 
	& \quad \times \frac{S_{qq;2}(z,\rho)  }{\zbar \lbar(1-\bar{\rho}\zbar)}\frac{B_2(z)}{2 z s} \mathcal J (p_1, z p_2, p_3, p_4) d \Phi_{Q(z, 1, \rho)\to\gamma^*\gamma^*}, \\
\sigma^{R;C}_{q \bar q} [\mathcal J] &=\frac{A(\epsilon)}{2s}  \int \frac{dz d\lambda ds_{g^*} d \Omega_{d-2}}{4 (2\pi)^{d}} \frac{s \bar z (s \bar z^2 \lambda \bar \lambda \rho)^{-\e}}{s_{g^*}^{1+\epsilon}} \nn \\
	& \quad \times \bigg[ |M_{q \bar q\to \gamma^*\gamma^* g} |^2_{(0)} \mathcal J (p_1, p_2, p_3, p_4, p_g) d \Phi_{Q(z, \lambda, 1)\to\gamma^*\gamma^*} \nn \\
	& \qquad \quad - 2 g_s^2 \mu^{2\e} \frac{S_{qq}(z)}{\zbar^2 \lambda} \frac{B_1(z)}{z s} \mathcal J (z p_1, p_2, p_3, p_4) d \Phi_{Q(z, 0, 1)\to\gamma^*\gamma^*} \nn \\ 
	& \qquad \quad - 2 g_s^2 \mu^{2\e} \frac{S_{qq}(z)  }{\zbar^2\lbar}\frac{B_2(z)}{z s} \mathcal J (p_1, z p_2, p_3, p_4) d \Phi_{Q(z, 1, 1)\to\gamma^*\gamma^*}\bigg],
\label{eq:sigmaRC}
\end{align}
where we have $p_g =\lim_{\rho \to 1} p_{g^*}$ such that the parameterization \eqref{eq:nnlo_g} tends to eq.~\eqref{eq:nlog} in the limit where $\rho \to 1$. 
As for the real radiation, the soft limit does not need to be subtracted explicitly. This can be checked by expanding the integrand of eq.~\eqref{eq:sigmaHH} around $z=1$.

We can extract the poles in $\e$ of $\sigma^{CC_{1}}_{q \bar q}$ and $\sigma^{CC_{2}}_{q \bar q}$ by integrating over the variables $\lambda$, $\rho$ and $e_T$.
The integration is slightly more complicated than for the real contributions, but the result can be written in terms of hypergeometric functions $_2F_1(a,b,c;\bar{z})$, where $a$ and $b$ depend on $\epsilon$. We have used the {\tt HypExp} \cite{Huber:2005yg} package to expand them in $\epsilon$ and then performed a plus-distribution expansion over $\bar{z}$ to extract the double soft singularity. We note here that the residue of the soft pole at $z=1$ is the same for both counterterms $\sigma^{CC_{1}}_{q \bar q}$ and $\sigma^{CC_{2}}_{q \bar q}$, such that the asymmetry due to the parameterization is limited to the regular coefficients and does not affect the delta- and plus-distribution terms. As for the real corrections, we can write the counterterms as
\begin{align}
	\sigma^{CC_{1}}_{q \bar q}[ \mathcal J ] &= \left(\frac{\alpha_s^b S_\e}{\pi}\right)^2\left(\frac{\mu^2}{s}\right)^{2\epsilon} \int dz\;  G^{(1)}_{qq;1} (z) \, \sigma^{(0)}_{q \bar q \to \gamma^* \gamma^*}[ \mathcal J ](zp_1, p_2), \label{eq:intctNNLO1} \\
	\sigma^{CC_{2}}_{q \bar q}[ \mathcal J ] &= \left(\frac{\alpha_s^b S_\e}{\pi}\right)^2\left(\frac{\mu^2}{s}\right)^{2\epsilon} \int dz\;  G^{(1)}_{qq;2} (z) \, \sigma^{(0)}_{q \bar q \to \gamma^* \gamma^*}[ \mathcal J ](p_1, zp_2), \label{eq:intctNNLO2}
\end{align}
where we have
\begin{eqnarray}
G^{(1)}_{qq; 1}(z)&=&
\frac{C_F}{48}\left\{
-\frac{\delta(\bar{z})}{\e^3}
+\frac{1}{\e^2}
\left[4\DD{0}{\zb}-\frac{5}{3}\delta(\bar{z})
-2(1+z)\right]\nonumber\right.\\&&
+\frac{1}{\e}
\left[
-16\DD{1}{\zb}
+\frac{20}{3}\DD{0}{\zb}
-\frac{1}{18}(56-21\pi^2)\delta(\bar{z})\right.\nn\\&&\left.
-\frac{10}{3}(1+z)+8(1+z)\ln\zb+2(1+z^2)\frac{\ln z}{\zb}\right]\nonumber\\
&&
+32\DD{2}{\zb}
-\frac{80}{3}\DD{1}{\zb}
+\frac{2}{9}(56-21\pi^2)\DD{0}{\zb}\nonumber\\
&&
-\frac{1}{54}(328-105\pi^2-1116\zeta_3)\delta(\zb)\nonumber\\
&&
-4(1+z^2)\frac{\Li_2(\zb)}{\zb}
-16(1+z)\ln^2\zb
-(1+z^2)\frac{\ln^2 z}{\zb}
-8(1+z^2)\frac{\ln z \ln\zb}{\zb}\nonumber\\
&&
+\frac{40}{3}(1+z)\ln\zb
+\frac{10}{3}(1+z^2)\frac{\ln z}{\zb}\nonumber\\
&&\left.
-\frac{1}{9}(38+74z-21\pi^2(1+z))
\right\}+\mathcal{O}(\e),\\
%%%%%%%%%%%%%%%%%%%%%%%%%%%%%%%%%%%%
G^{(1)}_{q  q; 2}(z)&=&
G^{(1)}_{q  q; 1}(z)
-\frac{C_F}{48}
\left(
4(1+z^2)\frac{\Li_2(\zb)}{\zb}
-4\ln z-4\zb
\right)
+\mathcal{O}(\e).
\end{eqnarray}

The pole in $\e$ of $\sigma^{R;C}_{q \bar q}$ is extracted by integrating over $\rho$ only, since the variables $z$ and $\lambda$ still parameterize the on-shell gluon in $\rho \to 1$ limit. Using
\beq
  \frac{A(\epsilon)}{2\pi} \int\frac{ds_{g^*}}{s_{g^*}^{1+\epsilon}}\rho^{-\epsilon} = -\frac{1}{6\epsilon} \left(\frac{\alpha_s^b S_\e}{\pi}\right) \left[ 1 + \frac{5}{3}\epsilon-\epsilon \ln\left( \frac{s}{\mu^2} \frac{\zbar^2\lambda\bar{\lambda}}{(1-\zbar\bar{\lambda})}\right) + \mathcal{O}(\epsilon^2) \right],
\eeq
we can expand as
\begin{align}
	\label{eq:sigmaRC_exp}
	\sigma^{R;C}_{q\bar q}[\mathcal J] = -\frac{1}{6 \e}\left( \frac{\alpha_s^b S_\e}{\pi} \right) \sigma^{H}_{q\bar q}[\mathcal J]+\sigma^{\tilde H}_{q\bar q}[\mathcal J] + \order{\e},
\end{align}
where $\sigma^{H}_{q\bar q}$ is our NLO expression  \eqref{eq:real}, and we defined
\begin{align}
\sigma^{\tilde H}_{q\bar{q}}[ \mathcal J ] &= -\frac{1}{6}\left( \frac{\alpha_s^b S_\e}{\pi} \right) \frac{1}{2 s}
  \int \frac{dz d\lambda d\Omega_{d-2}}{4 (2\pi)^{d-1}} s \bar z \left( s \zbar^2\lambda\bar{\lambda} \right)^{-\epsilon}  \nn \\ 
  & \quad \times \left( \frac{5}{3}- \ln\left( \frac{s}{\mu^2} \frac{\zbar^2\lambda\bar{\lambda}}{(1-\zbar\bar{\lambda})}\right) \right) \nn \\
  & \quad \times \Big[ |{M}_{q \bar q\to \gamma^*\gamma^* g}|^2_{(0)} \mathcal J (p_1, p_2, p_3, p_4, p_g) d\Phi_{Q(z ,\lambda)\to \gamma^*\gamma^*} \nn \\
	& \qquad \quad -  2 g_s^2 \mu^{2 \e} \frac{S_{qq}(z)}{\bar{z}^2 \lambda} \frac{B_{1} (z)}{z s} \mathcal J(z p_1, p_2, p_3, p_4) d\Phi_{Q(z, 0) \to \gamma^*\gamma^*} \nn \\ 
	& \qquad \quad -  2 g_s^2 \mu^{2 \e} \frac{S_{qq}(z)}{\bar{z}^2 \bar{\lambda} } \frac{B_{2} (z)}{z s} \mathcal J(p_1, z p_2, p_3, p_4) d\Phi_{Q(z, 1) \to \gamma^*\gamma^*} \Big]. \label{eq:realtilde}
\end{align}
After summing over the final-state quark flavours $q'$, the first term of eq.~\eqref{eq:sigmaRC_exp} will cancel the $\beta_0$ term coming from the renormalization of $\alpha^b_s$ in the large $N_F$ limit, given by eq.~\eqref{equ:catani_ren}, applied to the NLO contribution $\sigma^{H}_{q \bar q}$, given by eq.~\eqref{eq:real}. Also note that $\sigma^{\tilde H}_{q\bar{q}}$ has the exact the same structure as $\sigma^{H}_{q \bar q}$, except for the prefactor and the term on the second line.

At the partonic level, we finally obtain
\begin{align}
	& \int_0^1 d x_1 d x_2 \, f^b_q(x_1) f^b_{\bar q}(x_2) \sigma^{(0),int.}_{q \bar q \to \gamma^* \gamma^* q' \bar q'} [ \mathcal J ] = \nn \\ 
	& \qquad \qquad \qquad \int_0^1 d x_1 d x_2 \, f^b_q(x_1) f^b_{\bar q}(x_2) \left( \sigma^{HH}_{q \bar q} [ \mathcal J ] + \sigma^{\tilde H}_{q \bar q} [ \mathcal J ] \right) \nn \\
	& \qquad \qquad \qquad - \frac{1}{6 \e}\left( \frac{\alpha_s^b S_\e}{\pi} \right)  \int_0^1 d x_1 d x_2 \, f^b_q(x_1) f^b_{\bar q}(x_2) \sigma^{H}_{q \bar q} [ \mathcal J ] \nn \\
	& \qquad \qquad \qquad + \left(\frac{\alpha_s^b S_\epsilon}{\pi}\right)^2 \left(\frac{\mu^2}{s}\right)^{2\epsilon} \int_0^1 d x_1 d x_2\, [\, f^b_q\otimes G^{(1)}_{q q; 1}\,](x_1) f^b_{\bar q}(x_2)\,  \sigma^{(0)}_{q\bar q \to \gamma^* \gamma ^*} [ \mathcal J ] \nn \\
	& \qquad \qquad \qquad + \left(\frac{\alpha_s^b S_\epsilon}{\pi}\right)^2 \left(\frac{\mu^2}{s}\right)^{2\epsilon} \int_0^1 d x_1 d x_2\, f^b_q(x_1) [\, f^b_{\bar q}\otimes G^{(1)}_{q q; 2}\,](x_2)\,  \sigma^{(0)}_{q\bar q \to \gamma^* \gamma ^*} [ \mathcal J ],
\end{align}
where the first term is finite and the others contain all the poles in $\e$.

%%%%%%%%%%%%%%%%%%%%%%%%%%%%%%%%%%%%%%%%%%%%%%%%%%%%%%%%%%%%%%%%%%%%%%%%%%%%
%%%%%%%%%%%%%%%%%%%%%%%%%%%%%%%%%%%%%%%%%%%%%%%%%%%%%%%%%%%%%%%%%%%%%%%%%%%%
%%%%%%%%%%%%%%%%%%%%%%%%%%%%%%%%%%%%%%%%%%%%%%%%%%%%%%%%%%%%%%%%%%%%%%%%%%%%

\subsection{Fully differential subtraction} \label{sec:fullyX}
In this section, we show how to perform a similar subtraction in the case where the phase space of the two final state quarks 
cannot be integrated out. Hence we consider
\beq
	\sigma^{(0)}_{q\bar q \to \gamma^* \gamma^* q' \bar q'} [\mathcal J] = \frac{1}{2 s} \int d\Phi_{12\to \gamma^*\gamma^*q'\bar q '}\; \mathcal J (p_1, p_2, p_3, p_4, p_{q'}, p_{\bar q'}) |M_{q \bar q \to \gamma^*\gamma^*q'\bar{q}'}|^2_{(0)},
\eeq
where the function $\mathcal J$ depends now on all external momenta.

We extend our parameterization in a way similar to ref.~\cite{Anastasiou:2010pw}. In order to construct the momenta of the $q' \bar q'$ pair, we introduce another transverse unit vector $e'_T$, with the conditions $e'_T \cdot p_1 = e'_T \cdot p_2 = e'_T \cdot e_T = 0$ and ${e'_T}^2 = -1$.
\footnote{In d=4, these conditions actually fix $e'_T$ completely up to a reflection about the beam axis, in accordance with $ \Omega_{1} =2$. } The full phase space measure then reads
\begin{align} \label{eq:exclps}
  d\Phi_{12\to\gamma^*\gamma^* q'\bar{q}'} &= \frac{s^2 \bar z^3 \l \lbar}{1-\zbar \lbar}
    \left( \frac{ s^2 \zbar^4 \l^2 \lbar^2 \rho \bar \rho y_1 \bar y_1 \sin^2 \pi y_2}{1-\zbar \lbar} \right)^{-\e} \\ 
    & \quad \times \frac{d z d \lambda d \rho d \Omega_{d-2}}{4 (2\pi)^{d}} \frac{d y_1 d y_2 d \Omega_{d-3}}{8 (2 \pi)^{d-2}} d \Phi_{Q\to\gamma^*\gamma^*}
\end{align}
where $d \Omega_{d-2}$ and $d \Omega_{d-3}$ denote the integrals over $e_T$ and $e'_T$ respectively, and $Q = p_1+p_2-p_{q'}-p_{\bar q'}$. The new variables $y_1$, $y_2 \in [0,1]$ parameterize the phase space of the decay of the off-shell gluon. The expressions for the invariants $s_{1q'} = (p_1-p_{q'})^2$, $s_{1\bar q'} = (p_1-p_{\bar q'})^2$, $s_{2 q'} = (p_2-p_{q'})^2$ and $s_{2\bar q'} = (p_2-p_{\bar q'})^2$ can be found in the aforementioned reference.

The momenta of the two final-state quarks can now be fully reconstructed and read
\begin{align*}
	p_{q'} = \bar z \Bigg[ & \bar \lambda y_1\, p_1 %\\&  
	+ \lambda\left(y_1 \rho + \frac{\bar \rho \bar y_1}{1-\bar z \bar \lambda} -2 \cos \pi y_2 \sqrt{\frac{\rho \bar \rho y_1 \bar y_1}{1-\bar z \bar \lambda}} \right) p_2  \\ 
	& -\sqrt{s \frac{\lambda \bar \lambda}{\rho}}\left( y_1 \rho - \cos \pi y_2 \sqrt{\frac{\rho \bar \rho y_1 \bar y_1}{1-\bar z \bar \lambda}}\right) e_T % \\ &
	+\sin \pi y_2 \sqrt{s \frac{\lambda \bar \lambda \bar \rho y_1 \bar y_1}{1-\bar z \bar \lambda}}\; e'_T \Bigg], \\
	p_{\bar q'} = \bar z \Bigg[ & \bar \lambda \bar y_1 \, p_1 %\\&
	 + \lambda\left(\bar y_1 \rho + \frac{\bar \rho y_1}{1-\bar z \bar \lambda} + 2\cos \pi y_2 \sqrt{\frac{\rho \bar \rho y_1 \bar y_1}{1-\bar z \bar \lambda}} \right) p_2  \\ 
	& -\sqrt{s \frac{\lambda \bar \lambda}{\rho}}\left( \bar y_1 \rho + \cos \pi y_2 \sqrt{\frac{\rho \bar \rho y_1 \bar y_1}{1-\bar z \bar \lambda}}\right) e_T %\\&
	 -\sin \pi y_2 \sqrt{s \frac{\lambda \bar \lambda \bar \rho y_1 \bar y_1}{1-\bar z \bar \lambda}}\; e'_T \Bigg],
\end{align*}
such that the momentum of the parent gluon $p_{g^*} = p_{q'}+p_{\bar q'}$ is still given by our previous expression \eqref{eq:nnlo_g}.
Note that $p_{\bar q'}$ can be obtained from $p_{q'}$ by replacing $y_1 \leftrightarrow \bar y_1$, $y_2 \leftrightarrow \bar y_2$ and $e'_T \to -e'_T$.

There are no new singular limits to consider in this case. The singular limits where $p_{q'} + p_{\bar q'} \parallel p_1 $ and $p_{q'} + p_{\bar q'}\parallel p_2 $ have the same structure as in eqs.~\eqref{eq:triplecoll1} and \eqref{eq:triplecoll2}, and can be written as
\begin{align}
	\label{eq:nnlo134}
	|M_{q \bar q \to \gamma^*\gamma^* q' \bar q'}|^2_{(0)}  &= 4 g_s^4 \mu^{4\e} \frac{\tilde S_{qq;1}(z, \rho, y_1, y_2)}{\zbar^4 \lambda^2 \bar \rho} \frac{B_1(z)}{z s^2} + \order{\lambda^{-1}}, \\
	\label{eq:nnlo234}
	|M_{q \bar q \to \gamma^*\gamma^* q' \bar q'}|^2_{(0)}  &= 4 g_s^4 \mu^{4\e} \frac{\tilde S_{qq;2}(z, \rho, y_1, y_2)}{\zbar^4 \bar \lambda^2 (1-\zbar \bar \rho)^2 \bar \rho} \frac{B_2(z)}{z s^2} + \order{\bar \lambda^{-1}},
\end{align}
with new, fully differential, splitting kernels $\tilde S_{qq;1}$ and $\tilde S_{qq;2}$. Note that although the singularities are now quadratic at the level of the matrix element squared, the cross section diverges only logarithmically because the phase-space measure \eqref{eq:exclps} vanishes linearly in these limits.

The singular limit where $s_{g^*} =0$, such that $p_{q'} \parallel p_{\bar q'}$, is now non-trivial and involves spin correlations, and  reads
\begin{align}
	\label{eq:nnlo34}
	|M_{q \bar q \to \gamma^*\gamma^* q' \bar q'}|^2_{(0)}  &= 2 g_s^2 \mu^{2 \e} \frac{1-\zbar \bar \lambda}{s \zbar^2 \lambda \bar \lambda \bar \rho} \tilde S_{\mu \nu}(\lambda, y_1, y_2) \tilde M^{\mu\nu}_{q \bar q  \to \gamma^*\gamma^* g}  + \mathcal O (\bar \rho^0).
\end{align}
In the above we have defined
\[
	\tilde M^{\mu\nu}_{q \bar q \to \gamma^*\gamma^*g} = \left(\frac{1}{2 N_c}\right)^2 \sum_{pol.} \mathcal A_{q \bar q \to \gamma^*\gamma^*g^\mu} ( \mathcal A_{q \bar q \to  \gamma^*\gamma^*g^\nu} )^*,
\]
where $\sum_{pol.}$ denotes the sum over the polarizations of the photons and the spins of the quarks and $\mathcal A_{q \bar q \to \gamma^*\gamma^*g^\mu}$ is the tree amplitude for the process $q \bar q \to \gamma^*\gamma^*g$ where the gluon is not contracted with the corresponding polarization vector.

All the splitting kernels can be obtained from the universal limits given in ref.~\cite{Catani:1999ss}. They read
\begin{align}
	\label{eq:nnloSplittting1}
	\tilde S_{qq;1}(z, \rho, y_1, y_2) 
% Link with Catani+Grazzini
% & = z \zbar^2  \bar \rho \lim_{\lambda \to 0} C_F \frac{s_{134}}{4 s_{34}} \Bigg[-\frac{1}{s_{134} s_{34}}\left(\frac{2 (s_{14} z_3-s_{13} z_4)}{z_3+z_4}+\frac{s_{34} (z_3-z_4)}{z_3+z_4}\right)^2 \nn \\
% & \qquad \qquad \qquad \qquad  +(1-2 \epsilon ) \left(-\frac{s_{34}}{s_{134}}+z_3+z_4\right)+\frac{4 z_1+(z_3-z_4)^2}{z_3+z_4}\Bigg] \nn \\
	& = \frac{C_F z }{2 \bar \rho} \big[ \left( 1 + z^2 \right) \rho (1-2 y_1 \bar y_1) + 8 z \bar \rho y_1 \bar y_1 -\rho \bar z^2 \epsilon  - 4 z \rho y_1 \bar y_1 \cos (2 \pi  y_2) \nn  \\
 & \qquad \qquad  + 4  (1+z) (1-2 y_1)\sqrt{z \bar \rho \rho y_1 \bar y_1}   \cos (\pi y_2) \big], \\
 	\label{eq:nnloSplittting2} 
 	\tilde S_{qq;2} (z, \rho, y_1, y_2)
% Link with Catani+Grazzini
% &= z \zbar^2  \bar \rho  \lim_{\lambda \to 1} C_F \frac{s_{234}}{4 s_{34}} \Bigg[-\frac{1}{s_{234} s_{34}}\left(\frac{2 (s_{24} z_3-s_{23} z_4)}{z_3+z_4}+\frac{s_{34} (z_3-z_4)}{z_3+z_4}\right)^2 \nn \\
% & \qquad \qquad \qquad \qquad  +(1-2 \epsilon ) \left(-\frac{s_{34}}{s_{234}}+z_3+z_4\right)+\frac{4 z_1+(z_3-z_4)^2}{z_3+z_4}\Bigg] \nn \\
	& = \frac{C_F}{2 \bar \rho} \big[ 2 \bar \rho^2 \bar z (2-\bar \rho \bar z) (1-6 y_1 \bar y_1)+(1+\bar \rho) \left(1+z^2\right) (1-2 y_1 \bar y_1) \nn \\ 
	& \qquad \quad - 4 \bar \rho (1-2 y_1)^2- \epsilon \rho \bar z^2 - 4 (1-\bar \rho \bar z) (z-\bar \rho \bar z) \rho  y_1 \bar y_1  \cos (2 \pi  y_2) \nn \\ 
	& \qquad \quad + 4 (1+z-2 \bar \rho \bar z) (1-2 y_1) (1-\bar \rho \bar z) \sqrt{\bar \rho \rho y_1 \bar y_1} \cos (\pi  y_2) \big ],
\end{align}
%where we had
%\[
%	z_1 = z_2 \equiv \frac{E_1}{E_1-E_{34}}, \quad	z_3 \equiv  \frac{-E_3}{E_1-E_{34}}, \quad z_4 \equiv  \frac{-E_4}{E_1-E_{34}}\, ;
%\]
and
\beq
	\tilde S^{\mu \nu} = \frac 1 2 \left[ - g^{\mu\nu} + k^\mu k^\nu \right],
\eeq
where the dimensionless vector $k$ is given by
\beq
	k = -2 \sqrt{y_1 \bar y_1 } \Big[ \sqrt{\lambda \bar \lambda}\, 2 \cos \pi y_2\, \frac{p_1 - p_2}{\sqrt{s}}
	           + (1-2 \lambda) \cos \pi y_2\; e_T + \sin \pi y_2\; e'_T \Big],
\eeq
such that $k^2 =  - 4 y_1 \bar y_1$.
Double limits commute and can be obtained easily by extracting the pole at $\rho \to 1$ of the counterterms \eqref{eq:nnloSplittting1} and \eqref{eq:nnloSplittting2}.

Subtraction can be performed as in eq.~\eqref{eq:NNLOsubtraction}, by writing
\beq
	\sigma^{(0)}_{q\bar q \to \gamma^* \gamma^* q' \bar q'} [\mathcal J] =  \sigma^{HH}_{q\bar q} [\mathcal J] + \sigma^{R;C}_{q\bar q} [\mathcal J] + \sigma^{CC_1}_{q\bar q} [\mathcal J] + \sigma^{CC_2}_{q\bar q} [\mathcal J].
\eeq
Only the $\sigma^{HH}_{q \bar q}$ contribution needs to be modified, and now reads
\begin{align}
	\sigma^{HH}_{q \bar q}[\mathcal J] &= \frac 1 {2s} \int \frac{d z d \lambda d \rho d y_1 d y_2 d \Omega_{d-2} d \Omega_{d-3}}{32 (2 \pi)^{2d-2}} \left(\frac{s^2 \bar z^4 \lambda^2 \bar \lambda^2 \rho \bar \rho y_1\bar y_1 \sin^2 \pi y_2 }{1 -\bar z \bar \lambda}\right)^{-\epsilon} s^2 \bar z^3 \nn \\
	& \quad \times \Bigg[ 
	\frac{\lambda \bar \lambda}{1 - \bar z \bar \lambda} |M_{q \bar q \to \gamma^*\gamma^* q' \bar q'}|^2_{(0)} \mathcal J (p_1, p_2, p_3, p_4, p_{q'}, p_{\bar q'}) d \Phi_{Q(z, \lambda, \rho) \to \gamma^* \gamma^*} \nn \\
	& \qquad \quad - 4 g_s^4 \mu^{4\e} \frac{1}{z} \frac{\tilde S_{qq;1}(z, \rho, y_1, y_2)}{\zbar^4 \lambda \bar \rho} \frac{B_1(z)}{z s^2} \mathcal J (z p_1, p_2, p_3, p_4) d \Phi_{Q(z, 0, \rho) \to \gamma^* \gamma^*} \nn \\
	& \qquad \quad - 4 g_s^4 \mu^{4\e} \frac{\tilde S_{qq;2}(z, \rho, y_1, y_2)}{\zbar^4 \bar \lambda (1-\zbar \bar \rho)^2 \bar \rho} \frac{B_2(z)}{z s^2} \mathcal J (p_1, z p_2, p_3, p_4) d \Phi_{Q(z, 1, \rho) \to \gamma^* \gamma^*} \nn \\
	& \qquad \quad - 2 g_s^2 \mu^{2 \e} \frac{1}{s \zbar^2 \bar \rho} \tilde S_{\mu \nu}(\lambda, y_1, y_2) \tilde M^{\mu\nu}_{q \bar q  \to \gamma^*\gamma^* g} \mathcal J (p_1, p_2, p_3, p_4, p_{g}) d \Phi_{Q(z, \lambda, 1) \to \gamma^* \gamma^*}  \nn \\
	& \qquad \quad + 4 g_s^4 \mu^{4\e} \frac{1}{z} \frac{\tilde S_{qq;1}(z, 1, y_1, y_2)}{\zbar^4 \lambda \bar \rho} \frac{B_1(z)}{z s^2} \mathcal J (z p_1, p_2, p_3, p_4) d \Phi_{Q(z, 0, 1) \to \gamma^* \gamma^*} \nn \\
	& \qquad \quad + 4 g_s^4 \mu^{4\e} \frac{\tilde S_{qq;2}(z, 1, y_1, y_2)}{\zbar^4 \bar \lambda \bar \rho} \frac{B_2(z)}{z s^2} \mathcal J (p_1, z p_2, p_3, p_4) d \Phi_{Q(z, 1, 1) \to \gamma^* \gamma^*}
	\Bigg].
\end{align}
The contributions $\sigma^{CC_1}_{q \bar q}$, $\sigma^{CC_2}_{q \bar q}$, and $\sigma^{R;C}_{q \bar q}$ are again given by eqs.~\eqref{eq:sigmaCC1}, \eqref{eq:sigmaCC2} and \eqref{eq:sigmaRC} respectively, after integration over the variables $y_1$, $y_2$ and $e_T'$. In particular we have
\begin{align}
  2 g_s^2 \mu^{2\e} \int \frac{dy_1 dy_2 d \Omega_{d-3}}{8 (2 \pi)^{d-2}}
    \left( y_1 \bar y_1 \sin ^2 \pi y_2 \right)^{-\e}
    \frac{\tilde S_{qq;1}(z, \rho, y_1, y_2)}{z}
    &= \frac{A(\e)}{2\pi} S_{qq;1}(z, \rho), \\
  2 g_s^2 \mu^{2\e} \int \frac{dy_1 dy_2 d \Omega_{d-3}}{8 (2 \pi)^{d-2}}
    \left( y_1 \bar y_1 \sin ^2 \pi y_2 \right)^{-\e}
    \frac {\tilde S_{qq;2}(z, \rho, y_1, y_2)}{1-\zbar\bar\rho}
    &= \frac{A(\e)}{2\pi} S_{qq;2}(z, \rho). 
\end{align}

%%%%%%%%%%%%%%%%%%%%%%%%%%%%%%%%%%%%%%%%%%%%%%%%%%%%%%%%%%%%%%%%%%%%%%%%
%%%%%%%%%%%%%%%%%%%%%%%%%%%%%%%%%%%%%%%%%%%%%%%%%%%%%%%%%%%%%%%%%%%%%%%%

%% file: numerical_results.tex
\section{Numerical results}\label{sec:results}

We have implemented the various contributions to the differential cross section for the $N_F$ part of the process $pp\to\gamma^*\gamma^*+X$ up to the next-to-next-to leading order in the strong coupling expansion in two different programs. The virtual contributions are written in terms of master integrals which in turn are evaluated in terms of harmonic polylogarithms. In order to ensure the correct implementation of master integrals, various analytic and numerical checks were performed against published results in the literature as detailed in section \ref{sec:Fede}. We have used the program {\tt CHAPLIN}~\cite{Buehler:2011ev} for the numerical evaluation of the necessary harmonic polylogarithms in the physical region. The agreement of the poles of  the one- and two-loop virtual amplitudes, as predicted by ref.~\cite{Catani:1998bh} was checked both analytically and numerically, at the implementation level. The NLO contribution was checked against the MCFM~\cite{mcfm} implementation\footnote{The $pp\to\gamma^*\gamma^*$ without photon decays is not an out-of-the-box process in MCFM, but it was possible to compare our result with ${m_3}={m_4}=m_z$ against MCFM's $pp\to ZZ$ with modified couplings of the Z boson to quarks.}. 

The double-real contributions were implemented as described in sections \ref{sec:hierarchical} and \ref{sec:fullyX}, and double checked against another fully differential parameterization. Because the two parameterizations have different double-real counterterms, the numerical results for the double hard, the single hard and the integrated triple collinear counterterm cross sections are individually different. Only the sum of these contributions is physical, which provides a strong numerical check of our two implementations.  

In the following, we present indicatively some differential distributions of interest, including their factorization and renormalization scale dependence. Since we do not include the decay of the off-shell photons to leptons, or the single-resonant diagrams in this publication, we defer a more detailed phenomenological analysis to a future publication. 

In what follows, we use the central grid of the MSTW08 parton distribution functions~\cite{Martin:2009iq}, ignoring the uncertainties due to PDFs and the strong coupling constant. The strong coupling constant is run at the appropriate QCD order while the electromagnetic coupling constant is set to its value at $m_Z$, $a(m_Z) = 1/132.34$.  

\begin{figure}
\begin{center}
%\begin{minipage}[t]{\textwidth}
\includegraphics[scale=0.5]{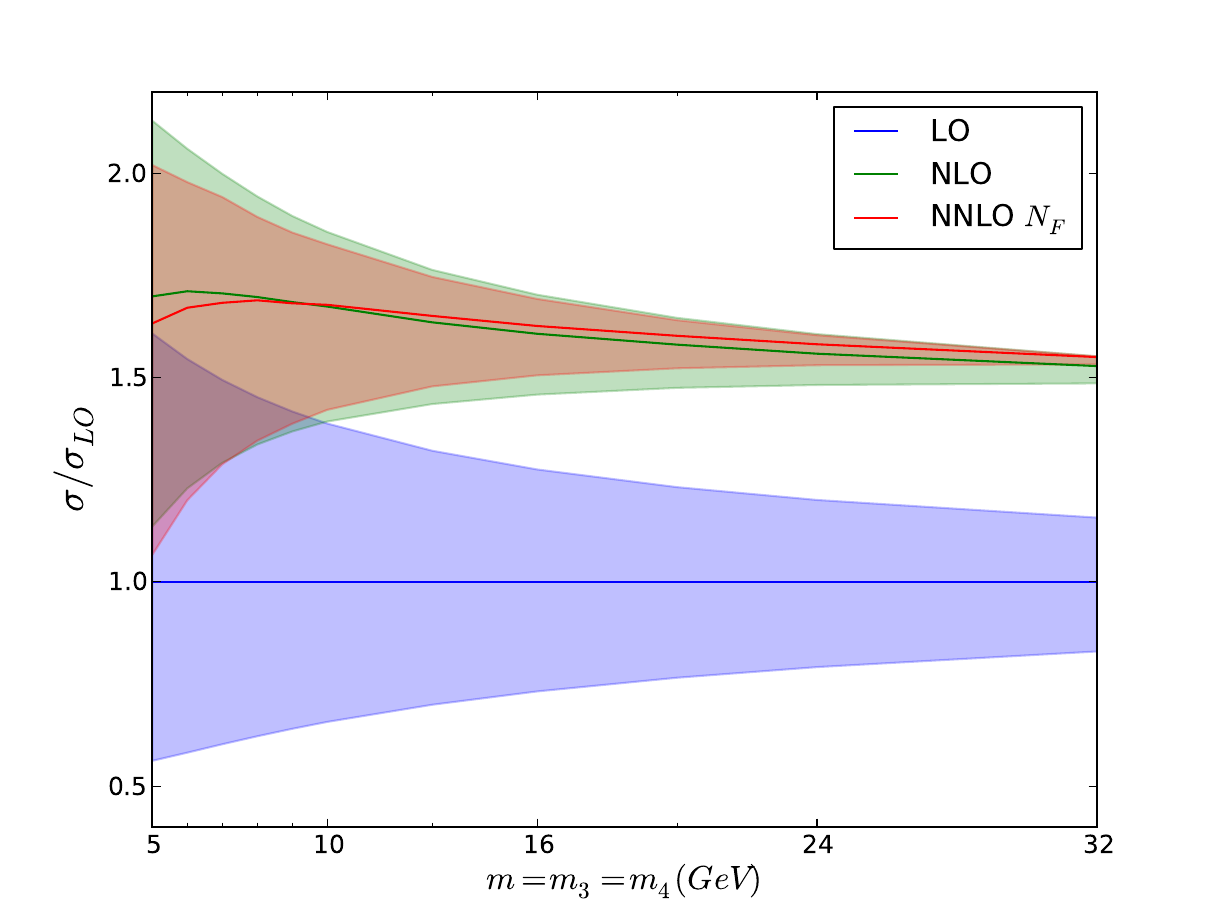}
%\end{minipage}
\caption{Scale variation at LO, NLO and NNLO as a function of the photon virtualities, here taken to be equal.}
\label{fig:scale}
\end{center}
\end{figure}

The total cross section depends on the virtualities of the off-shell photons, $m_3 = \sqrt {p_3^2}$, $m_4= \sqrt {p_4^2}$. 
 First, we set the virtualities of the two photons equal and study the scale uncertainty of the NLO and NNLO K-factors as a function of the common photon virtuality, in  fig.~\ref{fig:scale}. For photons that are widely off-shell, i.e. with $m_{3,4}>10$GeV, the NNLO corrections are at the per mille level and the NNLO scale uncertainty is reduced, implying a satisfactory perturbative convergence for the process. As the limit of on-shell photons is approached the LO cross-section blows up and so does its scale uncertainty. This is expected, since we do not impose any final-state cuts on the two photons.

\begin{figure}
\includegraphics[scale=0.41]{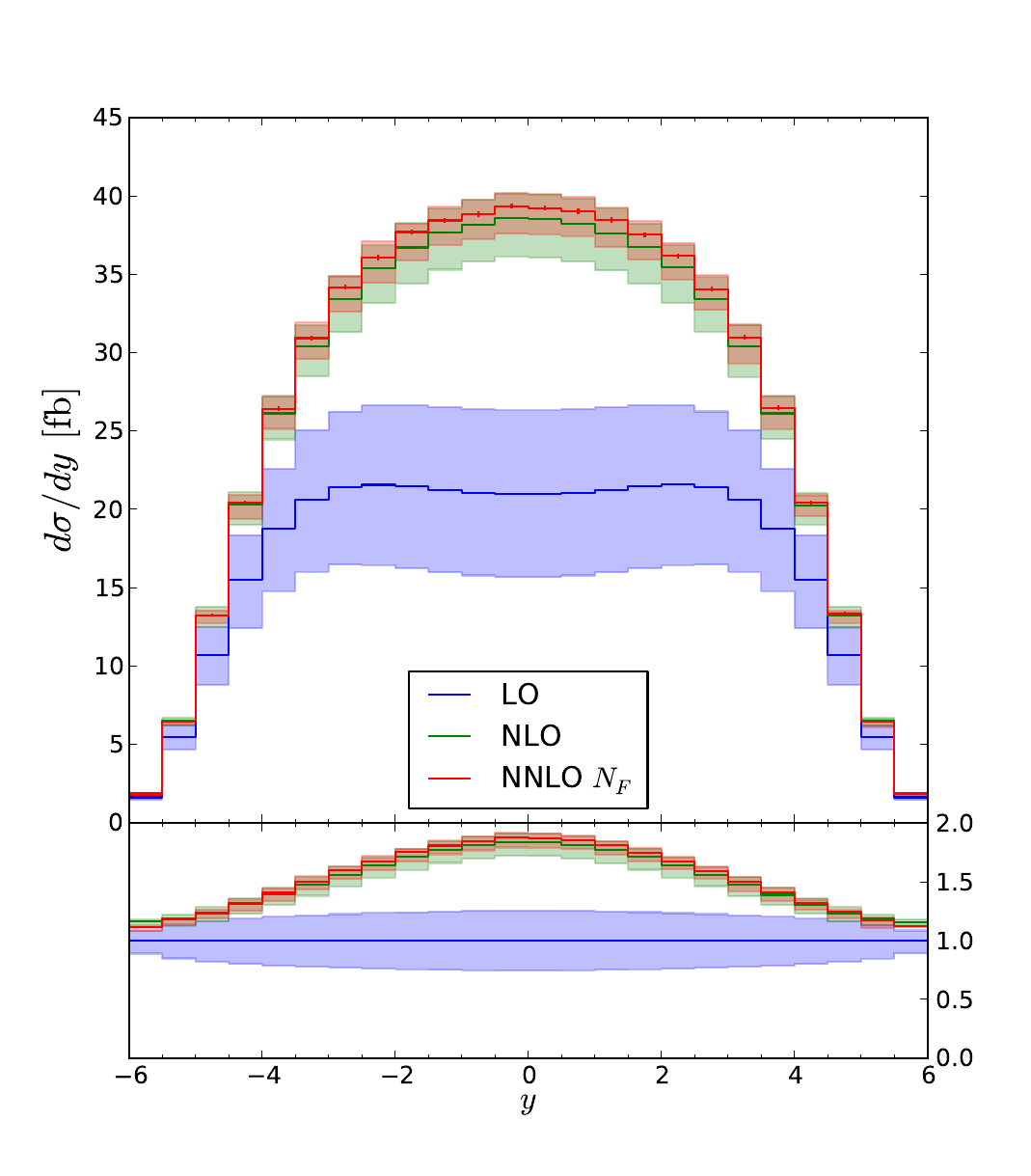}
\includegraphics[scale=0.41]{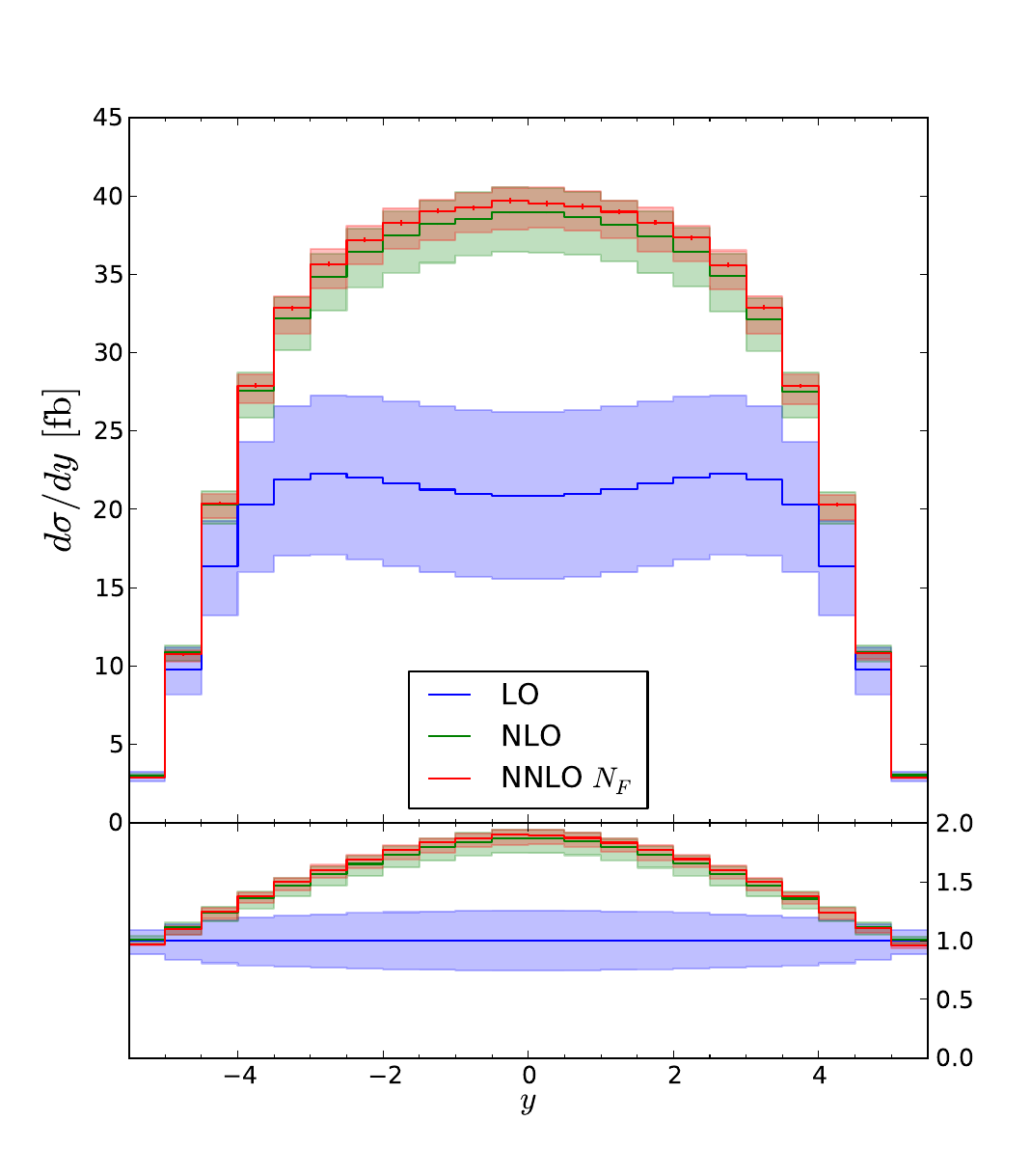}
\caption{Pseudo-rapidity distribution of the two off-shell photons with virtualities $m_3 = 30$GeV (left) and $m_4 = 15$GeV (right).}
\label{fig:rapidities}
\end{figure}

\begin{figure}
\center
\includegraphics[scale=0.41]{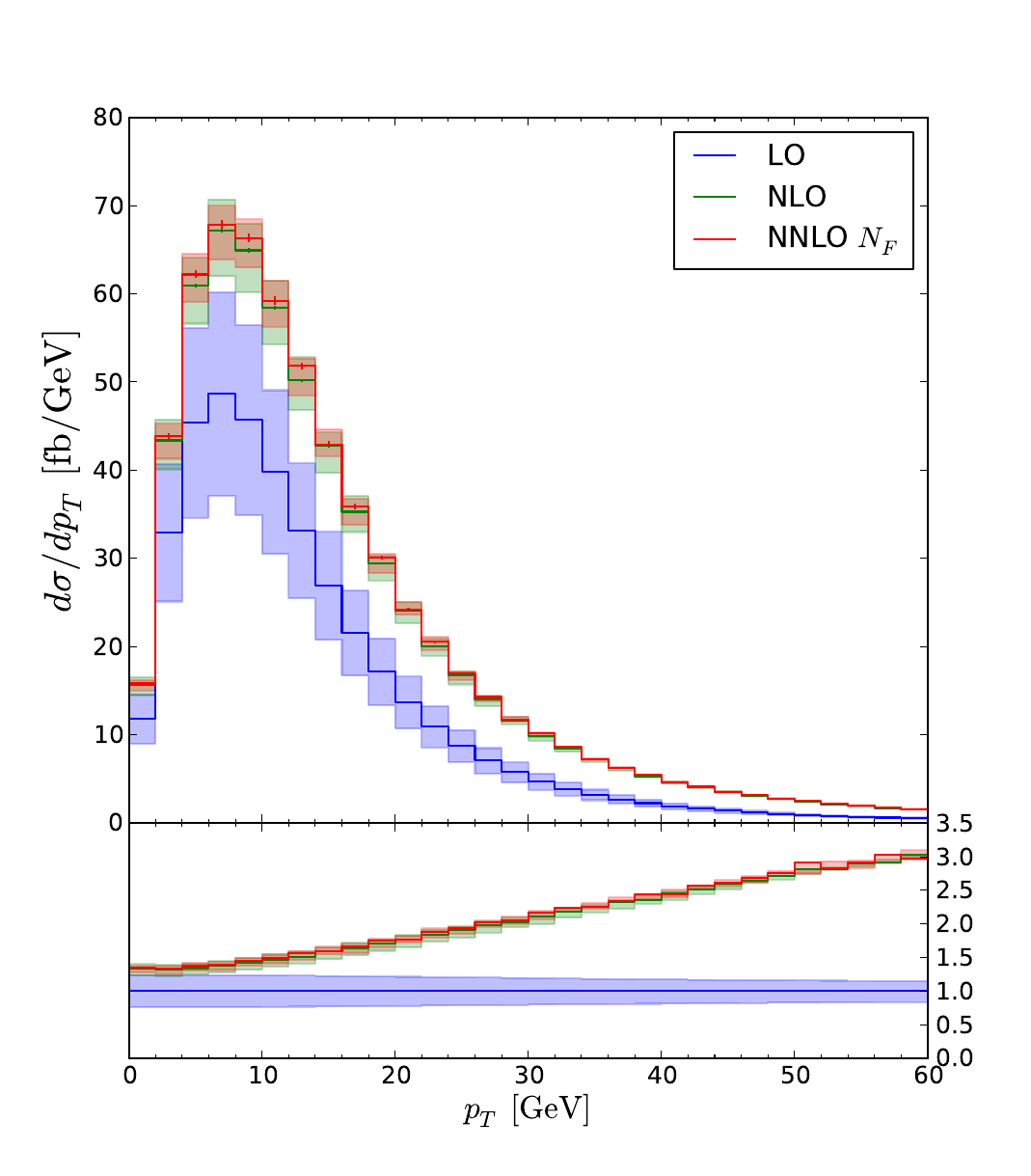}
\includegraphics[scale=0.41]{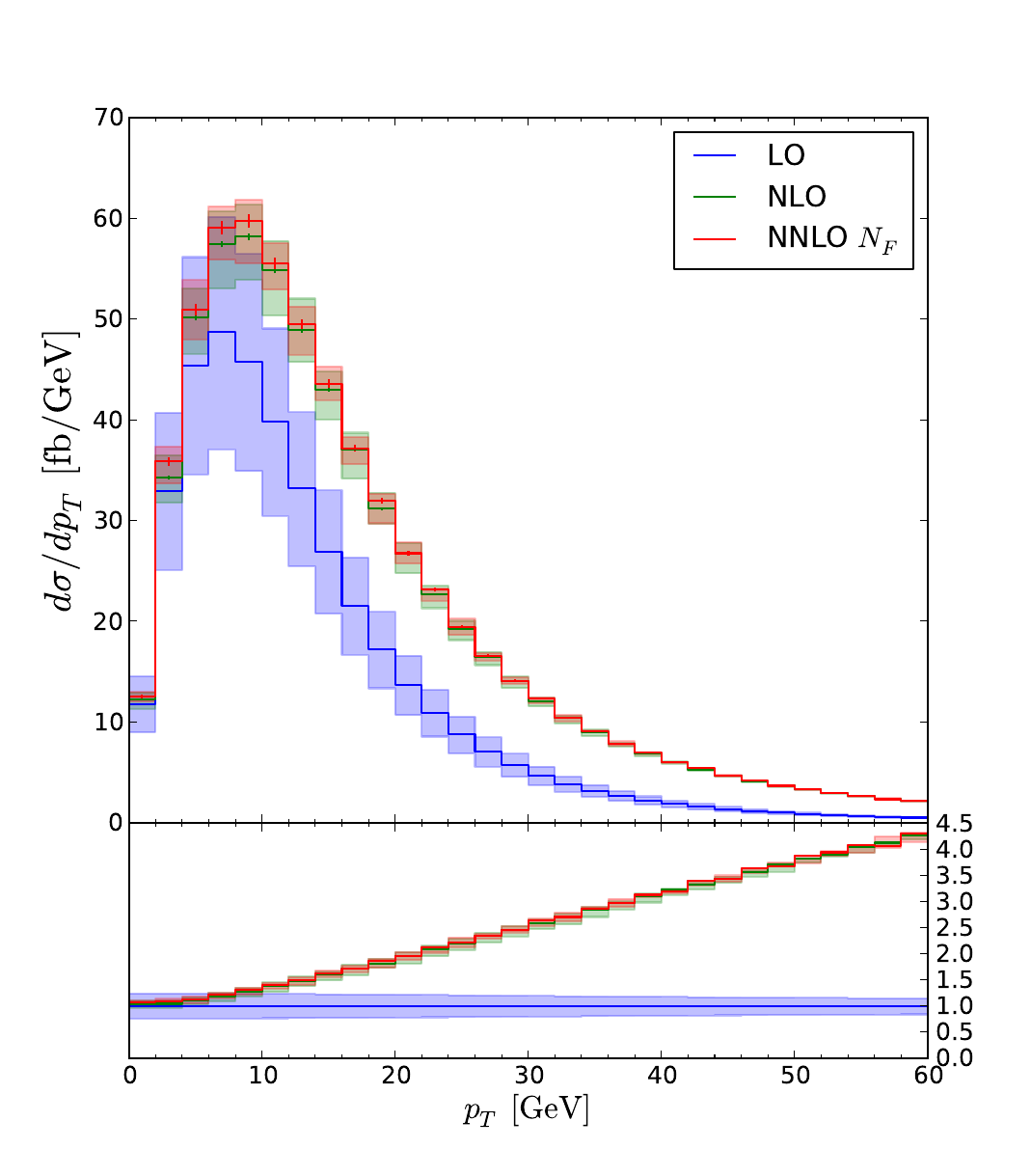}
\caption{Transverse momentum distribution of the two off-shell photons with virtualities $m_3 = 30$GeV (left) and $m_4 = 15$GeV (right).}
\label{fig:pts}
\end{figure}

Next we turn to differential distributions for unequal photon virtualities. We set 
\beq
{m_3} = 15\textrm{GeV},\qquad{m_4}=30\textrm{GeV}.
\eeq

In fig.~\ref{fig:rapidities}, we present the rapidity distributions of the two photons at each order in $\alpha_s$. The  transverse momentum distributions for the two photons can be seen in fig.~\ref{fig:pts}. The  uncertainty due to the renormalization and factorization scales is shown as shaded regions in the figures. The scales are kept equal and varied in the interval 
\beq
\mu_r=\mu_f \in [10,40]\;\textrm{GeV}.
\eeq
We note that while the NLO contribution changes the shape of the transverse momentum distributions, an effect that is more pronounced in the high transverse momentum region, the NNLO contribution does not induce any further changes. The rapidity distributions at NNLO also follow closely the NLO pattern.

\begin{figure}
%\begin{minipage}[t]{\textwidth}
\center
\includegraphics[scale=0.5]{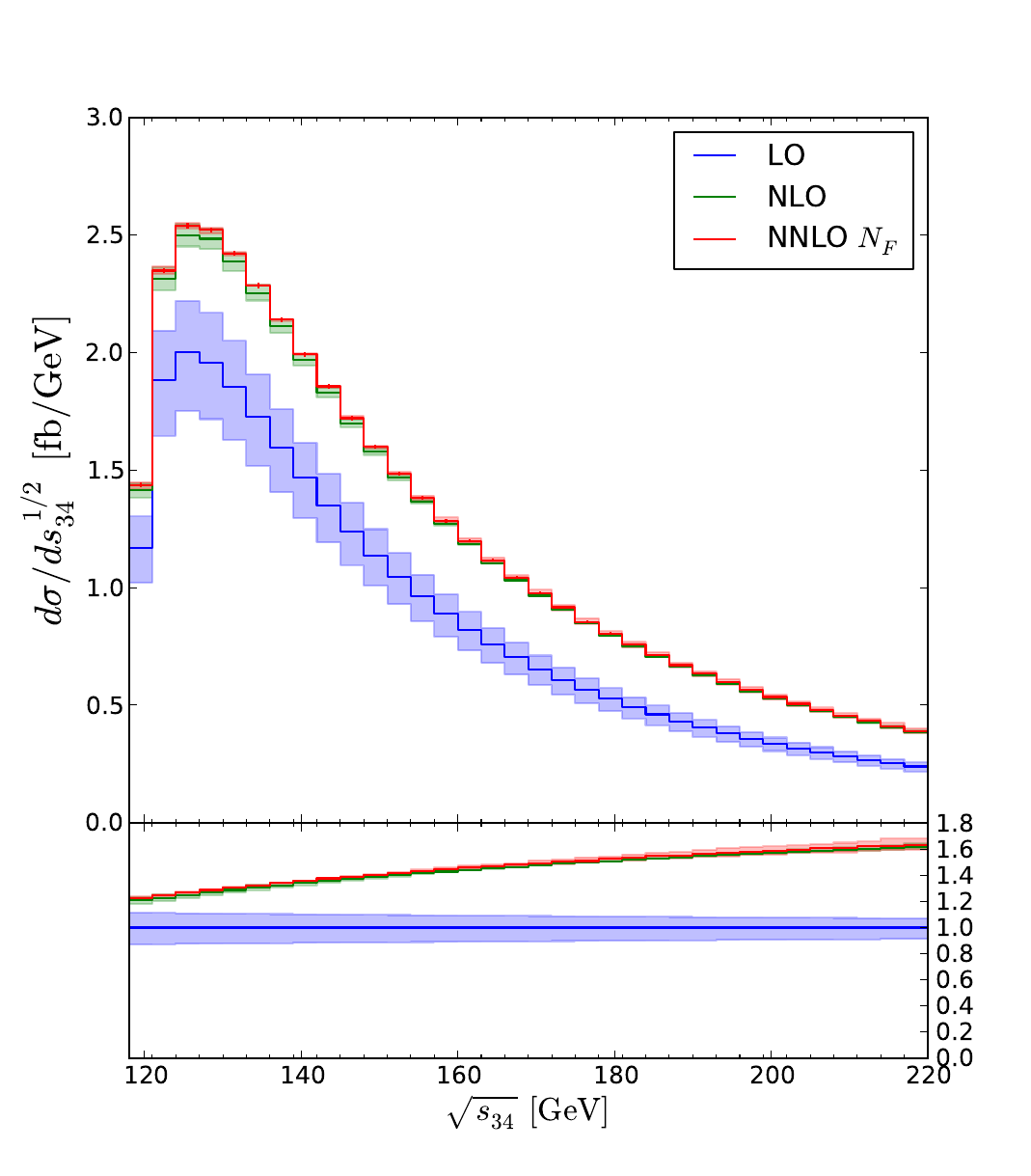}
%\end{minipage}
\caption{The invariant mass distribution of the diphoton pair, with ${m_3}=91.188$GeV and ${m_4}=27$GeV.}
\label{fig:bginvmasss}
\end{figure}

Off-shell diphoton production contributes as a background, along with Z pair production, to the Higgs boson measurements in the golden channel $pp\to H\to ZZ^*\to l_1^+l_1^-l_2^+l_2^-$. In that case the invariant mass of the two photons must be in a window of several GeV around the Higgs mass of $125$GeV. We therefore set the virtualities of the photons to ${m_3} = 91.188$GeV and ${m_4}=27$GeV, to simulate one on-shell and one off-shell Z boson.
The invariant mass distribution of the photon pair, shown at fig.~\ref{fig:bginvmasss}, has its peak in the $126$GeV region. The NNLO contributions are overall very small, but induce a slightly more pronounced correction at the region around the peak, and further stabilise the perturbative prediction there.

\begin{figure}
\center
\includegraphics[scale=0.36]{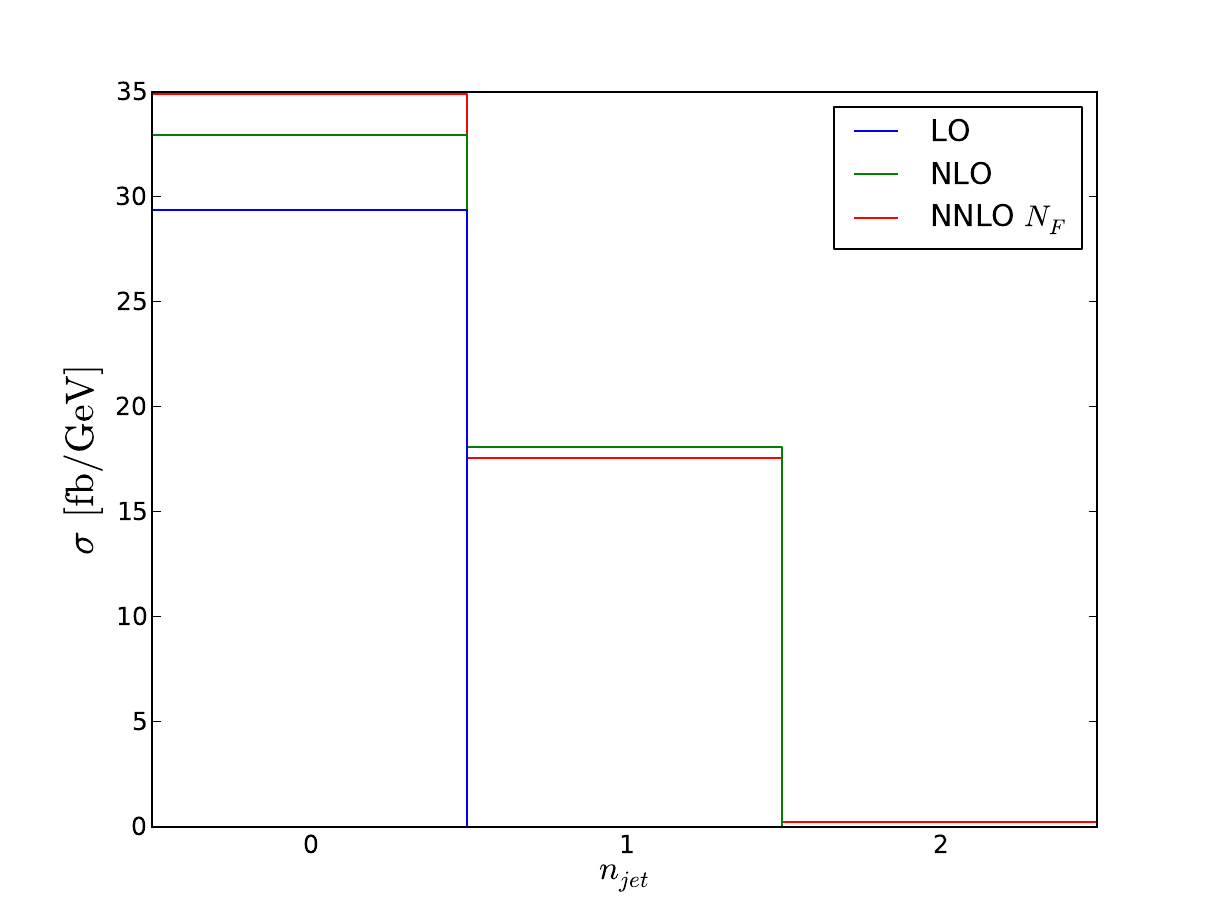}\hfill
\includegraphics[scale=0.36]{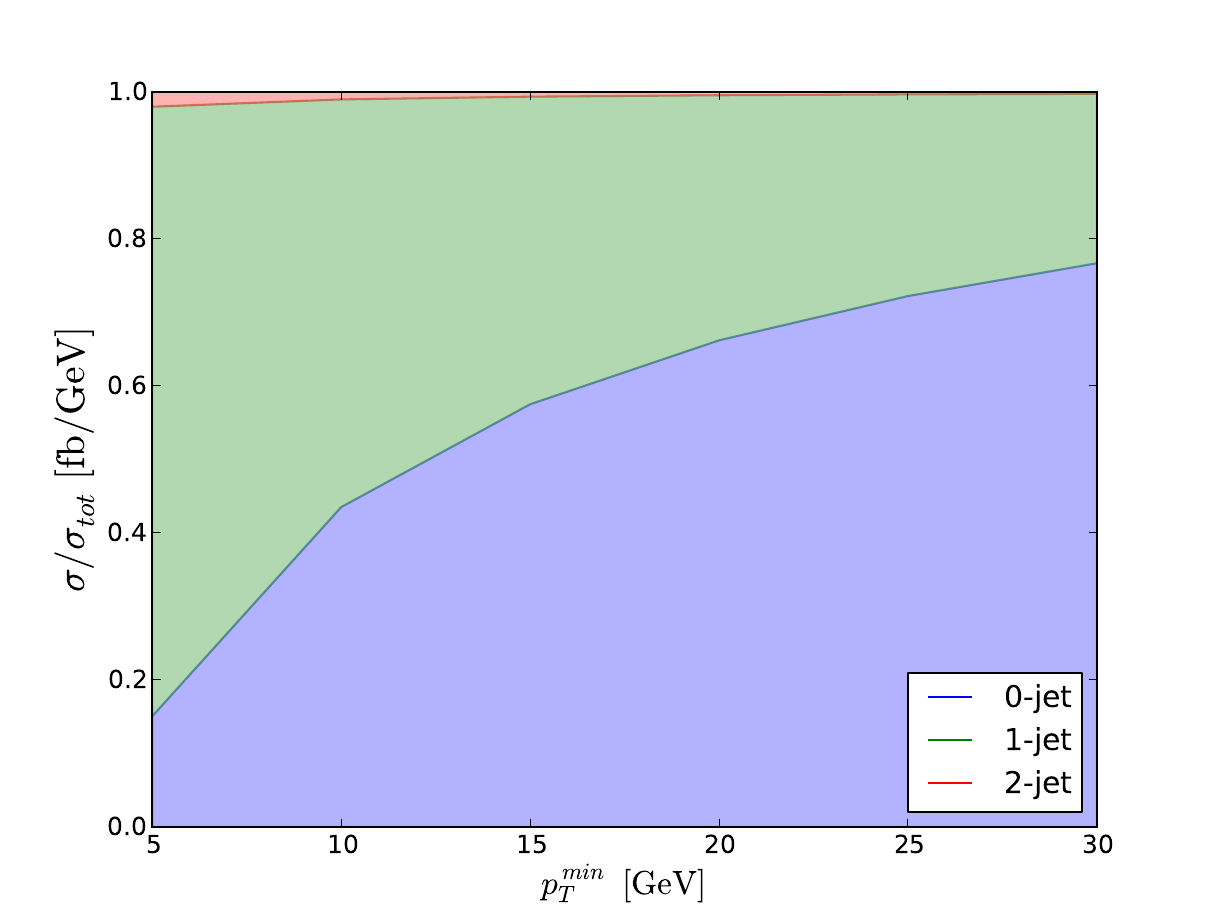} 
\caption{N-jets cross-section as a function of the perturbative order for $p_T^{min}=20$GeV (left) and as a function of $p^{min}_T$  (right), for $ {m_3} = 91.188$GeV, $ {m_4} = 27$GeV.}
\label{fig:njet}
\end{figure}

The N-jet cross section is shown in fig.~\ref{fig:njet}. We have implemented the anti-$k_T$ algorithm with cone size $R=0.7$ and a $p_T^{min}$ that defines how soft the jet is allowed to be. The 0-,1- and 2-jet cross sections for $p_T^{min}=20$GeV are shown in the left panel of fig.~\ref{fig:njet}. We observe that there is a migration of events away from the 1-jet bin at NNLO.  On the right panel we show the dependence of the size of the 0-, 1- and 2-jet bins as a function of $p_T^{min}$. In general the contribution of the NNLO cross section is at the percent level or lower.  

\begin{figure}
\center
\includegraphics[scale=0.41]{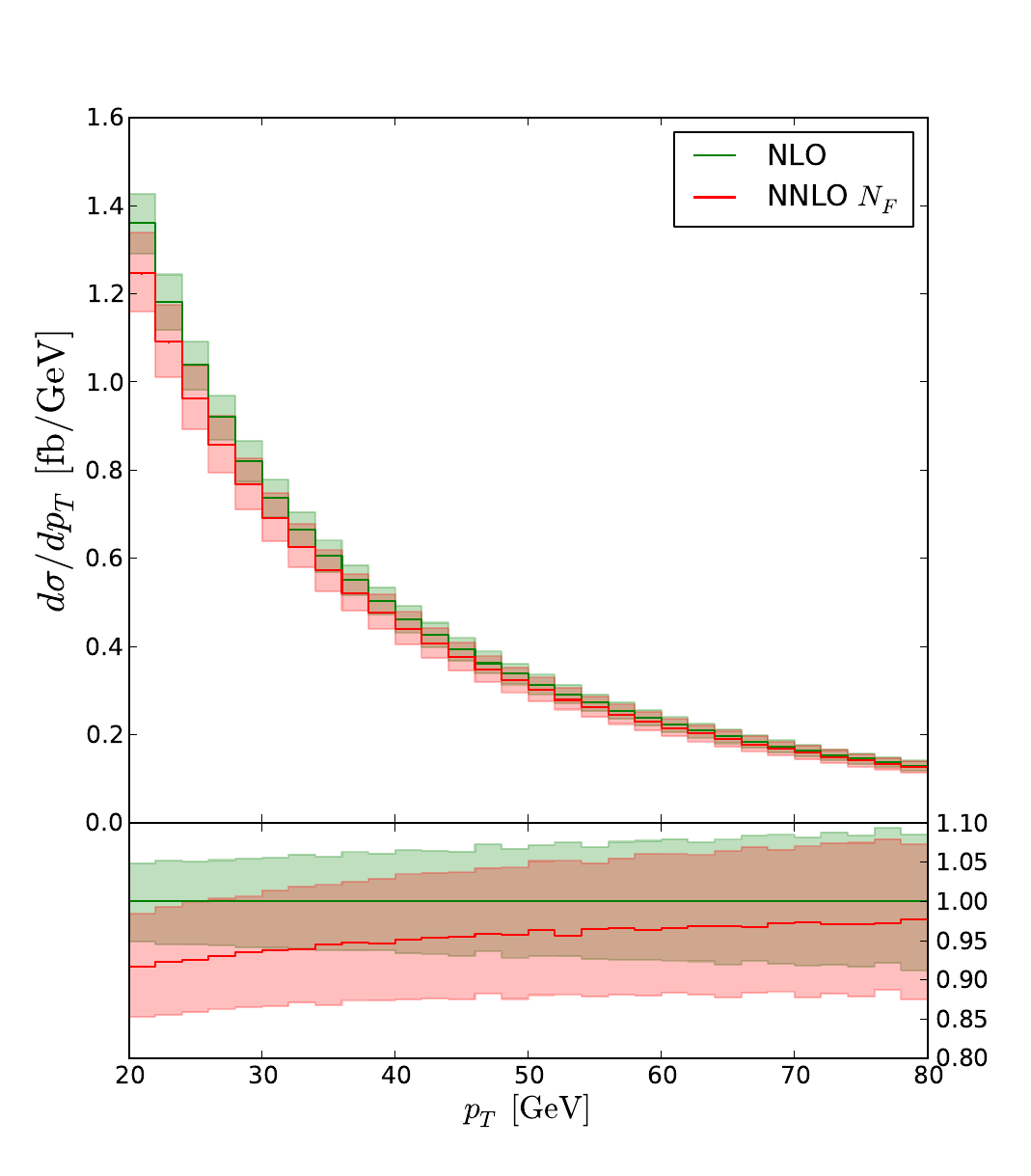} \hfill
\includegraphics[scale=0.41]{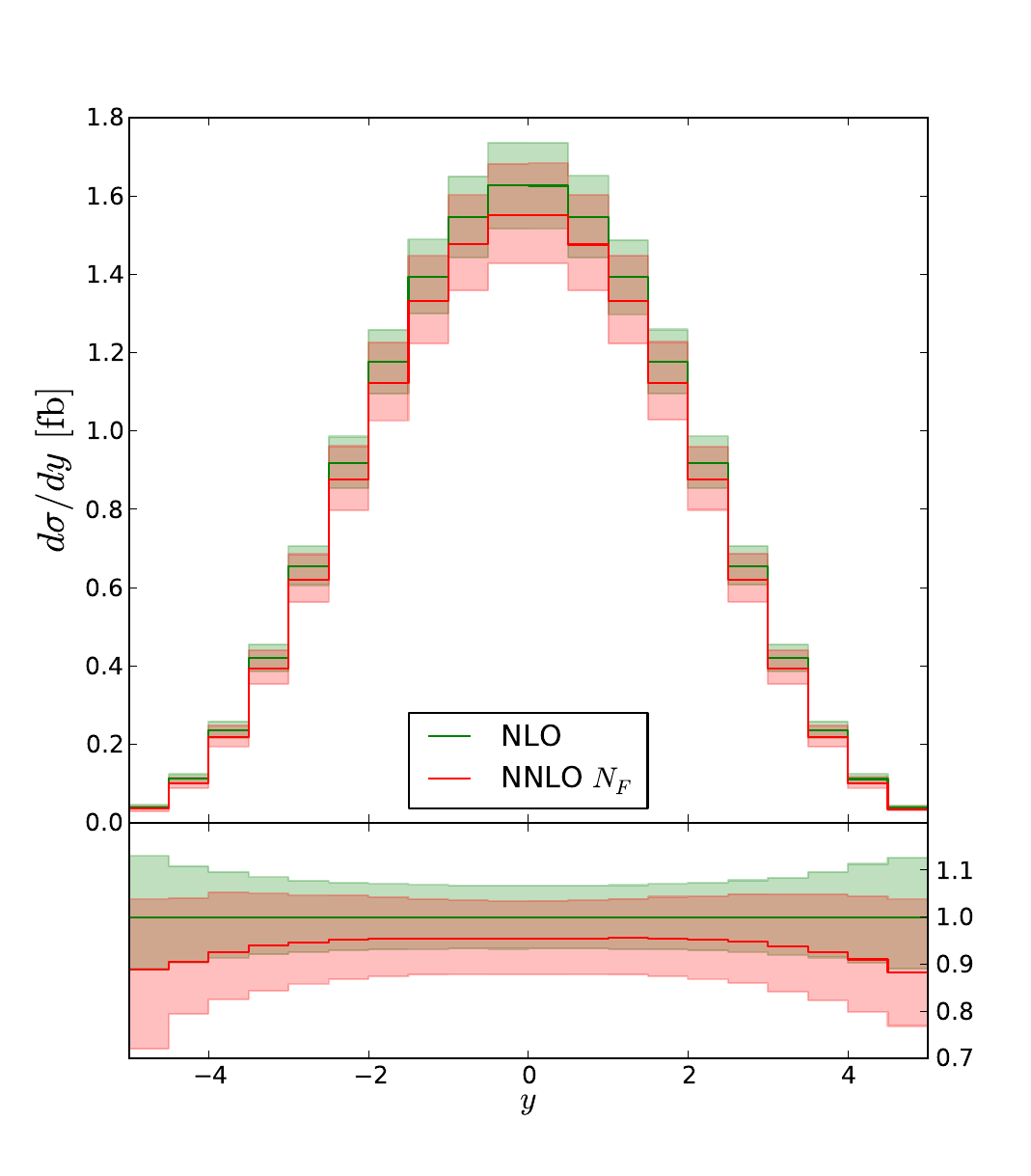}
\caption{Transverse momentum (left) and rapidity (right) of the leading jet for $ {m_3} = 91.188$GeV, $ {m_4} = 27$GeV, and $p^{min}_T = 20$GeV.}
\label{fig:leadingjet}
\end{figure}

Finally the transverse momentum and rapidity distributions of the leading jet when the jet algorithm is defined with $p_T^{min}=20$GeV is shown in fig.~\ref{fig:leadingjet}. This observable starts at NLO and the NNLO corrections are seen to be small and negative.

%% file: conclusions.tex
\section{Conclusions}\label{sec:conclusions}
We have computed the NNLO corrections to off-shell diphoton production at the large $N_F$ limit, as a first step towards a complete, fully differential NNLO computation of off-shell diboson production that is necessary for improving the simulation of backgrounds for Higgs production in the four-lepton channel at the LHC. 

We have provided explicit analytic expressions for the necessary two-loop master integrals in terms of classical polylogarithms, using direct integration methods along the lines of ref.~\cite{Chavez:2012kn}. 

We have treated the double-real radiation with a direct subtraction method where all subtraction counterterms are analytically integrated, thanks to the factorized structure of the singular limits in this process. The approach described is currently restricted to $N_F$-type contributions, but is independent of the specific, colourless  Born-level final state. 

We have implemented the NNLO corrections in a fully differential partonic Monte Carlo code and provided selected differential distributions, demonstrating the numerical stability of both the double virtual and the double real contributions, in anticipation of a complete diboson computation.

%% file: acknowledgments.tex
\section*{Acknowledgements}
We are grateful to Raoul Rontsch for his assistance in comparisons with MCFM. This work was supported by the ERC Starting Grant IterQCD and the FP7 Marie Curie Initial Training Network ``LHCPhenoNet'' (PITN-GA- 2010-264564).

%% file: Appendix_Basis.tex
\section{Construction of the set of basis functions}
\label{app:basis} 
In this appendix we discuss the construction of the set of basis functions introduced in Section~\ref{sec:Fede}. More precisely, we construct a linearly independent set of polylogarithmic functions (up to weight four) with prescribed branch cuts reflecting the branch cut structure of Feynman integrals through which all the integrals considered in this paper can be expressed. The construction of this basis of functions follows closely the discussion in ref.~\cite{Chavez:2012kn}, where this procedure was already carried out for the subset of three-point functions (see also ref.~\cite{Dixon:2013eka} for a similar discussion in the context of so-called \emph{hexagon functions}). In order to rigorously define the notion of `basis' and `linearly independence' in the context of polylogarithmic functions, we first give a very brief review of the mathematical properties of polylogarithmic functions in general before discussing the construction of the basis.

\subsection{A lighting review of the Hopf algebra of multiple polylogarithms}
In this section we provide a lightning review of the Hopf algebra of multiple polylogarithms, as it plays a central role in the construction of the basis. First, multiple polylogarithms form a shuffle algebra, 
\beq
G(\vec a;\zz)\,G(\vec b;\zz) = \sum_{\vec c\in \vec a\shuffle \vec b}G(\vec c;\zz)\,,
\eeq
where $\vec a\shuffle\vec b$ denotes the set of all \emph{shuffles} of $\vec a$ and $\vec b$, i.e., the set of all mergers of $\vec a$ and $\vec b$ that preserve the relative orderings inside $\vec a$ and $\vec b$. In the following we denote this algebra by $\mathcal{H}$. The algebra $\mathcal{H}$ is obviously graded by the weight,
\beq
\mathcal{H} = \bigoplus_{n=0}^\infty\mathcal{H}_n\,,\qquad \textrm{with }\mathcal{H}_m\cdot\mathcal{H}_n\subset\mathcal{H}_{m+n}\,,
\eeq
where $\mathcal{H}_n$ denotes the $\mathbb{Q}$-vector space spanned by all multiple polylogarithms of weight $n$, and we set $\mathcal{H}_0=\mathbb{Q}$.

Moreover, $\mathcal{H}$ can be equipped with a coproduct, turning it into a Hopf algebra. In the following we refrain from giving a detailed discussion of the Hopf algebra structure and only concentrate on the essentials that we will need in the following. In a nutshell (and very loosely speaking), a coproduct is a linear map $\Delta: \mathcal{H} \to \mathcal{H}\otimes\mathcal{H}$ that preserves the weight and the algebra structure\footnote{In practise, $\mathcal{H}$ is only defined modulo $i\pi$, and we are working with the $\mathcal{H}$-comodule $\mathcal{A}=\mathbb{Q}[i\pi]\otimes_{\mathbb{Q}}\mathcal{H}$, and $\Delta$ is a comodule map $\Delta:\mathcal{A}\to\mathcal{A}\otimes\mathcal{H}$ with $\Delta(i\pi) = i\pi\otimes 1$~\cite{Brown:2011ik,Duhr:2012fh}. Since this distinction does not change the discussion in the following, we prefer not to make this technical distinction at this point in order not to clutter the discussion.}. For example, for the classical polylogarithms and the ordinary logarithms we have
\beq\begin{split}
\Delta(\log \zz) &\,= 1\otimes \log \zz + \log \zz\otimes 1\,, \\ 
\Delta(\Li_n(\zz)) &\,= 1\otimes \Li_n(\zz) + \sum_{k=0}^{n-1}\Li_{n-k}(\zz)\otimes \frac{\log^k\zz}{k!}\,.
\end{split}\eeq
The advantage of the coproduct lies in the fact that it allows one to decompose a multiple polylogarithm of a specific weight into pairs of lower weight objects, for which properties like functional equations are already known. In addition, this decomposition can be iterated $\mathcal{H}\to \mathcal{H}\otimes\mathcal{H}\to \mathcal{H}\otimes\mathcal{H}\otimes\mathcal{H}\to\ldots$, allowing one to decompose the functions into more and more combinations of functions of lower weight (which we will consider `simpler' in the following). In the following we denote the by $\Delta_{n_1,\ldots,n_k}$ the component of the the coproduct in $\mathcal{H}_{n_1}\otimes\ldots\otimes \mathcal{H}_{n_k}$. For a multiple polylogarithm of weight $n$ this decomposition naturally stops when the function has been decomposed into an $n$-fold tensor product of functions of weight one, i.e., ordinary logarithms for which all identities are known. This maximal iteration of the coproduct is known as the symbol map in the literature~\cite{Goncharov.A.B.:2009tja, Chen:1977oja, Brown:2009qja, Goncharov:2010jf,Duhr:2011zq}.

The coproduct also encodes information on the discontinuities and the derivatives of a function. More precisely, discontinuities are encoded in the first factor of the coproduct, while derivatives only act on the second factor~\cite{Duhr:2012fh},
\beq\label{eq:disc_der}
\Delta(\textrm{Disc}F) = (\textrm{Disc}\otimes \textrm{id})\,\Delta(F) {\rm~~and~~}
\Delta\left(\frac{\partial}{\partial \zz}F\right) = \left( \textrm{id}\otimes \frac{\partial}{\partial \zz}\right)\,\Delta(F) \,.
\eeq

\subsection{Construction of the basis}
We now exploit the concepts reviewed in the previous section to construct a basis of functions through which all the integrals presented in this paper can be expressed. The discussion follows very closely the discussion in ref.~\cite{Chavez:2012kn}, so we will be brief and online outline the main steps. Either by analysing explicit results for the integrals or by analysing the singularities of the differential equations satisfied by the master integrals, we see that the symbols of the master integrals have all their entries drawn from the set
\beq
A_4=\{\zz,\,\zp,\,w,\,1-\zz,\,1-\zp,\,\zz-\zp,\,u-w,\,v-w,\,w+\zz-\zz\zp,\,w+\zp-\zz\zp\}\,,
\eeq
where $u=\zz\zp$, $v=(1-\zz)(1-\zp)$ and $w$ were defined in Section~\ref{sec:Fede}. Note that $A_4$ contains a subset $A_3=\{\zz,\,\zp,\,1-\zz,\,1-\zp,\,\zz-\zp\}$, which corresponds to the case of the three-point functions considered in ref.~\cite{Chavez:2012kn}. Moreover, Cutkosky's rules imply that the Feynman integrals considered in this paper can only have branch cuts starting at point where $u,v,w=0$. Let from now on $\mathcal{H}$ denote the Hopf algebra of all polylogarithmic functions whose symbols have all their entries drawn from the set $A_4$, and $\mathcal{H}'$ its subalgebra consisting of all functions having at most the branch cuts prescribed by Cutkosky's rule. Note that $\mathcal{H}'$ is manifestly graded by the weight. Our goal is to find for every weight $n$ (up to weight four) a basis for $\mathcal{H}'_n$. In addition, we require this basis to be as `simple as possible', i.e., we require that the product of two basis functions of weight $m$ and $n$ be an element of the basis of weight $m+n$.

A basis for $\mathcal{H}'$ can now be constructed recursively in the weight. Indeed, since we know from eq.~\eqref{eq:disc_der} that discontinuities are encoded in the first entry of the coproduct, we conclude that\footnote{Technically speaking, $\mathcal{H}'$ is an $\mathcal{H}$-comodule.}
\beq\label{eq:first_entry}
\Delta(\mathcal{H}') \subset \mathcal{H}'\otimes\mathcal{H}\,.
\eeq
Equation~\eqref{eq:first_entry} is known as the \emph{first entry condition}~\cite{Chen:1977oja}. In the rest of this section we discuss how the first entry condition can be used to construct a basis for $\mathcal{H}'$ recursively in the weight, following the procedure of ref.~\cite{Chavez:2012kn} (see also ref.~\cite{Brown:2004}).

Let us start with weight one. It is easy to see that a basis for $\mathcal{H}_1$ is given by 
\beq\begin{split}
\mathcal{B}_1=\{&\log \zz,\,\log \zp,\,\log w,\,\log(1-\zz),\,\log(1-\zp),\,\log(\zz-\zp),\\
&\log(u-w),\,\log(v-w),\,\log(w+\zz-\zz\zp),\,\log(w+\zp-\zz\zp)\}\,,
\end{split}\eeq
and a basis for the subspace $\mathcal{H}'_1$ is
\beq
\cB'_1 = \{\ln{u},\,\ln{v},\,\ln{w}\}.
\eeq

Next, we want to construct a basis for $\mathcal{H}'_2$. From eq.~\eqref{eq:first_entry} we know that
\beq
\Delta_{1,1}(\mathcal{H}'_2) \subset \mathcal{H}'_1\otimes \mathcal{H}_1\,,
\eeq
and it is clear that a basis for $\mathcal{H}'_1\otimes \mathcal{H}_1$ is given by
\beq
\cB_{1,1} = \{b'\otimes b\,|\,b'\in \mathcal{B}'_1{\rm~and~} b\in\mathcal{B}_1\}\,.
\eeq
However, not every element of $\mathcal{H}'_1\otimes \mathcal{H}_1$ corresponds to a function in $\mathcal{H}'_2$. Let us illustrate this with an example. Consider the element $\log (\zz\zp)\otimes\log \zz \in \mathcal{H}'_1\otimes \mathcal{H}_1$, and suppose that there is a function $f \in \mathcal{H}'_2$ such that $\Delta_{1,1}(f) = \log (\zz\zp)\otimes\log \zz$. Using eq.~\eqref{eq:disc_der} and the fact that for the total differential $d^2=0$, we obtain a contradiction, because
\beq
0 = \Delta_{1,1}(d^2f) = d\log(\zz\zp)\wedge d\log \zz = d\log\zp\wedge d\log \zz \neq 0\,.
\eeq
It can however be shown that
\beq
\Delta_{1,1}(\mathcal{H}'_2) = \{\xi\in\mathcal{H}'_1\otimes \mathcal{H}_1\,|\,(d\wedge d)\xi=0\}\,.
\eeq
This is known as the \emph{integrability condition}. We can thus write down the most general linear combination of elements in $\cB_{1,1}$ and then solve the integrability condition. The a basis for the solution space of this problem is at the same time a basis for $\Delta_{1,1}(\mathcal{H}'_2)$. Every basis element of corresponds to a basis element in $\mathcal{H}'_2$, and it is straightforward to find the corresponding function. Note that we also need to add all those elements $\xi$ such that $\Delta_{1,1}(\xi) = 0$. In our case there is just one such element, namely $\zeta_2$. Carrying out this procedure at weight two, we find that, besides all possible products of elements of $\cB'_1$, there are 4 new basis elements of weight two, which we choose as
\beq\label{eq:B2}
\zeta_2\,, \quad \cP_2(\zz)\,,\quad \textrm{Li}_2\left(1-
\frac{u}{w}\right)\,,\quad \textrm{Li}_2\left(1-
\frac{u}{w}\right)\,,
\eeq
in agreement with the result quoted in section~\ref{sec:Fede}.

This procedure immediately carries over to higher weight. Indeed, assume that we have constructed a basis $\cB'_{n-1}$ of $\mathcal{H}'_{n-1}$. The first entry condition and the integrability conditions imply that
\beq
\Delta_{n-1,1}(\mathcal{H}'_n) = \{\xi\in\mathcal{H}'_{n-1}\otimes \mathcal{H}_1\,|\,(d\wedge d)\xi=0\}\subset \mathcal{H}'_{n-1}\otimes \mathcal{H}_1\,,
\eeq
and a basis for $\mathcal{H}'_{n-1}\otimes \mathcal{H}_1$ is given by
\beq
\cB_{n-1,1} = \{b'\otimes b\,|\,b'\in \mathcal{B}'_{n-1}{\rm~and~} b\in\mathcal{B}_1\}\,.
\eeq
Starting from the most general linear combination of elements in $\cB_{n-1,1}$, we can solve the integrability condition and obtain a basis for $\Delta_{n-1,1}(\mathcal{H}'_n)$. To every basis element corresponds a function in $\mathcal{H}'_n$ that can easily be constructed. We find that (up to weight four) the basis elements can be expressed in terms of multiple polylogarithms $G(\vec a;x)$ with
\begin{enumerate}
\item
 $a_i\in\left\lbrace 0,u,v,\zz(-1+\zp),\zp(-1+\zz) \right\rbrace$ and $x=w$, 
 \item
$a_i\in \left\lbrace 0,\frac{1}{\zz},\frac{1}{\zp} \right\rbrace$ and $x=1$, 
\item $a_i\in\left\lbrace 0,\frac{1}{1-\zz},\frac{1}{1-\zp} \right\rbrace$ and $x=1$.
\end{enumerate}
Up to weight four, and omitting products of lower weight functions, we find the following basis functions,
\beq\begin{split}\label{eq:B3B4}
\cB_3 = &\Big\lbrace \cR_3^+(\zz,w), \cR_3^-(\zz,w), \cQ_3(\zz), \cP_3(\zz), \cP_3(1-\zz), \zeta_3  \\ & \;\Li_3\left(1-\frac{u}{w}\right), \Li_3\left(1-\frac{v}{w}\right), \Li_3\left(1-\frac{w}{u}\right), \Li_3\left(1-\frac{w}{v}\right) \Big\rbrace\,,\\
\cB_4 = &\Big\lbrace \mathcal{L}i_4(u,w),\mathcal{L}i_4(v,w),
 \\ & \; \cR_4^{+;1\ldots,5}(\zz,w), \cR_4^{-;1\ldots,4}(\zz,w), \cR_4^{+;1,2}(1-\zz,w), \cR_4^{-;1}(1-\zz,w),  \\ & \; \cQ_4^+(\zz), \cQ_4^+(1-\zz), \cQ_4^-(\zz), \cP_4(\zz), \cP_4(1-\zz), \cP_4(1-1/\zz),   \\ & \;  \Li_4\left(1-\frac{u}{w}\right), \Li_4\left(1-\frac{v}{w}\right), \Li_4\left(1-\frac{w}{u}\right), \Li_4\left(1-\frac{w}{v}\right) \Big\rbrace.
\end{split}\eeq
The basis functions of weight three were already defined in section~\ref{sec:Fede}. The new basis functions of weight four are rather lengthy, and are given as ancillary files attached to the {\tt arXiv} submission.

Let us conclude this section by making some comments about our choice of basis functions. 
\begin{enumerate}
\item All basis functions are chosen such that they are manifestly real in the Euclidean region where $\lambda(1,u,v)<0$, and thus $\zz$ and $\zp$ complex conjugate to each other. Note that, similar to the case of the three-point functions considered in ref.~\cite{Chavez:2012kn},  this implies for fixed values of $w$ all basis functions are single-valued in the complex $\zz$ plane. The analytic continuation to other regions can be performed using the techniques described in section~\ref{sec:analytic_cont}.
\item We already discussed that our basis is `as simple as possible', in the sense that at every weight we have to add all possible products of lower weight basis function the new \emph{indecomposable} functions defined in eqs.~\eqref{eq:B2} and~\eqref{eq:B3B4}. One could ask whether the inverse is also true, i.e., whether it is possible to find a linear combination of indecomposable functions which can be expressed in terms of products of lower weights (not necessarily \emph{basis} functions of lower weight). It can be checked that this is not so. Indeed, in ref.~\cite{Ree,Griffing,Duhr:2011zq} a set of projectors (acting on symbols) was defined whose kernels are precisely generated by products of lower weights. It is then easy to check that there is no non-trivial decomposable linear combination of indecomposable basis functions.
\item The parameterization~\eqref{eq:uvdef} induces a $\mathbb{Z}_2$ symmetry on the space of functions which acts by interchanging $\zz$ and $\zp$, or, equivalently, changes the sign of the square root $\sqrt{\lambda(1,u,v)}$. All our basis functions are eigenfunctions under this symmetry.
\item We already noted that we have the inclusion $A_3\subset A_4$, corresponding to the fact that the massless three point functions are subtopologies of the four-point functions considered here. In ref.~\cite{Chavez:2012kn} a basis up to weight four for these three-point functions was constructed. Our basis has been chosen such that all basis functions of ref.~\cite{Chavez:2012kn} appear explicitly as basis elements in our basis. 
\end{enumerate}

%% file: Appendix_Masters.tex
\section{Computation of the master integrals}
\label{app:masters_computation}

In this appendix we illustrate how we computed the four-point master integrals defined in section~\ref{sec:Fede}. The method used to compute the integrals follows the algorithm introduced in ref.~\cite{Brown:2008um} (see also ref.~\cite{Ablinger:2012qm, Bogner:2012dn, Anastasiou:2013srw,Panzer:2014gra,Panzer:2014caa,Ablinger:2014yaa,Bogner:2014mha}), which, under certain conditions which are always satisfied in the following, allows to perform the integrations one at the time. 

In a nutshell, the general procedure is the following. 
If the integral is finite as $\eps\to 0$, we can expand the Feynman parameterized integral in $\ep$ under the integration sign. At each order in $\eps$ we then obtain integrands composed of (logarithms of) rational functions of the Feynman parameters and the external scales. If in addition we find an ordering of the Feynman parameters such that, after integrating over the first $k$ Feynman parameters, all the polynomials appearing in the integrand are linear in the next Feynman parameter, then we can perform the next integration trivially using the definition of multiple polylogarithms, eq.~\eqref{G_def}. Several explicit algorithms to perform the integrations exist~\cite{Brown:2008um,Ablinger:2012qm, Bogner:2012dn, Anastasiou:2013srw,Panzer:2014gra,Panzer:2014caa,Ablinger:2014yaa,Bogner:2014mha}, and we refer to literature for the details. In the following we content ourselves to discuss the example of the four-point function $B_{2a}$ defined in section~\ref{sec:Fede}.

\subsection{A representative example: the integral $B_{2a}$}
Let us illustrate the algorithm on the representative example of the integrals $B_{2a}$, corresponding to the integral
\be
B_{2a} = e^{2\gamma\ep} \int\frac{\ud^dk\,\ud^dl}{(i\pi^{d/2})^2}\frac{1}{(k+l)^4(k+p_1)^2(k+p_1+p_3)^2(k+p_1+p_3+p_4)^2l^2} 
\ee
The integral over $l$ is a massless bubble integral and can be done in closed form
\be
\int\frac{\ud^dl}{i\pi^{d/2}}\frac{1}{[l^2]^{\nu_a}[(k+l)^2]^{\nu_b}} = (-1)^{\frac{d}{2}}(k^2)^{\frac{d}{2}-\nu_a-\nu_b}\frac{\Gamma(\nu_a+\nu_b-\frac{d}{2})\Gamma(\frac{d}{2}-\nu_a)\Gamma(\frac{d}{2}-\nu_b)}{\Gamma(\nu_a)\Gamma(\nu_b)\Gamma(d-\nu_a-\nu_b)},
\ee
After integration over the bubble, we obtain effectively a one-loop box integral where one of the propagators is raised to 
an $\ep$-dependent power. Note that this applies to all the two-loop four-point integrals considered in section~\ref{sec:Fede} and is not specific to $B_{2a}$. After Feynman parameterization of the remaining one-loop integral, we are left with the following integrals to compute
\beq
\int\limits_0^1\left(\prod_{i=1}^4\ud x_i\right)\mathrm{fp}(\nu_1,\nu_2,\nu_3,\nu_4,d)\,,
\ee
$\mathrm{fp}(\nu_1,\nu_2,\nu_3,\nu_4,d)$ is the usual Feynman parameterization of the 1-loop box integral for arbitrary powers $\nu_i$ of propagators in $d$ dimension
\eq{
\mathrm{fp}(\nu_1,\nu_2,\nu_3,\nu_4,d) &\,= \frac{(-1)^{\frac{d}{2}}\Gamma(\nu-\frac{d}{2})}{\prod_i\Gamma(\nu_i)}\delta(1-\sum_i x_i)x_1^{\nu_1-1}x_2^{\nu_2-1}x_3^{\nu_3-1}x_4^{\nu_4-1} \\ &\times(x_1+x_2+x_3+x_4)^{\nu-d}(sx_2x_4+tx_1x_3+m_3^2x_2x_3+m_4^2x_3x_4)^{\frac{d}{2}-\nu}. \nn
}
In the case of $B_{2a}$, the propagator between the legs $p_1$ and $p_2$ is raised to the power $1+\ep$,
\eq{\bs \label{fp_B2a}
\mathrm{fp}(1+\ep,1,1,1,4-2\ep) = &\,\frac{(-1)^{2-\ep}\Gamma(2+2\ep)}{\Gamma(1+\ep)}\delta\left(1-\sum x_i\right)\left(\sum x_i\right)^{3\ep} \\ 
&\times x_1^{\ep}(m_4^2x_3x_4+m_3^2x_2x_3+x_1x_3t+x_2x_4s)^{-2-2\ep}.
\es}
Next, we would like to compute this remaining integral using the algorithm outlined at the beginning of this section. The integral is, however, divergent in $d=4$ dimensions, and so are cannot naively expand in $\eps$ under the integration sign, but we first need to extract all the singularities. We first describe our method to extract the singularities, and then we illustrate the aforementioned algorithm on the resulting finite integrals.

\paragraph{Extraction of the singularities.} The integral contains overlapping singularities that need to be factorized. After all singularities are factored, we can expand the singular terms in the integrand in terms of plus-distributions, obtaining a set of finite integrals that can be expanded in $\eps$ under the integration sign. In order to extract the singularities, we use the method of non-linear mappings introduced in ref.~\cite{Anastasiou:2010pw}, which we review in the following.

We start by considering the mapping,
\be
x_i \rightarrow \frac{x_i}{\sum_j x_jA_j},
\ee
where the $A_j$ are constants.
We then obtain for the integrand in eq.~\eqref{fp_B2a},
\be
\frac{(-1)^{2-\ep}\Gamma(2+2\ep)}{\Gamma(1+\ep)}\frac{\delta\left(1-\sum x_i\right)A_1^{1+\ep}A_2A_3A_4\left(\sum x_iA_i\right)^{3\ep}x_1^{\ep}}{\left(sA_2A_4x_2x_4 + tA_1A_3x_1x_3 + m_3^2A_2A_3x_2x_3 + m_4^2A_3A_4x_3x_4\right)^{2+\ep}}.
\ee
It is possible to remove all the kinematical dependencies from the denominator by solving the system of equations \cite{Bern:1993kr}
\be
A_2A_4 = 1/s, \quad A_1A_3 = 1/t, \quad A_2A_3 = 1/m_3^2, \quad A_3A_4 = 1/m_4^2.
\ee
We obtain the solution for $s>0$
\be\label{As}
A_1 = \sqrt{\frac{m_3^2m_4^2}{st^2}}, \quad A_2 = \sqrt{\frac{m_4^2}{sm_3^2}}, \quad A_3 = \sqrt{\frac{s}{m_3^2m_4^2}}, \quad A_4 = \sqrt{\frac{m_3^2}{sm_4^2}},
\ee
and we get
\be
\frac{(-1)^{2-\ep}\Gamma(2+2\ep)}{\Gamma(1+\ep)}\frac{\delta\left(1-\sum x_i\right)A_1^{1+\ep}A_2A_3A_4\left(\sum x_iA_i\right)^{3\ep}x_1^{\ep}}{\left(x_2x_4+x_3(x_1+x_2+x_4)\right)^{2+2\ep}}.
\ee
The $\delta$ distribution can for example be solved by change of variables
\be\label{deltasubs}
x_3 = y_1, \quad x_2 = (1-y_1)y_2, \quad x_4 = (1-y_1)(1-y_2)y_3, \quad x_1 = (1-y_1)(1-y_2)(1-y_3),
\ee
where the Jacobian of the transformation is $(1-y_1)^2(1-y_2)$. Writing $\bar{y}_i = 1-y_i$ we arrive at
\be
\frac{(-1)^{2-\ep}\Gamma(2+2\ep)}{\Gamma(1+\ep)}\frac{A_1^{1+\ep}A_2A_3A_4\left(\sum x_iA_i\right)^{3\ep}\bar{y}_1^{-\ep}\bar{y}_2^{1+\ep}\bar{y}_3^{\ep}}{\left(\bar{y}_1y_2\bar{y}_2y_3+y_1\right)^{2+2\ep}},
\ee
where for the moment we did not apply the change of variables in the sum $\left(\sum x_iA_i\right)^{3\ep}$ for better readability. We obtain overlapping singularities that can be factorized completely by the following non-linear mapping
\be\label{alpha}
y_1 \mapsto \frac{y_1y_2(1-y_2)y_3}{y_1y_2(1-y_2)y_3 + (1-y_1)}.
\ee
The Jacobian is cancelled entirely and we end up with a integral free of overlapping singularities. Putting everything together,  we obtain,
\eq{\bs
B_{2a} = &-(-1)^{4-2\ep}\frac{c^2_{\Gamma}}{\ep}\frac{\Gamma(1-2\ep)\Gamma(2+2\ep)}{\Gamma(1-\ep)^2\Gamma(1+\ep)^2}\int\limits_0^1\left(\prod_{n=1}^3\ud y_i\right)y_2^{-1-2\ep}y_3^{-1-2\ep} \\ 
&\times A_1^{1+\ep}A_2A_3A_4\,\bar{y}_1^{-\ep}\bar{y}_2^{-\ep}\bar{y}_3^{\ep}\left(\bar{y}_1y_2A_2+y_1y_2\bar{y}_2y_3A_3+\bar{y}_1\bar{y}_2y_3A_4+\bar{y}_1\bar{y}_2\bar{y}_3A_1\right)^{3\ep}\,,
\es}
where the $A_i$'s are given in eq.~\eqref{As}.
Substituting the functions for the $A$'s \eqref{As} and trading the invariants $t,m_3^2,m_4^2$ for the variables $u,v,w$ we finally obtain the following representation for $B_{2a}$ (writing the $y_i$ again as $x_i$),
\eq{\bs \label{B2a_para}
B_{2a} = -&\frac{c^2_{\Gamma}}{\ep}\frac{\Gamma(1-2\ep)\Gamma(2+2\ep)}{\Gamma(1-\ep)^2\Gamma(1+\ep)^2}(-s)^{-2-2\ep}u^{-\ep}v^{-\ep}w^{-1-4\ep} \\ 						&\times\int\limits_0^1\left(\prod_{i=1}^3\ud x_i\right)b_{2a}(x_1,x_2,x_3)x_2^{-1-2\ep}x_3^{-1-2\ep},
\es}
where the function $b_{2a}(x_1,x_2,x_3)$ is non-singular inside the integration region and given by
\be
b_{2a}(x_1,x_2,x_3) = \bar{x}_1^{-\ep}\bar{x}_2^{-\ep}\bar{x}_3^{\ep}(w(u\bar{x}_1\bar{x}_2x_3+v\bar{x}_1x_2+x_1x_2\bar{x}_2x_3)+uv\bar{x}_1\bar{x}_2\bar{x}_3)^{3\ep}.
\ee
The two singularities are located in the variables $x_2$ and $x_3$ in a factorized form as intended. We then perform the expansion in $\ep$ with the help of the plus-distribution, i.e. by substituting 
\be
x_i^{-1+a_i\ep} = \frac{\delta(x_i)}{a_i\ep} + \left[\frac{1}{x_i}\right]_+ + \ord(\ep),
\ee
and we obtain four finite integrals
\eq{\bs
I_{2a}[1] &= \int\limits_0^1\ud x_1\frac{b_{2a}(x_1,0,0)}{4\ep^2}, \\
I_{2a}[2] &= -\int\limits_0^1\ud x_1\ud x_3\frac{b_{2a}(x_1,0,x_3) - b_{2a}(x_1,0,0)}{2\ep x_3^{1+2\ep}}, \\
I_{2a}[3] &= -\int\limits_0^1\ud x_1\ud x_2\frac{b_{2a}(x_1,x_2,0) - b_{2a}(x_1,0,0)}{2\ep x_2^{1+2\ep}}, \\
I_{2a}[4] &= \int\limits_0^1\ud x_1\ud x_2\ud x_3\frac{b_{2a}(x_1,x_2,x_3)- b_{2a}(x_1,0,x_3) - b_{2a}(x_1,x_2,0) + b_{2a}(x_1,0,0)}{x_2^{1+2\ep}x_3^{1+2\ep}}\,.
\es}
The sum of the four integrals represents the integral in eq.~\eqref{B2a_para} up to order $\ord(\ep)$. As each of these integrals is finite, they can be computed using the algorithm outlined at the beginning of this section. This will be illustrated in the rest of this section.

\paragraph{Doing the integrals.}
Let us do some of the integration explicitly to give a taste of the integration using multiple polylogarithms. The integral $I_{2a}[1]$ is trivial and can be integrated directly without having to expand the integrand in $\ep$. Also the integration over $x_1$ in $I_{2a}[2]$ can be performed without any trouble, but let us not do this for the sake of illustration. The coefficient of $\ep^0$ of $I_{2a}[2]$ is given by
\be
I_{2a}[2]\left(\ord(\ep^0)\right) = -\frac{1}{2}\int\limits_0^1\ud x_3\frac{3\ln{(-vx_3+wx_3+v)}+\ln{(1-x_3)}-3\ln{(v)}}{x_3},
\ee
where the dependence on $x_1$ dropped out and we are left with the integration over $x_3$. The integrand can be written in terms of multiple polylogarithms%
\be
I_{2a}[2]\left(\ord(\ep^0)\right) = -\frac{1}{2}\int\limits_0^1\ud x_3\frac{3\,G\left(\frac{v}{v-w};x_3\right) + G(1;x_3)}{x_3},
\ee
and we can readily integrate over $x_3$ using the definition of multiple polylogarithms, eq.~\eqref{G_def}. We obtain
\be
I_{2a}[2]\left(\ord(\ep^0)\right) = -\frac{3}{2}\,G\left(0,\frac{v}{v-w};1\right) - \frac{1}{2}G(0,1;1).
\ee
All the other integrals can be done in this manner.

%% file: citations.tex
\providecommand{\href}[2]{#2}\begingroup\raggedright\endgroup